\PassOptionsToPackage{table}{xcolor}
\documentclass{article}
\usepackage{iclr2027_conference,times}

\usepackage{amsmath,amsfonts,bm}

\def\eqref#1{equation~\ref{#1}}

\def\1{\bm{1}}

\DeclareMathAlphabet{\mathsfit}{\encodingdefault}{\sfdefault}{m}{sl}
\SetMathAlphabet{\mathsfit}{bold}{\encodingdefault}{\sfdefault}{bx}{n}

\usepackage[T1]{fontenc}
\usepackage[utf8]{inputenc}
\usepackage{amsmath,amssymb}
\usepackage{graphicx}
\usepackage{booktabs}
\usepackage[table]{xcolor}
\definecolor{FLHeader}{HTML}{EDF2F7}
\definecolor{FLBlue}{HTML}{5065AD}
\definecolor{FLTeal}{HTML}{398C80}
\definecolor{FLWarm}{HTML}{CD796D}
\definecolor{FLTotal}{HTML}{E6F1EE}
\definecolor{FLStripe}{HTML}{F5F8FB}
\usepackage{array,tabularx,longtable}
\newcolumntype{Y}{>{\raggedright\arraybackslash}X}
\newcolumntype{L}[1]{>{\raggedright\arraybackslash}p{#1}}
\newcolumntype{C}[1]{>{\centering\arraybackslash}p{#1}}
\usepackage{float}
\usepackage{placeins}
\usepackage{needspace}
\usepackage{microtype}
\usepackage{wrapfig}
\usepackage{xspace}
\usepackage{hyperref}
\usepackage{xurl}
\usepackage{hunyuan_preprint}

\hypersetup{
  colorlinks=true,
  linkcolor=blue!45!black,
  citecolor=green!35!black,
  urlcolor=blue!45!black,
  pdftitle={Benchmarking Coding Agents under Evolving Requirements},
  pdfauthor={Jia Feng, Alyssa Wu, Cuiyun Gao, Yunlong Wang, Songyuanyi Lu, Zhichao Hu}
}

\newcommand{\benchmark}{FeatureLoop\xspace}

\title{Benchmarking Coding Agents\\under Evolving Requirements}
\author{%
  \textbf{Jia Feng}\textsuperscript{1},
  \textbf{Alyssa Wu}\textsuperscript{1,*},
  \textbf{Cuiyun Gao}\textsuperscript{2,*},\\[3pt]
  \textbf{Yunlong Wang}\textsuperscript{1},
  \textbf{Songyuanyi Lu}\textsuperscript{1},
  \textbf{Zhichao Hu}\textsuperscript{1}\\[7pt]
  \textsuperscript{1}Hunyuan Team, Tencent\quad
  \textsuperscript{2}The Chinese University of Hong Kong\\[5pt]
  {\footnotesize\texttt{\{plusfeng, alyssawwu, joeyolwang, alinasyylu, frankiehu\}@tencent.com}}\\[2pt]
  {\footnotesize\href{mailto:cygao@cse.cuhk.edu.hk}{\texttt{cygao@cse.cuhk.edu.hk}}}\\[3pt]
  {\footnotesize\textsuperscript{*}Corresponding author}%
}

\begin{document}

\maketitle
\enlargethispage{-25.6pt}
\raggedbottom
\setcounter{topnumber}{3}
\setcounter{bottomnumber}{2}
\renewcommand{\topfraction}{0.90}
\renewcommand{\bottomfraction}{0.85}
\renewcommand{\textfraction}{0.08}
\renewcommand{\floatpagefraction}{0.72}
\makeatletter
\setlength{\@fptop}{0pt}
\setlength{\@fpsep}{18pt}
\setlength{\@fpbot}{0pt plus 1fil}
\makeatother

\begin{abstract}

Software development with coding agents often begins with a rough request that users refine after trying the implementation.
Final correctness alone does not capture the repeated guidance this process requires.
We introduce \benchmark{}, a benchmark of 61 human-reviewed tasks from 34 repositories across five primary programming languages and seven software forms.
Each task follows one development goal in a persistent workspace through \emph{evolution feedback}, which adds or clarifies requirements, and \emph{completion feedback}, which reports failures of stated requirements.
Cumulative executable checks track new behavior, repair, and preservation of earlier functionality.
We evaluate 12 frontier models and find that the strongest completes 72.1\% of tasks with both feedback types.
Across models, however, only 15.4\% of observed failed stages recover within two corrective revisions.
Further analysis reveals incomplete repairs, regressions, and model-specific differences in feedback use and token efficiency, identifying obstacles to reliable, low-intervention development.
\end{abstract}

\section{Introduction}
\label{sec:introduction}

Coding agents combine repository inspection and code editing \citep{yang2024sweagent} with command execution in sandboxed environments \citep{wang2025openhands}. Real-user records show sustained development with repeated corrections and failure reports \citep{baumann2026swechat}. Initial requests can also omit information needed for implementation \citep{vijayvargiya2026ambigswe}. Delegating the opening request therefore leaves the user with further work: trying the implementation, explaining additional requirements, and reporting errors. Turning those reports into working code remains a separate challenge, since repair success depends on the quality of the feedback \citep{olausson2024selfrepair}. The practical question is: \emph{how much repeated guidance does an agent need to turn an initial request into a complete, working implementation?}

\begin{wrapfigure}{R}{0.49\textwidth}
  \centering
  \includegraphics[width=\linewidth]{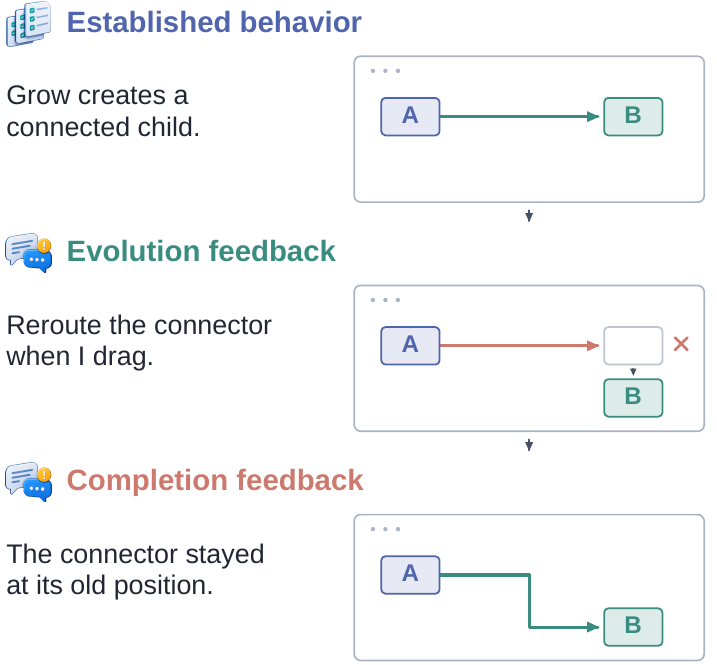}
  \caption{An Excalidraw interaction excerpt. The canvases show successive implementations; the faded shape marks the position before dragging.}
  \label{fig:excalidraw-case}
\end{wrapfigure}

Figure~\ref{fig:excalidraw-case} illustrates two reasons for returning to a coding agent. After a Grow menu can create connected shapes, the user asks for connectors that reroute during dragging. This changes what the software should do. Once rerouting has been requested, a report that a connector stays in its old position instead identifies unfinished implementation work. We call the first kind of intervention \emph{evolution feedback} and the second \emph{completion feedback}. Across both, progress must accumulate: creating a correct connector is insufficient if the next edit breaks child creation or loses its bindings. Rechecking earlier behavior after changes is the purpose of regression testing \citep{yoo2012regression}.

Existing benchmarks establish several parts of this evaluation. Executable tasks cover repository issue resolution \citep{jimenez2024swebench}, complex feature implementation \citep{zhou2026featurebench}, technical terminal workflows \citep{merrill2026terminalbench}, and vulnerability reproduction \citep{wang2026cybergym}. Interactive evaluation measures responses to user requests and corrections \citep{wu2026swetogether}, while sequential evaluation checks earlier requirements after continued modification \citep{shen2026evocodebench}. Conversational code generation further shows that solving more problems and solving them in fewer turns are distinct outcomes \citep{han2025convcodeworld}. Our focus is how these demands interact: \emph{when a user adds requirements or reports failures, does the next revision improve the complete implementation, and how much more guidance is still needed?}

We introduce \benchmark{} to make this development path measurable. Each task follows one user-facing development goal from an initial request through successive refinements of the same repository implementation. Later requests specify its capabilities, constraints, and interactions, while each revision must retain the requirements already established. Executable checks determine which later requirements still need to be requested, which stated requirements need correction, and whether earlier functionality remains correct. The resulting record separates success on the latest request from success on everything requested so far. It also distinguishes eventual delivery from a development path that repeatedly leaves the user with an incomplete or broken implementation.

\benchmark{} contains 61 tasks from 34 repositories, covering five primary programming languages and seven software forms. Its construction connects public development requests to staged, executable acceptance criteria. Tests and reference implementations are authored independently, each stage is checked for both added behavior and preservation of earlier requirements, and human reviewers inspect task quality before and after trial agent runs. This yields tasks with explicit user goals, feasible implementations, and requirement-level evidence across command-line tools, services, interactive applications, and libraries. The same feedback loop can therefore be evaluated across a broad range of software-engineering settings.

We evaluate 12 frontier models with complete requirements upfront, progressive requests alone, and progressive requests with corrective feedback. The results expose three complementary weaknesses. First, a model's final success rate does not describe how much correction it needs. Second, failure reports rarely yield a complete repair: only 15.4\% of observed failed states recover within two revisions. Third, local success does not ensure reliable cumulative progress; even successful individual repairs leave other stated requirements unfinished in 20.9\% of cases. Model-level analysis further separates responsiveness to correction, preservation of working behavior, and resource use. These differences identify where continued user intervention remains necessary.


Our contributions are threefold:
\begin{itemize}

    \item 
    We introduce an executable evaluation of iterative repository development that connects requirement evolution and corrective feedback to cumulative correctness and user guidance.

    \item 
    We build FeatureLoop, a quality-reviewed benchmark of 61 multi-stage tasks from 34 repositories across five primary programming languages and seven software forms, with acceptance checks retained throughout development.

    \item 
    We evaluate 12 frontier models and identify common gaps in complete repair and user-facing verification, alongside model-specific differences in preservation, failure patterns, and token efficiency.
\end{itemize}

\section{FeatureLoop}
\label{sec:benchmark}

\subsection{Task Formulation}
\label{sec:task-formulation}

A \benchmark{} task develops one user-facing goal within an existing repository. The initial request establishes the direction; later requests specify further capabilities, constraints, and interactions that refine the target software. All stages build on the same implementation, and each revision must satisfy the requirements accumulated so far.

Let $W_0$ be the initial repository. A \emph{requirement group} collects related behaviors evaluated together, such as creating a child shape and binding its connector. The groups form a \emph{directed acyclic graph} (DAG), where an edge $g\to h$ means that $h$ depends on $g$. We unfold this graph into a fixed sequence of stages that respects its dependencies. At stage $i$, the newly introduced groups are $A_i$, and the \emph{requirement state} $Q_i$ contains all groups introduced so far:
\begin{equation}
  Q_1=A_1,\qquad Q_{i+1}=Q_i\cup A_{i+1},\qquad Q_1\subset\cdots\subset Q_K.
  \label{eq:cumulative-expectation}
\end{equation}
Each stage adds at least one group. Every earlier group remains part of the task through the final stage $K$.

\paragraph{Checking the complete implementation.}
Let $Z_g(W)$ indicate whether the implementation in workspace $W$ passes the checks for group $g$. Let $b(W)$ indicate whether it passes the \emph{baseline suite}, which checks the repository behavior that was working before the task. We evaluate the requested functionality and the baseline separately:
\begin{equation}
  P_i(W)=\mathbf{1}[\forall g\in Q_i,\ Z_g(W)=1],
  \qquad C_i(W)=b(W)P_i(W).
  \label{eq:cumulative-closure}
\end{equation}
Here, $P_i=1$ means that all requirements introduced so far pass; $C_i=1$ also requires the original repository behavior to pass. For example, rerouting a connector correctly is only local progress if creating its bound child still fails. Regression testing checks that modifications preserve established behavior \citep{yoo2012regression}. Here, the retained obligations include both the original baseline and every requirement introduced during the task.

\paragraph{Two forms of feedback.}
Evolution feedback introduces missing behavior from the next requirement group or groups $A_{i+1}$. Completion feedback reports a failure among requirements already in $Q_i$ and asks the agent to repair its implementation. Before sending a later request, the evaluator checks whether the corresponding behavior already works and omits the parts already satisfied. Revisions then update the existing workspace. This makes feedback depend on the agent's progress while keeping the development target fixed.

Final task success requires every requested behavior and the baseline to pass:
\begin{equation}
  \operatorname{Success}(W_T)=C_K(W_T),
  \label{eq:terminal-success}
\end{equation}
where $W_T$ is the final workspace. Intermediate checks identify requirements completed immediately, repaired after feedback, left unfinished, or broken by later edits. Appendix~\ref{app:construction-mapping} defines the stored task components and the indexing of revisions. The complete six-stage Excalidraw walkthrough appears in Appendix~\ref{app:excalidraw-case}.

\Needspace{195pt}
\subsection{Three Development Scenarios}
\label{sec:protocols}

\begin{wrapfigure}[15]{r}{0.50\textwidth}
  \centering
  \includegraphics[width=\linewidth]{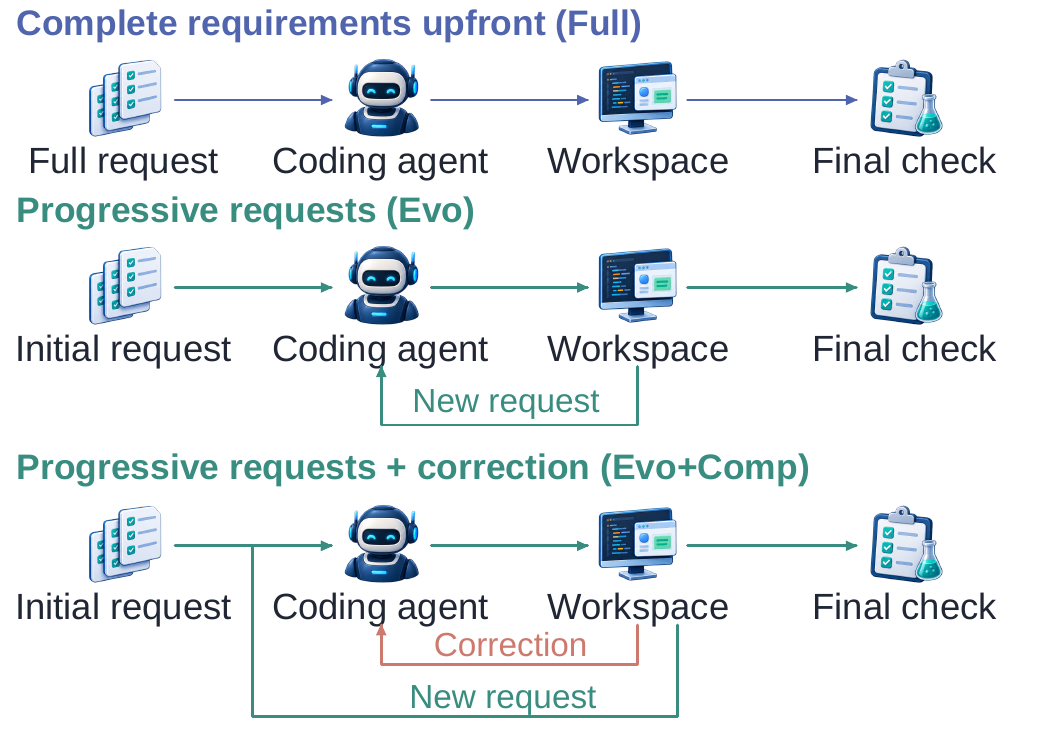}
  \caption{Three development scenarios. Green: new requests; orange: corrections.}
  \label{fig:experiment-protocols}
\end{wrapfigure}

We compare three ways a user can delegate the same development task (Figure~\ref{fig:experiment-protocols}). They differ in when the user provides requirements and whether the user returns to report implementation problems.

\paragraph{Complete requirements upfront (Full).}
The user specifies the entire target at the outset and asks the agent to deliver it in one coding session. The agent inspects the repository, edits code, and uses its tools to check the result. This scenario evaluates implementation when the user has already described everything needed.

\paragraph{Progressive requests (Evo).}
The user starts with an initial request and introduces further requirements as development proceeds. Each new request gives the agent another opportunity to extend the existing implementation. The user supplies additional requirements but leaves debugging to the agent. This scenario evaluates whether the agent can integrate each request and preserve earlier functionality using its own verification.

\paragraph{Progressive requests with correction (Evo+Comp).}
The user also reports problems encountered in the current implementation. The agent can respond to those reports before receiving the next requirement. This is the main evaluation scenario: the user continues refining the task and checking the delivered software, while the benchmark measures how effectively the agent uses both kinds of feedback.

To keep the task comparable, all scenarios start from the same repository checkpoint and use the same final requirements and acceptance checks. The two progressive scenarios also share the requirement order. Within that order, the current implementation determines which requests are still needed and which failures can be reported. Figure~\ref{fig:protocols-full} in Appendix~\ref{app:protocol-implementation} gives the full protocol diagram, including disclosure checks and corrective-feedback paths.

\Needspace{100pt}
\subsection{Benchmark Construction}
\label{sec:construction}

\begin{wrapfigure}{R}{0.49\textwidth}
  \centering
  \includegraphics[width=\linewidth]{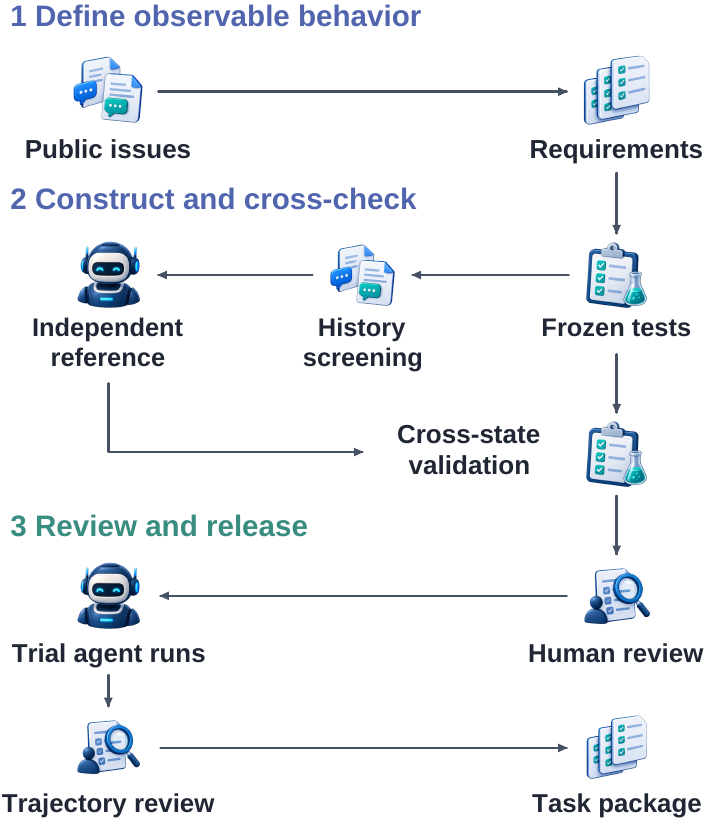}
  \caption{Construction stages and their output artifacts.}
  \label{fig:construction-pipeline}
\end{wrapfigure}

Construction must connect a realistic development goal to runnable stages and acceptance checks that faithfully test the requested behavior. Stronger tests can expose errors missed by weaker evaluation suites \citep{liu2023evalplus}, motivating checks of both the new behavior and its interaction with earlier requirements. Our pipeline combines agent-assisted authoring, executable validation, and human review. Figure~\ref{fig:construction-pipeline} gives a simplified overview; the complete author--critic workflow, including information boundaries and revision paths, appears in Figure~\ref{fig:construction-detail} of Appendix~\ref{app:construction-process}.

\paragraph{Ground the development goal in real requests.}
Each candidate begins with an exact repository checkpoint and at least two complete public issues supporting a common user-facing direction. These issues anchor the motivation and initial scope; repository evidence supports compatible extensions. We organize this shared goal into a dependency DAG of requirement groups, then select a stage sequence that respects the graph and follows the use scenarios that motivate further detail. A behavioral specification records the actions or inputs, expected outcomes, preserved behavior, and public observation points for each group. The corresponding user requests introduce that behavior in order. Appendices~\ref{app:source-grounding}--\ref{app:frozen-contract} describe source selection, requirement planning, environment setup, and request authoring.

\paragraph{Check behavior and exclude known complete solutions.}
To evaluate functionality across different implementations, the Test author derives executable checks from the behavioral specification. These checks cover each requirement group, the requirements accumulated at each stage, and the repository baseline. Before constructing a reference solution, we run the frozen checks on later public repository history and remove candidates whose complete target is already implemented. This targets publicly available solutions while retaining the original repository context. The availability of public solutions is a source of evaluation contamination \citep{badertdinov2025swerebench}; screening later history addresses this concern for the recorded repository window. Appendices~\ref{app:test-construction} and~\ref{app:history-screening} specify test construction and the history searched.

\paragraph{Verify that each stage adds feasible behavior and preserves earlier work.}
An independent Reference author implements the same specification, producing a working checkpoint $W_i^*$ for each stage. Test and Reference authors use separate workspaces and review sessions, so the tests assess the shared specification rather than being copied from a reference patch. For every new group $g\in A_i$, its checks must fail on $W_{i-1}^*$ and pass on $W_i^*$ and all later checkpoints. Every checkpoint must also pass the baseline and earlier requirements. These checks establish both a meaningful increment and its continued compatibility with the accumulated implementation. Selected candidates additionally use valid alternative implementations and deliberately faulty variants to check acceptance and failure detection. Appendices~\ref{app:reference-construction} and~\ref{app:cross-validation} give the construction and validation rules.

\paragraph{Review task quality before and after trial agent runs.}
Human reviewers first inspect the requirements and requests, run every reference stage, and verify that the tests enforce the intended user-visible behavior. Candidates that pass this review are then implemented by four coding models. Reviewers examine the resulting conversations, edits, and execution evidence to identify task defects exposed by implementations beyond the reference solution. Defects return to the responsible construction stage for correction and revalidation. Final acceptance retains 61 tasks. Appendix~\ref{app:review-admission} specifies the human review, and Appendix~\ref{app:pre-freeze-validation} records the candidate counts, trial models, and selection outcomes.

\subsection{Dataset Characteristics}
\label{sec:dataset-analysis}

\begin{figure}[t]
  \centering
  \begin{minipage}[t]{0.44\linewidth}
    \centering\includegraphics[width=\linewidth]{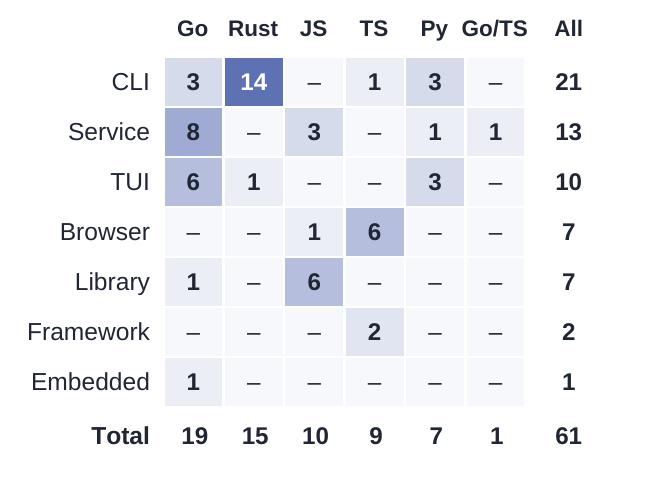}\\[-2pt]\textbf{(a)}
  \end{minipage}\hfill
  \begin{minipage}[t]{0.53\linewidth}
    \centering\includegraphics[width=\linewidth]{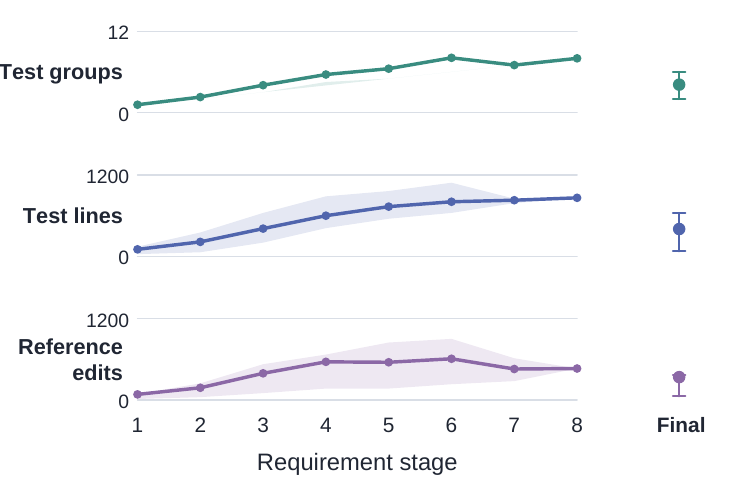}\\[-2pt]\textbf{(b)}
  \end{minipage}
  \caption{Dataset composition and cumulative implementation size. (a) Task counts by programming language and software form, including one mixed Go/TypeScript task. (b) Means and interquartile ranges of cumulative test groups, nonblank test lines, and reference-code additions plus deletions. Stage $i$ includes tasks that reach it; the final column includes all 61 tasks.}
  \label{fig:dataset-overview}
\end{figure}

\paragraph{Broad software-engineering coverage.}
The 61 tasks cover 34 repositories and five primary programming languages: Go, Rust, JavaScript, TypeScript, and Python. They include command-line tools, terminal interfaces, browser applications, services, libraries, frameworks, and embedded software (Figure~\ref{fig:dataset-overview}(a)). This coverage includes backend services, developer tooling, interactive software, and reusable components, each evaluated through its public commands, APIs, or user interface. Appendices~\ref{app:dataset-coverage} and~\ref{app:repository-coverage} give the complete composition.

\paragraph{Successive stages add continuing obligations.}
Tasks contain two to eight stages. Each stage adds behavior that must continue to work as the implementation changes. Across the 61 tasks, the average number of active test groups grows from 1.15 at the first stage to 4.11 at the final stage (Figure~\ref{fig:dataset-overview}(b)). Each new stage therefore expands the set of behaviors that must work together in the same implementation. Appendix~\ref{app:trajectory-metrics} reports stage counts, test sizes, and reference-edit volumes; Appendix~\ref{app:requirement-horizon} defines the counting units.

\paragraph{Later requirements include both foreseeable extensions and new information.}
The 61 tasks contain 147 later requirement groups. Of these, 42 follow a direction supported by earlier requests. For 16 of these groups, earlier information also constrains the exact expected behavior, and implementing it early carries limited risk. Another 85 depend on new user information or a discovery unavailable in the earlier interaction. Underspecified requests can omit information that materially changes the correct implementation \citep{vijayvargiya2026ambigswe}. We therefore distinguish two capabilities: completing well-supported behavior before another request is needed, and incorporating new requirements correctly once the user supplies them. The semantic annotation procedure, examples, and full labels are in Appendix~\ref{app:requirement-semantics}.

\section{Experimental Setup and Evaluation}
\label{sec:experiments}\label{sec:experimental-setup}\label{sec:main-metrics}

\paragraph{Setup.}
We evaluate 12 frontier models on all 61 tasks under the three development scenarios, yielding 2,196 runs. All models use Claude Code as the shared coding harness. Progressive runs retain the workspace and conversation; the corrective scenario uses a fixed feedback generator and permits at most two corrective revisions per stage. Each revision is a complete coding-agent response with its own tool calls. Appendix~\ref{app:experimental-details} gives the models, runtime configuration, and execution rules.

\paragraph{Measures.}
StrictPass measures final completion of all requirements and the repository baseline. 
We report new-requirement and corrective message counts ($N_E$, $N_C$), together with Feedback-Discounted Agent Completion (FDAC), which discounts successful delivery according to when intermediate states become correct; a state left incomplete within its response budget yields zero. 
Intermediate checks distinguish new-requirement uptake, complete repair, and preservation of earlier success. Agent Tokens measures reported input and output tokens. Trajectory analyses use both feedback types unless stated otherwise. Appendix~\ref{app:experiment-metrics} provides metric definitions, observation windows, token accounting, and uncertainty estimates.

\section{Results}
\label{sec:results}

\subsection{Intervention Changes the Model Comparison, Not Just the Average Score}
\label{sec:main-results}\label{sec:protocol-results}
\paragraph{The development process is part of the difficulty.}
Providing the entire specification at the outset does not produce the highest completion rate. Across models, final success rises from 40.3\% with an upfront specification (Full), to 51.6\% with progressive requests (Evo), and to 64.1\% with progressive requests and correction (Evo+Comp). The target and final checks are the same, but the opportunities to focus on a new requirement or revisit a failure differ. Full-specification performance therefore does not adequately characterize iterative development.

\begin{table}[!htbp]
  \centering
  \small
  \setlength{\tabcolsep}{6pt}
  \renewcommand{\arraystretch}{1.13}
  \caption{Main results, ordered by Evo+Comp FDAC. StrictPass and FDAC are percentages; $N_E$ and $N_C$ are mean messages per task, including failed runs. Each model is evaluated on the same 61 tasks. Best observed outcome values and the lowest correction count are bold.}
  \label{tab:main-results}
  \begin{tabular}{@{}lrrrrrr@{}}
    \toprule
\rowcolor{FLHeader}
& \multicolumn{3}{c}{StrictPass $\uparrow$} & \multicolumn{3}{c}{Evo+Comp path} \\
    \cmidrule(lr){2-4}\cmidrule(l){5-7}
    \rowcolor{FLHeader}
    Model & Full & Evo & Evo+Comp & FDAC $\uparrow$ & $N_E$ & $N_C$ \\
    \midrule
    GPT-6 Astra & \textbf{59.0} & \textbf{68.9} & \textbf{72.1} & \textbf{46.5} & 2.25 & 2.00 \\
\rowcolor{FLStripe}
    GPT-5.6 Sol & 52.5 & 62.3 & 70.5 & 45.4 & 2.23 & 1.89 \\
    Kimi K3 & 39.3 & 54.1 & 70.5 & 44.2 & 2.25 & \textbf{1.87} \\
\rowcolor{FLStripe}
    Hy4-Preview & 47.5 & 52.5 & 68.9 & 42.7 & 2.28 & 2.08 \\
    DeepSeek V4 Flash & 37.7 & 57.4 & 70.5 & 41.2 & 2.25 & 1.97 \\
\rowcolor{FLStripe}
    DeepSeek V4.1 Flash & 42.6 & 57.4 & 70.5 & 40.8 & 2.28 & 2.07 \\
    GLM 5.3 & 44.3 & 50.8 & 65.6 & 40.5 & 2.23 & 2.02 \\
\rowcolor{FLStripe}
    Qwen 3.8 Flash & 39.3 & 47.5 & 57.4 & 37.3 & 2.30 & 2.59 \\
    GLM 5.3 Flash & 36.1 & 49.2 & 62.3 & 37.2 & 2.28 & 2.48 \\
\rowcolor{FLStripe}
    Grok 4.6 & 27.9 & 37.7 & 63.9 & 37.1 & 2.31 & 2.82 \\
    DeepSeek V4 Pro & 36.1 & 49.2 & 60.7 & 36.7 & 2.30 & 2.44 \\
\rowcolor{FLStripe}
    Hy3 & 21.3 & 32.8 & 36.1 & 21.2 & 2.36 & 4.03 \\
    \midrule
    \rowcolor{FLTotal}
    Pooled & 40.3 & 51.6 & 64.1 & 39.2 & 2.27 & 2.35 \\
    \bottomrule
  \end{tabular}
\end{table}

\paragraph{Small gains can reflect very different starting points.}
GPT-6 Astra and Hy3 each gain only 3.3 percentage points when correction is enabled. For GPT-6 Astra, this follows already high success without correction; for Hy3, it accompanies low success in both settings (Table~\ref{tab:main-results}). A small workflow gain alone cannot distinguish limited need for intervention from failure to make useful progress despite it. Models also respond differently to the type of intervention: DeepSeek V4 Flash gains most from staged requests, whereas Grok 4.6 gains most from the correction-enabled workflow. Appendix~\ref{app:paired-transitions} separates these outcome changes from direct repairs within individual runs.

\paragraph{Repeated correction is the larger source of variation in guidance.}
The shared requirement sequence produces similar numbers of new requests across models, while corrective messages range from fewer than two to about four per task. Even the four models tied at 43 successful deliveries receive different feedback-discounted scores. Thus, matching final success does not imply matching user involvement or timely intermediate completion. Appendix~\ref{app:state-outcome-results} reports the message counts together with stalled and regressed states.

\subsection{Local Progress Often Fails to Produce a Complete, Lasting Repair}
\label{sec:feedback-conversion}\label{sec:cumulative-integration}\label{sec:interaction-reliability}
\paragraph{Providing the missing information does not remove the implementation bottleneck.}
We first examine requirements that introduce information unavailable earlier in the task. Agents implement the newly requested behavior on the first response in 77.3\% of observed cases. Yet some of those locally successful responses still leave earlier requirements unfinished. The difficulty is not simply understanding the latest request: the new behavior must coexist with everything the user has already asked for. Appendix~\ref{app:group-prefix-results} separates local success from complete requirement-state success.

\paragraph{An explicit failure report rarely restores the whole state.}
We next follow stages that fail their checks after the first response and receive corrective feedback. Only 15.4\% become fully correct within the two permitted revisions. Moreover, fixing an individual requirement is not always enough: 20.9\% of successful individual repairs still leave another stated requirement unmet. Earlier work may have remained unfinished or been broken by a subsequent edit. The relevant stopping condition is therefore the complete set of active requirements, not disappearance of the latest symptom. Appendix~\ref{app:feedback-recovery-results} gives state-level recovery, and Appendix~\ref{app:group-prefix-results} gives the individual-repair breakdown.

\paragraph{Correct behavior must survive the next request.}
To examine preservation, we continue checking requirements that have already passed and are observed again during later stages. Of these previously working requirements, 11.4\% fail at a subsequent check. In the NextChat example, an early repair is followed by a runtime error that persists while further capabilities are added. Continued feature development can thus conceal a broken execution path rather than repair it. The case appears in Appendix~\ref{app:failure-cases}; the complete retention counts and observation conditions are in Table~\ref{tab:feedback-recovery}. Together, these findings locate three separate demands on an agent: implement the new behavior, restore all stated behavior, and preserve that result through further edits.

\subsection{Similar Final Scores Conceal Different Recovery and Preservation Profiles}
\label{sec:model-profiles}
\begin{wrapfigure}{r}{0.53\textwidth}
  \centering
  \includegraphics[width=\linewidth]{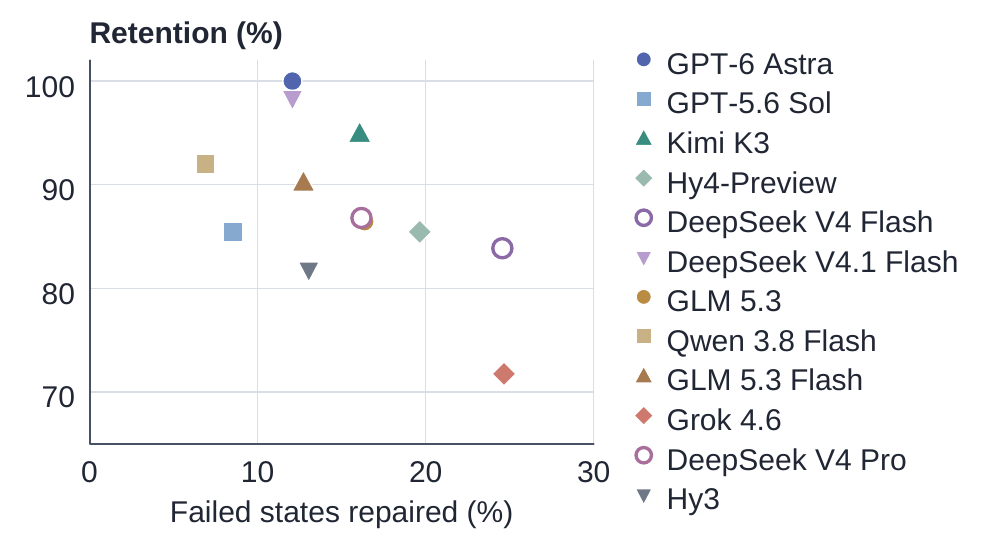}
  \caption{Model-level recovery and retention under both feedback types. Recovery uses failed states; retention uses previously passing requirements with later observations.}
  \label{fig:model-tradeoffs}
\end{wrapfigure}

\paragraph{Responsiveness to correction and preservation are distinct strengths.}
The two DeepSeek Flash models have the same final completion rate, but V4 repairs roughly twice the fraction of failed states that V4.1 repairs. V4.1, in contrast, preserves a larger fraction of its previously working requirements (Figure~\ref{fig:model-tradeoffs}). Grok 4.6 shows another imbalance: it has the largest correction-enabled workflow gain and the highest observed direct-recovery fraction, yet the lowest observed retention. These profiles explain why final success alone cannot reveal whether a model reaches its result through stable progress or repeated repair. Each conditional rate follows the failures or previously passing requirements actually encountered by that model; Appendix~\ref{app:feedback-recovery-results} provides the full model-level counts.

\paragraph{Tied scores do not imply the same unresolved tasks.}
The four models that each complete 43 tasks jointly cover 51 tasks, meaning that a task may fail for one of these models and succeed for another despite their tied aggregate results. This union measures the coverage of their independently recorded runs. Their failure profiles identify different obstacles: GPT-5.6 Sol more often leaves build or runtime failures than Kimi K3, while DeepSeek V4.1 Flash's remaining failures concentrate on behavioral compatibility rather than broken builds. Appendix~\ref{app:failure-model-profiles} reports the task-overlap calculation and all model profiles. Improving iterative coding consequently requires identifying which stage of implementation fails for a model, rather than treating a leaderboard score as a uniform capability.

\FloatBarrier
\subsection{Verification Often Misses the User's Actual Execution Path}
\label{sec:failure-analysis}

\Needspace{200pt}
\begin{wrapfigure}{r}{0.49\textwidth}
  \centering
  \includegraphics[width=\linewidth]{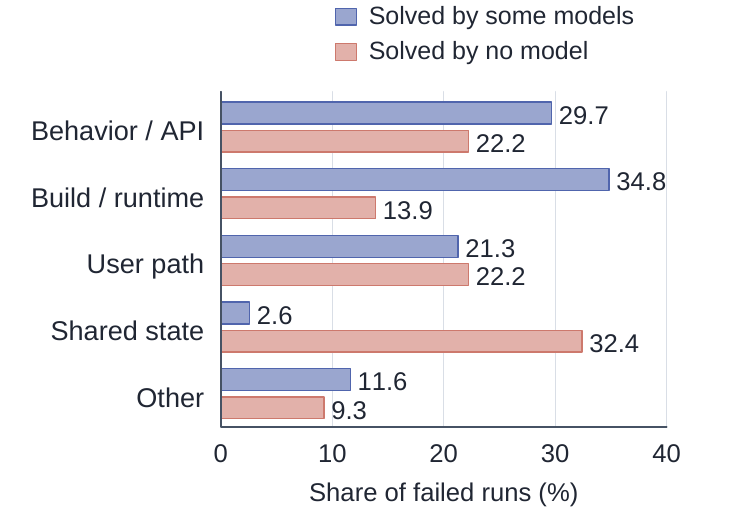}
  \caption{Primary diagnoses within each task set's final failures. ``Other'' groups the remaining diagnoses.}
  \label{fig:failure-strata}\label{fig:failure-diagnosis}
\end{wrapfigure}

\paragraph{Checking a convenient component is not checking the requested behavior.}
We inspect the requests, edits, actions, and execution evidence of every run that ends unsuccessfully. In 75.3\% of these failures, the agent either omits a relevant user-path check or validates a helper, stub, or proxy instead. For example, an internal merge routine can pass a custom test while the live endpoint continues to use the old behavior. This explains how apparently reasonable code and repeated tool use coexist with an unresolved user report: the agent's check does not exercise the behavior being evaluated. Appendix~\ref{app:failure-attribution} documents the coding procedure, and Appendix~\ref{app:failure-cases} traces representative examples.

\paragraph{Shared-state consistency characterizes the common hard cases.}
On tasks that no evaluated model completes, inconsistent state across components accounts for 32.4\% of final failures, compared with 2.6\% on tasks solved by some models (Figure~\ref{fig:failure-strata}). These traces involve registries, stores, parsers, renderers, or lifecycle handlers that do not agree on the same state. A local patch can leave that disagreement intact even when its own helper works. The common difficulty is therefore broader than writing a missing function: the change must reach the public entry point and remain consistent across the components that consume it.

\par\WFclear
\Needspace{205pt}
\begin{wrapfigure}{r}{0.50\textwidth}
  \centering
  \includegraphics[width=\linewidth]{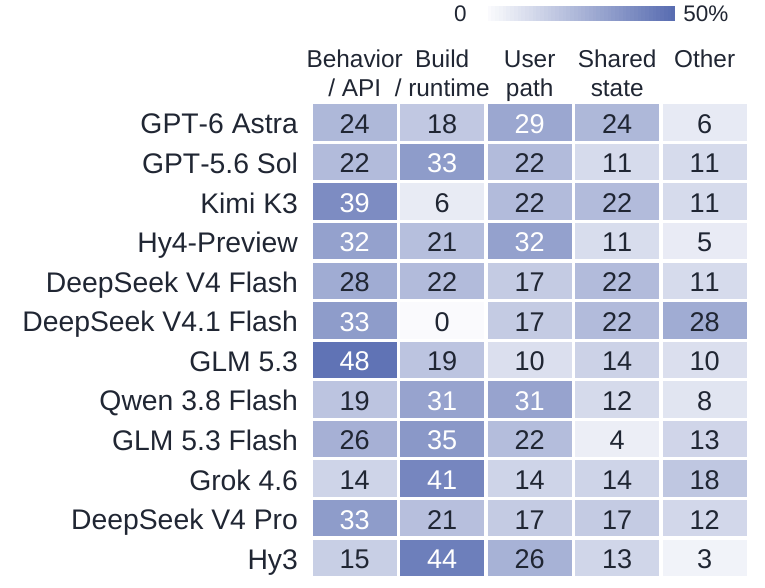}
  \caption{Primary diagnoses within each model's final failures; cell values are percentages.}
  \label{fig:failure-models}
\end{wrapfigure}

\paragraph{Model-specific failures distinguish broken execution from incorrect behavior.}
Build and runtime problems form a large share of Hy3 and Grok 4.6 failures, while GLM 5.3 more often produces a runnable implementation with incorrect behavior or API details (Figure~\ref{fig:failure-models}). DeepSeek V4.1 Flash has no final failure primarily attributed to a broken build, yet still misses legacy calling forms, timing bounds, or output representations. These are different verification needs: restoring executability, checking exact public behavior, and preserving compatibility. The model-level cases in Appendix~\ref{app:failure-model-profiles} identify the corresponding commands, interfaces, and state transitions. Appendix~\ref{app:dimension-results} reports how these patterns are distributed across the sampled languages and software forms.

\par\WFclear
\Needspace{215pt}
\subsection{Comparable Delivery Quality Can Carry Very Different Token Costs}
\label{sec:token-results}

\begin{wrapfigure}[16]{R}{0.53\textwidth}
  \centering
  \includegraphics[width=\linewidth]{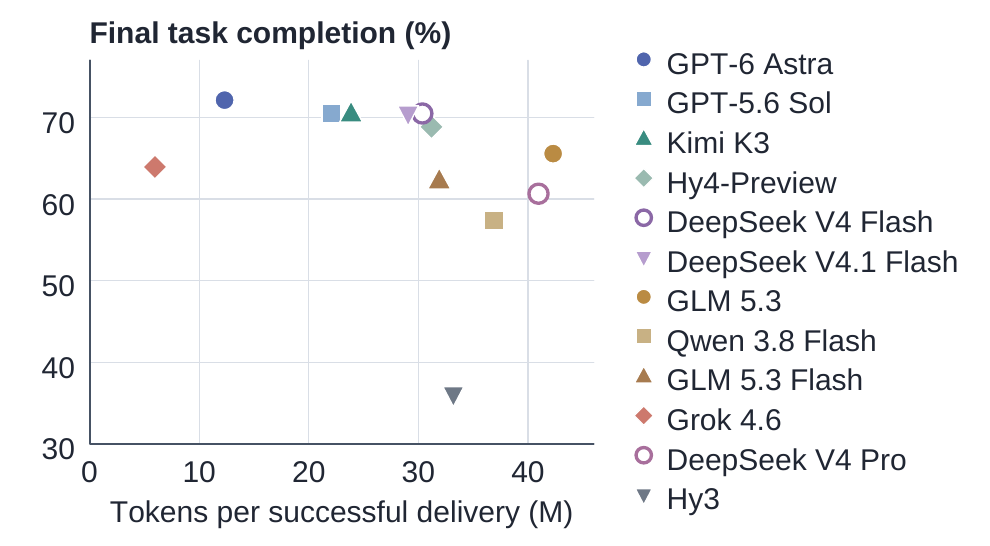}
  \caption{Model-level completion and token expenditure under both feedback types. Cost includes tokens spent on failed runs.}
  \label{fig:model-cost-delivery}
\end{wrapfigure}

Enabling correction increases average reported tokens per run by 47.2\%. More successful final deliveries therefore come with substantially greater token expenditure during development. To compare models at the level of delivered outcomes, we charge all recorded runs, including failures, to their successful deliveries. Grok 4.6 and GLM 5.3 achieve similar final completion rates, but the latter uses about 7.1 times as many tokens per successful delivery (Figure~\ref{fig:model-cost-delivery}). Similar outcome quality can consequently arise at very different resource expenditure.

This cost difference must be read alongside the development profile. Grok's lower token use coexists with weaker preservation of earlier behavior; GLM 5.3's greater expenditure still leaves exact interface requirements unresolved in several traces. Neither a low token total nor a high final score describes reliable, low-intervention development on its own. Appendix~\ref{app:cost-results} provides the complete token accounting, while Appendix~\ref{app:stability-results} reports the reliability of the full intermediate trajectory.

\FloatBarrier

\section{Scope and Limitations}
\label{sec:limitations}

\benchmark{} focuses on cumulative, executable requirements in existing repositories, with one predefined continuation per task.
We use a fixed feedback policy and at most two corrective revisions per stage.
The findings characterize the sampled tasks under these interaction settings; human collaboration may involve different requirement changes and feedback patterns.
Correctness is assessed through executable checks, whose coverage is bounded by the specified behaviors.

\section{Conclusion}
\label{sec:conclusion}

We introduced \benchmark{}, a benchmark for iterative software development under evolving requirements.
Each task follows one development goal in a persistent workspace, combining new requests and corrective feedback with cumulative executable checks.
Our evaluation of 12 frontier models shows that interactive workflows improve final completion, while incomplete repairs and regressions remain recurring obstacles.
Models with similar final success rates also differ in their reliance on correction, preservation of working functionality, and token expenditure.
By evaluating both final outcomes and progress across revisions, \benchmark{} provides a testbed for advancing reliable coding agents that require less repeated user guidance.

\bibliography{references,references_verified_legacy}

@misc{yang2024sweagent,
  title = {{SWE-agent: Agent-Computer Interfaces Enable Automated Software Engineering}},
  author = {John Yang and Carlos E. Jimenez and Alexander Wettig and Kilian Lieret and Shunyu Yao and Karthik Narasimhan and Ofir Press},
  year = {2024},
  eprint = {2405.15793},
  archivePrefix = {arXiv},
  primaryClass = {cs.SE},
  url = {https://arxiv.org/abs/2405.15793},
}

@misc{wang2025openhands,
  title = {{OpenHands: An Open Platform for AI Software Developers as Generalist Agents}},
  author = {Xingyao Wang and Boxuan Li and Yufan Song and Frank F. Xu and Xiangru Tang and Mingchen Zhuge and Jiayi Pan and Yueqi Song and Bowen Li and Jaskirat Singh and Hoang H. Tran and Fuqiang Li and Ren Ma and Mingzhang Zheng and Bill Qian and Yanjun Shao and Niklas Muennighoff and Yizhe Zhang and Binyuan Hui and Junyang Lin and Robert Brennan and Hao Peng and Heng Ji and Graham Neubig},
  year = {2025},
  eprint = {2407.16741},
  archivePrefix = {arXiv},
  primaryClass = {cs.SE},
  url = {https://arxiv.org/abs/2407.16741},
}

@misc{baumann2026swechat,
  title = {{SWE-chat: Coding Agent Interactions From Real Users in the Wild}},
  author = {Joachim Baumann and Vishakh Padmakumar and Xiang Li and John Yang and Diyi Yang and Sanmi Koyejo},
  year = {2026},
  eprint = {2604.20779},
  archivePrefix = {arXiv},
  primaryClass = {cs.AI},
  url = {https://arxiv.org/abs/2604.20779},
}

@misc{jimenez2024swebench,
  title = {{SWE-bench: Can Language Models Resolve Real-World GitHub Issues?}},
  author = {Carlos E. Jimenez and John Yang and Alexander Wettig and Shunyu Yao and Kexin Pei and Ofir Press and Karthik Narasimhan},
  year = {2024},
  eprint = {2310.06770},
  archivePrefix = {arXiv},
  primaryClass = {cs.CL},
  url = {https://arxiv.org/abs/2310.06770},
}

@misc{zhou2026featurebench,
  title = {{FeatureBench: Benchmarking Agentic Coding for Complex Feature Development}},
  author = {Qixing Zhou and Jiacheng Zhang and Haiyang Wang and Rui Hao and Jiahe Wang and Minghao Han and Yuxue Yang and Shuzhe Wu and Feiyang Pan and Lue Fan and Dandan Tu and Zhaoxiang Zhang},
  year = {2026},
  eprint = {2602.10975},
  archivePrefix = {arXiv},
  primaryClass = {cs.SE},
  url = {https://arxiv.org/abs/2602.10975},
}

@misc{merrill2026terminalbench,
  title = {{Terminal-Bench: Benchmarking Agents on Hard, Realistic Tasks in Command Line Interfaces}},
  author = {Mike A. Merrill and Alexander G. Shaw and Nicholas Carlini and Boxuan Li and Harsh Raj and Ivan Bercovich and Lin Shi and Jeong Yeon Shin and Thomas Walshe and E. Kelly Buchanan and Junhong Shen and Guanghao Ye and Haowei Lin and Jason Poulos and Maoyu Wang and Marianna Nezhurina and Jenia Jitsev and Di Lu and Orfeas Menis Mastromichalakis and Zhiwei Xu and Zizhao Chen and Yue Liu and Robert Zhang and Leon Liangyu Chen and Anurag Kashyap and Jan-Lucas Uslu and Jeffrey Li and Jianbo Wu and Minghao Yan and Song Bian and Vedang Sharma and Ke Sun and Steven Dillmann and Akshay Anand and Andrew Lanpouthakoun and Bardia Koopah and Changran Hu and Etash Guha and Gabriel H. S. Dreiman and Jiacheng Zhu and Karl Krauth and Li Zhong and Niklas Muennighoff and Robert Amanfu and Shangyin Tan and Shreyas Pimpalgaonkar and Tushar Aggarwal and Xiangning Lin and Xin Lan and Xuandong Zhao and Yiqing Liang and Yuanli Wang and Zilong Wang and Changzhi Zhou and David Heineman and Hange Liu and Harsh Trivedi and John Yang and Junhong Lin and Manish Shetty and Michael Yang and Nabil Omi and Negin Raoof and Shanda Li and Terry Yue Zhuo and Wuwei Lin and Yiwei Dai and Yuxin Wang and Wenhao Chai and Shang Zhou and Dariush Wahdany and Ziyu She and Jiaming Hu and Zhikang Dong and Yuxuan Zhu and Sasha Cui and Ahson Saiyed and Arinbjörn Kolbeinsson and Jesse Hu and Christopher Michael Rytting and Ryan Marten and Yixin Wang and Alex Dimakis and Andy Konwinski and Ludwig Schmidt},
  year = {2026},
  eprint = {2601.11868},
  archivePrefix = {arXiv},
  primaryClass = {cs.SE},
  url = {https://arxiv.org/abs/2601.11868},
}

@misc{wu2026swetogether,
  title = {{SWE-Together: Evaluating Coding Agents in Interactive User Sessions}},
  author = {Yifan Wu and Zhuokai Zhao and Songlin Li and Ho Hin Lee and Jiacheng Zhu and Shirley Wu and Tianhe Yu and Serena Li and Lizhu Zhang and Xiangjun Fan and Shengzhi Li},
  year = {2026},
  eprint = {2606.29957},
  archivePrefix = {arXiv},
  primaryClass = {cs.SE},
  url = {https://arxiv.org/abs/2606.29957},
}

@misc{wang2026asuka,
  title = {{Asuka-Bench: Benchmarking Code Agents on Underspecified User Intent and Multi-Round Refinement}},
  author = {Xin Wang and Liangtai Sun and Yaoming Zhu and Shuang Zhou and Jiaxing Liu and Fengjiao Chen and Lin Qiu and Xuezhi Cao and Xunliang Cai and Licheng Zhang and Zhendong Mao},
  year = {2026},
  eprint = {2606.05920},
  archivePrefix = {arXiv},
  primaryClass = {cs.SE},
  url = {https://arxiv.org/abs/2606.05920},
}

@misc{shen2026evocodebench,
  title = {{EvoCode-Bench: Evaluating Coding Agents in Multi-Turn Iterative Interactions}},
  author = {Haiyang Shen and Xuanzhong Chen and Wendong Xu and Yun Ma and Liang Chen and Kuan Li},
  year = {2026},
  eprint = {2605.24110},
  archivePrefix = {arXiv},
  primaryClass = {cs.AI},
  url = {https://arxiv.org/abs/2605.24110},
}

@misc{li2026loopsbench,
  title = {{LoopsBench: From Harness Engineering to Loop Engineering in Coding Agent Evaluation}},
  author = {Han Li and Zhemin Fang and Rili Feng and Yingqi Zhao and Jiaheng Liu and Pengfei Gao and He Ye and Dayi Lin and Qingwei Lin and Saravan Rajmohan and Dongmei Zhang},
  year = {2026},
  eprint = {2608.00267},
  archivePrefix = {arXiv},
  primaryClass = {cs.SE},
  url = {https://arxiv.org/abs/2608.00267},
}

@misc{zhang2025swelive,
  title = {{SWE-bench Goes Live!}},
  author = {Linghao Zhang and Shilin He and Chaoyun Zhang and Yu Kang and Bowen Li and Chengxing Xie and Junhao Wang and Maoquan Wang and Yufan Huang and Shengyu Fu and Elsie Nallipogu and Qingwei Lin and Yingnong Dang and Saravan Rajmohan and Dongmei Zhang},
  year = {2025},
  eprint = {2505.23419},
  archivePrefix = {arXiv},
  primaryClass = {cs.SE},
  url = {https://arxiv.org/abs/2505.23419},
}

@misc{badertdinov2025swerebench,
  title = {{SWE-rebench: An Automated Pipeline for Task Collection and Decontaminated Evaluation of Software Engineering Agents}},
  author = {Ibragim Badertdinov and Alexander Golubev and Maksim Nekrashevich and Anton Shevtsov and Simon Karasik and Andrei Andriushchenko and Maria Trofimova and Daria Litvintseva and Boris Yangel},
  year = {2025},
  eprint = {2505.20411},
  archivePrefix = {arXiv},
  primaryClass = {cs.SE},
  url = {https://arxiv.org/abs/2505.20411},
}

@misc{liu2023evalplus,
  title = {{Is Your Code Generated by ChatGPT Really Correct? Rigorous Evaluation of Large Language Models for Code Generation}},
  author = {Jiawei Liu and Chunqiu Steven Xia and Yuyao Wang and Lingming Zhang},
  year = {2023},
  eprint = {2305.01210},
  archivePrefix = {arXiv},
  primaryClass = {cs.SE},
  url = {https://arxiv.org/abs/2305.01210},
}

@misc{mundler2024swtbench,
  title = {{SWT-Bench: Testing and Validating Real-World Bug-Fixes with Code Agents}},
  author = {Niels Mündler and Mark Niklas Müller and Jingxuan He and Martin Vechev},
  year = {2025},
  eprint = {2406.12952},
  archivePrefix = {arXiv},
  primaryClass = {cs.SE},
  url = {https://arxiv.org/abs/2406.12952},
}

@misc{olausson2024selfrepair,
  title = {{Is Self-Repair a Silver Bullet for Code Generation?}},
  author = {Theo X. Olausson and Jeevana Priya Inala and Chenglong Wang and Jianfeng Gao and Armando Solar-Lezama},
  year = {2024},
  eprint = {2306.09896},
  archivePrefix = {arXiv},
  primaryClass = {cs.CL},
  url = {https://arxiv.org/abs/2306.09896},
}

@misc{orlanski2026slopcodebench,
  title = {{SlopCodeBench: Benchmarking How Coding Agents Degrade Over Long-Horizon Iterative Tasks}},
  author = {Gabriel Orlanski and Devjeet Roy and Alexander Yun and Changho Shin and Alex Gu and Albert Ge and Dyah Adila and Nicholas Roberts and Frederic Sala and Aws Albarghouthi},
  year = {2026},
  eprint = {2603.24755},
  archivePrefix = {arXiv},
  primaryClass = {cs.SE},
  url = {https://arxiv.org/abs/2603.24755},
}

@misc{wang2026cybergym,
  title = {{CyberGym: Evaluating AI Agents' Real-World Cybersecurity Capabilities at Scale}},
  author = {Zhun Wang and Tianneng Shi and Jingxuan He and Matthew Cai and Jialin Zhang and Dawn Song},
  year = {2026},
  eprint = {2506.02548},
  archivePrefix = {arXiv},
  primaryClass = {cs.CR},
  url = {https://arxiv.org/abs/2506.02548},
}

@misc{zan2025multiswebench,
  title = {{Multi-SWE-bench: A Multilingual Benchmark for Issue Resolving}},
  author = {Daoguang Zan and Zhirong Huang and Wei Liu and Hanwu Chen and Linhao Zhang and Shulin Xin and Lu Chen and Qi Liu and Xiaojian Zhong and Aoyan Li and Siyao Liu and Yongsheng Xiao and Liangqiang Chen and Yuyu Zhang and Jing Su and Tianyu Liu and Rui Long and Kai Shen and Liang Xiang},
  year = {2025},
  eprint = {2504.02605},
  archivePrefix = {arXiv},
  primaryClass = {cs.SE},
  url = {https://arxiv.org/abs/2504.02605},
}

@misc{rashid2025swepolybench,
  title = {{SWE-PolyBench: A multi-language benchmark for repository level evaluation of coding agents}},
  author = {Muhammad Shihab Rashid and Christian Bock and Yuan Zhuang and Alexander Buchholz and Tim Esler and Simon Valentin and Luca Franceschi and Martin Wistuba and Prabhu Teja Sivaprasad and Woo Jung Kim and Anoop Deoras and Giovanni Zappella and Laurent Callot},
  year = {2025},
  eprint = {2504.08703},
  archivePrefix = {arXiv},
  primaryClass = {cs.SE},
  url = {https://arxiv.org/abs/2504.08703},
}

@misc{yang2025swemultimodal,
  title = {{SWE-bench Multimodal: Do AI Systems Generalize to Visual Software Domains?}},
  author = {John Yang and Carlos E. Jimenez and Alex L. Zhang and Kilian Lieret and Joyce Yang and Xindi Wu and Ori Press and Niklas Muennighoff and Gabriel Synnaeve and Karthik R. Narasimhan and Diyi Yang and Sida I. Wang and Ofir Press},
  year = {2024},
  eprint = {2410.03859},
  archivePrefix = {arXiv},
  primaryClass = {cs.CL},
  url = {https://arxiv.org/abs/2410.03859},
}

@misc{miserendino2025swelancer,
  title = {{SWE-Lancer: Can Frontier LLMs Earn \$1 Million from Real-World Freelance Software Engineering?}},
  author = {Samuel Miserendino and Michele Wang and Tejal Patwardhan and Johannes Heidecke},
  year = {2025},
  eprint = {2502.12115},
  archivePrefix = {arXiv},
  primaryClass = {cs.LG},
  url = {https://arxiv.org/abs/2502.12115},
}

@misc{vijayvargiya2026ambigswe,
  title = {{Ambig-SWE: Interactive Agents to Overcome Underspecificity in Software Engineering}},
  author = {Sanidhya Vijayvargiya and Xuhui Zhou and Akhila Yerukola and Maarten Sap and Graham Neubig},
  year = {2026},
  eprint = {2502.13069},
  archivePrefix = {arXiv},
  primaryClass = {cs.AI},
  url = {https://arxiv.org/abs/2502.13069},
}

@misc{raghavendra2026sweinteract,
  title = {{SWE-INTERACT: Reimagining SWE Benchmarks as User-Driven Long-Horizon Coding Sessions}},
  author = {Mohit Raghavendra and Anisha Gunjal and Aakash Sabharwal and Yunzhong He},
  year = {2026},
  eprint = {2606.30573},
  archivePrefix = {arXiv},
  primaryClass = {cs.LG},
  url = {https://arxiv.org/abs/2606.30573},
}

@misc{peng2026icaebench,
  title = {{ICAE-Bench: Evaluating Coding Agents as Interactive Project Builders}},
  author = {Zhongyuan Peng and Dan Huang and Chuyu Zhang and Caijun Xu and Changyi Xiao and Shibo Hong and David Lo and Lin Qiu and Xuezhi Cao and Jiyuan He and Yixin Cao},
  year = {2026},
  eprint = {2607.21217},
  archivePrefix = {arXiv},
  primaryClass = {cs.AI},
  url = {https://arxiv.org/abs/2607.21217},
}

@misc{han2025convcodeworld,
  title = {{ConvCodeWorld: Benchmarking Conversational Code Generation in Reproducible Feedback Environments}},
  author = {Hojae Han and Seung-won Hwang and Rajhans Samdani and Yuxiong He},
  year = {2025},
  eprint = {2502.19852},
  archivePrefix = {arXiv},
  primaryClass = {cs.SE},
  url = {https://arxiv.org/abs/2502.19852},
}

@misc{zhan2026sreval,
  title = {{SR-Eval: Evaluating LLMs on Code Generation under Stepwise Requirement Refinement}},
  author = {Zexun Zhan and Shuzheng Gao and Ruida Hu and Cuiyun Gao},
  year = {2026},
  eprint = {2509.18808},
  archivePrefix = {arXiv},
  primaryClass = {cs.SE},
  url = {https://arxiv.org/abs/2509.18808},
}

@misc{yan2026slump,
  title = {{When the Specification Emerges: Benchmarking Faithfulness Loss in Long-Horizon Coding Agents}},
  author = {Lu Yan and Xuan Chen and Xiangyu Zhang},
  year = {2026},
  eprint = {2603.17104},
  archivePrefix = {arXiv},
  primaryClass = {cs.SE},
  url = {https://arxiv.org/abs/2603.17104},
}

@misc{le2026sweevo,
  title = {{SWE-EVO: Benchmarking Coding Agents in Long-Horizon Software Evolution Scenarios}},
  author = {Tue Le and Minh V. T. Thai and Dung Nguyen Manh and Huy Phan Nhat and Nghi D. Q. Bui},
  year = {2026},
  eprint = {2512.18470},
  archivePrefix = {arXiv},
  primaryClass = {cs.SE},
  url = {https://arxiv.org/abs/2512.18470},
}

@misc{madaan2023selfrefine,
  title = {{Self-Refine: Iterative Refinement with Self-Feedback}},
  author = {Aman Madaan and Niket Tandon and Prakhar Gupta and Skyler Hallinan and Luyu Gao and Sarah Wiegreffe and Uri Alon and Nouha Dziri and Shrimai Prabhumoye and Yiming Yang and Shashank Gupta and Bodhisattwa Prasad Majumder and Katherine Hermann and Sean Welleck and Amir Yazdanbakhsh and Peter Clark},
  year = {2023},
  eprint = {2303.17651},
  archivePrefix = {arXiv},
  primaryClass = {cs.CL},
  url = {https://arxiv.org/abs/2303.17651},
}

@misc{shinn2023reflexion,
  title = {{Reflexion: Language Agents with Verbal Reinforcement Learning}},
  author = {Noah Shinn and Federico Cassano and Edward Berman and Ashwin Gopinath and Karthik Narasimhan and Shunyu Yao},
  year = {2023},
  eprint = {2303.11366},
  archivePrefix = {arXiv},
  primaryClass = {cs.AI},
  url = {https://arxiv.org/abs/2303.11366},
}

@misc{pan2025swegym,
  title = {{Training Software Engineering Agents and Verifiers with SWE-Gym}},
  author = {Jiayi Pan and Xingyao Wang and Graham Neubig and Navdeep Jaitly and Heng Ji and Alane Suhr and Yizhe Zhang},
  year = {2025},
  eprint = {2412.21139},
  archivePrefix = {arXiv},
  primaryClass = {cs.SE},
  url = {https://arxiv.org/abs/2412.21139},
}

@misc{jain2025r2egym,
  title = {{R2E-Gym: Procedural Environments and Hybrid Verifiers for Scaling Open-Weights SWE Agents}},
  author = {Naman Jain and Jaskirat Singh and Manish Shetty and Liang Zheng and Koushik Sen and Ion Stoica},
  year = {2025},
  eprint = {2504.07164},
  archivePrefix = {arXiv},
  primaryClass = {cs.SE},
  url = {https://arxiv.org/abs/2504.07164},
}

@misc{yang2025swesmith,
  title = {{SWE-smith: Scaling Data for Software Engineering Agents}},
  author = {John Yang and Kilian Lieret and Carlos E. Jimenez and Alexander Wettig and Kabir Khandpur and Yanzhe Zhang and Binyuan Hui and Ofir Press and Ludwig Schmidt and Diyi Yang},
  year = {2025},
  eprint = {2504.21798},
  archivePrefix = {arXiv},
  primaryClass = {cs.SE},
  url = {https://arxiv.org/abs/2504.21798},
}

@misc{zhang2025sweflow,
  title = {{SWE-Flow: Synthesizing Software Engineering Data in a Test-Driven Manner}},
  author = {Lei Zhang and Jiaxi Yang and Min Yang and Jian Yang and Mouxiang Chen and Jiajun Zhang and Zeyu Cui and Binyuan Hui and Junyang Lin},
  year = {2025},
  eprint = {2506.09003},
  archivePrefix = {arXiv},
  primaryClass = {cs.CL},
  url = {https://arxiv.org/abs/2506.09003},
}

@misc{chen2024selfdebug,
  title = {{Teaching Large Language Models to Self-Debug}},
  author = {Xinyun Chen and Maxwell Lin and Nathanael Schärli and Denny Zhou},
  year = {2023},
  eprint = {2304.05128},
  archivePrefix = {arXiv},
  primaryClass = {cs.CL},
  url = {https://arxiv.org/abs/2304.05128},
}

@article{yoo2012regression,
  title = {{Regression testing minimization, selection and prioritization: A survey}},
  author = "S. Yoo and M. Harman",
  year = "2012",
  month = mar,
  doi = {10.1002/stvr.430},
  language = "English",
  volume = "22",
  pages = "67--120",
  journal = "Software Testing Verification and Reliability",
  issn = "0960-0833",
  number = "2",
}

@article{wilson1927probable,
  title={Probable Inference, the Law of Succession, and Statistical Inference},
  author={Wilson, Edwin B.},
  journal={Journal of the American Statistical Association},
  volume={22}, number={158}, pages={209--212}, year={1927},
  doi={10.1080/01621459.1927.10502953},
  url={https://doi.org/10.1080/01621459.1927.10502953}
}

@ARTICLE{segura2016metamorphic,
  author={Segura, Sergio and Fraser, Gordon and Sanchez, Ana B. and Ruiz-Cortés, Antonio},
  journal={IEEE Transactions on Software Engineering}, 
  title={A Survey on Metamorphic Testing}, 
  year={2016},
  volume={42},
  number={9},
  pages={805-824},
  doi={10.1109/TSE.2016.2532875}}
\bibliographystyle{iclr2027_conference}

\clearpage
\appendix
\section{Related Work}
\label{sec:related-work}\label{app:related-work}\label{app:benchmark-comparison}

\subsection{Executable Benchmarks and Task Construction}
\label{app:executable-related-work}

\paragraph{Executable software-engineering benchmarks.}
SWE-bench \citep{jimenez2024swebench} evaluates patches for issues in existing repositories. Multi-SWE-bench \citep{zan2025multiswebench} and SWE-PolyBench \citep{rashid2025swepolybench} extend repository-level evaluation across languages, while FeatureBench \citep{zhou2026featurebench} targets complex feature implementation. SWE-bench Multimodal \citep{yang2025swemultimodal} includes visual software domains, and SWE-Lancer \citep{miserendino2025swelancer} grounds tasks in paid freelance work. Terminal-Bench \citep{merrill2026terminalbench} broadens evaluation to hard technical tasks in command-line environments. CyberGym \citep{wang2026cybergym} verifies vulnerability reproduction by executing generated proof-of-concept inputs. These benchmarks connect realistic work to checkable outcomes. \benchmark{} measures the path to an outcome as requirements accumulate, retaining intermediate implementation, correction, and preservation evidence.

\paragraph{Task construction and verifier quality.}
SWE-Gym \citep{pan2025swegym} provides executable environments for training agents and verifiers. R2E-Gym \citep{jain2025r2egym} combines procedural environments with hybrid verification; SWE-smith \citep{yang2025swesmith} scales repository-based training tasks; and SWE-Flow \citep{zhang2025sweflow} synthesizes tasks through test-driven construction. EvalPlus \citep{liu2023evalplus} demonstrates the effect of stronger test coverage on code evaluation, while SWT-Bench \citep{mundler2024swtbench} evaluates tests that reproduce bugs and validate fixes. \benchmark{} independently constructs tests and reference checkpoints, checks every new behavior across adjacent and later states, and combines executable evidence with human review. Its historical screening complements the fresh-task collection used by SWE-bench-Live \citep{zhang2025swelive} and SWE-rebench \citep{badertdinov2025swerebench}.

\subsection{Interaction and Continued Development}
\label{app:interaction-comparison}

\begin{table}[!htbp]
  \centering
  \small
  \setlength{\tabcolsep}{4pt}
  \renewcommand{\arraystretch}{1.17}
  \caption{Interaction and evaluation mechanisms in closely related benchmarks. Cumulative checks retain earlier requirements as active correctness obligations.}
  \label{tab:benchmark-comparison}
  \begin{tabularx}{\linewidth}{@{}l YYY@{}}
    \toprule
\rowcolor{FLHeader}
Benchmark & Later information & Behavioral evaluation & Interaction measure \\
    \midrule
    SWE-Together & Session-conditioned requests and corrections & Final rubric with execution evidence & User Correction \\
\rowcolor{FLStripe}
    Asuka-Bench & Feedback from browser checks & Repeated UI criteria & Per-round fix rates \\
    EvoCode-Bench & Fixed instruction sequence & Cumulative tests & Per-round outcomes \\
\rowcolor{FLStripe}
    LoopsBench & Upfront dependent units & Retained regression tests & Scheduling and loop diagnostics \\
    \rowcolor{FLTotal}
    \benchmark{} & State-conditioned disclosure and correction & Group, prefix, and baseline checks & Message counts and FDAC \\
    \bottomrule
  \end{tabularx}
\end{table}

Table~\ref{tab:benchmark-comparison} contrasts how benchmarks obtain later information, check evolving implementations, and measure interaction. The following definitions distinguish information acquisition, corrective feedback, and continued correctness.

\paragraph{Information and feedback.}
A clarification supplies missing information about an established objective. A new request adds or changes desired behavior. A corrective message reports a failure of already stated behavior. We call feedback \emph{execution-grounded} when the user-facing message is derived from behavioral checks of the current implementation. This identifies the evidence supplied to the agent, separately from whether its final output is evaluated by execution.

\paragraph{Persistent correctness.}
Cumulative evaluation carries earlier, still-active requirements forward as correctness obligations. In \benchmark{}, requirement sets grow monotonically, so every introduced behavior remains active. Local checks identify whether the latest behavior works; cumulative checks require all active groups to work together; the baseline suite additionally checks retained repository behavior.

\paragraph{Ambig-SWE and ICAE-Bench: obtaining missing information.}
Ambig-SWE \citep{vijayvargiya2026ambigswe} removes information from repository issues and lets an agent ask a user proxy for clarification. ICAE-Bench \citep{peng2026icaebench} starts from underspecified product requests and uses a constrained router to reveal information from a fixed product contract. Both make information acquisition part of the agent's task. In \benchmark{}, the controller visits a fixed state sequence and discloses requirements according to the behavior still missing from the current workspace.

\paragraph{SWE-Interact and SWE-Together: state-dependent user steering.}
SWE-Interact \citep{raghavendra2026sweinteract} creates user-driven sessions from benchmark tasks using a simulator that can inspect repository content and changes. SWE-Together \citep{wu2026swetogether} conditions interventions on real-session evidence and the current trajectory. It distinguishes new requests from corrections, reports User Correction, and uses task-specific final rubrics with an agentic judge that can gather execution evidence. \benchmark{} ties each requirement group to fixed executable checks. Those checks determine both the missing information to disclose and the active failures to report, while intermediate evaluation records completion of individual requirements, all stated requirements, and the repository baseline.

\paragraph{Asuka-Bench: feedback from browser use.}
Asuka-Bench \citep{wang2026asuka} starts with an underspecified web-development request, checks the resulting interface through browser interaction, and converts the findings into feedback for another revision. Repeated evaluation measures improvement and repair over the interaction. \benchmark{} extends the evaluation setting across existing repositories and multiple software forms, with explicit cumulative requirement states and corrective revisions within each requirement stage. These states connect each message to the behavior that is newly requested or remains unfinished.

\paragraph{EvoCode-Bench and LoopsBench: continued development.}
EvoCode-Bench \citep{shen2026evocodebench} presents extensions, corrections, conflicts, and supersessions in a persistent workspace. Its verifier retains requirements that remain valid, and its reference-predecessor comparison separates single-round implementation from persistent execution. LoopsBench \citep{li2026loopsbench} organizes work into dependent development units whose activated tests remain regression obligations. \benchmark{} couples continued development to workspace-dependent user intervention: already implemented later behavior skips disclosure, whereas failed active behavior can receive completion feedback. The resulting trajectory records direct repairs and the preservation of earlier success.

\paragraph{Other forms of iterative development.}
SR-Eval \citep{zhan2026sreval} studies stepwise requirement refinement. SlopCodeBench \citep{orlanski2026slopcodebench} examines code degradation over iterative development, while SLUMP \citep{yan2026slump} studies faithfulness as specifications emerge. SWE-EVO \citep{le2026sweevo} constructs multi-step, multi-file tasks from project release notes. At the program level, ConvCodeWorld \citep{han2025convcodeworld} compares reproducible feedback environments and distinguishes conversational efficiency from overall task completion. \benchmark{} connects an accumulating requirement sequence to state-dependent disclosure and correction, recording whether each response completes the active requirements and preserves earlier success.

\paragraph{Comparing complete development workflows.}
Our upfront, progressive, and corrective workflows target the same final contract from the same checkpoint. Each model--task pair is independently executed under all three workflows. A paired comparison measures the change in final outcome between those complete workflows; direct recovery is measured separately within the trajectory that receives the failure report. FDAC summarizes final completion and intermediate timeliness under a fixed discount schedule, while message counts report the actual number of delivered interventions.

\subsection{Feedback-Driven Refinement and Repair}
\label{app:repair-related-work}

Self-Refine \citep{madaan2023selfrefine} iteratively revises outputs using model-generated feedback. Reflexion \citep{shinn2023reflexion} stores verbal feedback to guide later attempts, and Self-Debugging \citep{chen2024selfdebug} combines execution results with explanations of generated code. \citet{olausson2024selfrepair} find that self-repair gains depend on feedback quality and the computation spent obtaining them. These methods motivate treating feedback use as a measurable capability. In \benchmark{}, the repair target is the cumulative repository state: a correction succeeds only at the level specified by its local, cumulative, or baseline-inclusive check. Previously satisfied behavior remains testable throughout later development.

\FloatBarrier

\section{Construction Details and Excalidraw Case Study}
\label{app:construction}

\subsection{From Task Formulation to the Stored Task Record}
\label{app:construction-mapping}

We extend the notation in Section~\ref{sec:task-formulation} to identify individual revisions. Let $W_{i,0}$ be the first evaluated workspace at requirement state $i$: either a workspace that already passes before disclosure or the result of the first response. Let $W_{i,r}$ be the workspace after corrective revision $r$. The task record fixes the complete requirement sequence, and the evaluator delivers requests and feedback according to the current implementation. Table~\ref{tab:formalization-artifacts} maps the formal objects to the stored components.

\begin{table}[htbp]
  \centering
  \footnotesize
  \setlength{\tabcolsep}{4pt}
  \renewcommand{\arraystretch}{1.12}
  \caption{Mapping the task formulation to the stored task components used during evaluation.}
  \label{tab:formalization-artifacts}
  \begin{tabular}{@{}L{0.19\linewidth}L{0.34\linewidth}L{0.38\linewidth}@{}}
    \toprule
\rowcolor{FLHeader}
Formal object & Stored task component & Role during evaluation \\
    \midrule
    $W_0$ & Repository checkpoint, frozen environment, and baseline &
      Gives every evaluated agent the same starting state. \\
\rowcolor{FLStripe}
    $Q_1$ & First active set of requirement groups &
      Produces the initial request. \\
    $A_{i+1}=Q_{i+1}\setminus Q_i$ &
      Groups activated at the next state and their use scenario &
      Provides candidate evolution feedback. \\
\rowcolor{FLStripe}
    Public observations for groups in $Q_i$ &
      Observation-to-check mapping &
      Grounds completion feedback when expressed behavior fails. \\
    $Z_g(W)$ & Frozen behavioral checks for group $g$ &
      Determines whether a requirement is already satisfied. \\
\rowcolor{FLStripe}
    $b(W)$ & Baseline suite for the initial repository &
      Detects regressions in retained repository behavior. \\
    $W_i^*$ & Reference checkpoint &
      Demonstrates at least one feasible implementation of $Q_i$. \\
    \bottomrule
  \end{tabular}
\end{table}

For $i<K$, let $A_{i+1}$ denote the groups first activated by the next state. The frozen task record allows the controller to decide whether either type of user intervention is still needed in the current workspace:
\begin{align}
  \operatorname{Need}_{\mathrm C}(Q_i,W_{i,r})
  &=\mathbf{1}\!\left[\exists g\in Q_i,\ Z_g(W_{i,r})=0\right],
  \nonumber\\
  \operatorname{Need}_{\mathrm E}(A_{i+1},W_{i,r})
  &=\mathbf{1}\!\left[\exists g\in A_{i+1},\ Z_g(W_{i,r})=0\right].
  \label{eq:feedback-necessity}
\end{align}
The first indicator identifies failures of active requirements and the evidence used for completion feedback. The second identifies later groups that still need disclosure. The controller applies the protocol's revision budget and advances through the state sequence while retaining earlier obligations. Already implemented behavior is skipped, and partially satisfied groups produce a message covering the remaining behavior. The same development target can therefore require different messages as agents make different progress. Appendix~\ref{app:event-accounting} specifies the treatment of unavailable observations.

\subsection{The Complete Excalidraw Task}
\label{app:excalidraw-case}

\subsubsection{Provenance and Complete Goal}
\label{app:excalidraw-source}

The case in Figure~\ref{fig:excalidraw-case} is a \benchmark{} task titled \emph{Grow and restyle connected Excalidraw shapes}. It is anchored to the public Excalidraw repository at checkpoint \nolinkurl{a2ec2889babf7d2295469c6d90ebe77fae57df84} and task revision R003. Its source record joins five public requests: faster node and tree creation (\href{https://github.com/excalidraw/excalidraw/issues/11508}{\#11508}), converting existing shapes (\href{https://github.com/excalidraw/excalidraw/issues/11585}{\#11585}), improved arrows (\href{https://github.com/excalidraw/excalidraw/issues/11283}{\#11283}), blur effects (\href{https://github.com/excalidraw/excalidraw/issues/11228}{\#11228}), and using a shape as a clipping mask (\href{https://github.com/excalidraw/excalidraw/issues/11349}{\#11349}).

The scenario combines these public requests into a diagram-editing workflow: creating connected nodes, changing their types and relations, and preparing the result for sharing. The complete goal is to create a bound child from a selected shape, keep its connector bound and rerouted during dragging, convert shapes in place while preserving text, style, and connections, and compose blur and clipping on the canvas and in standard exports.

\begin{figure}[htbp]
  \centering
  \includegraphics[width=0.92\linewidth]{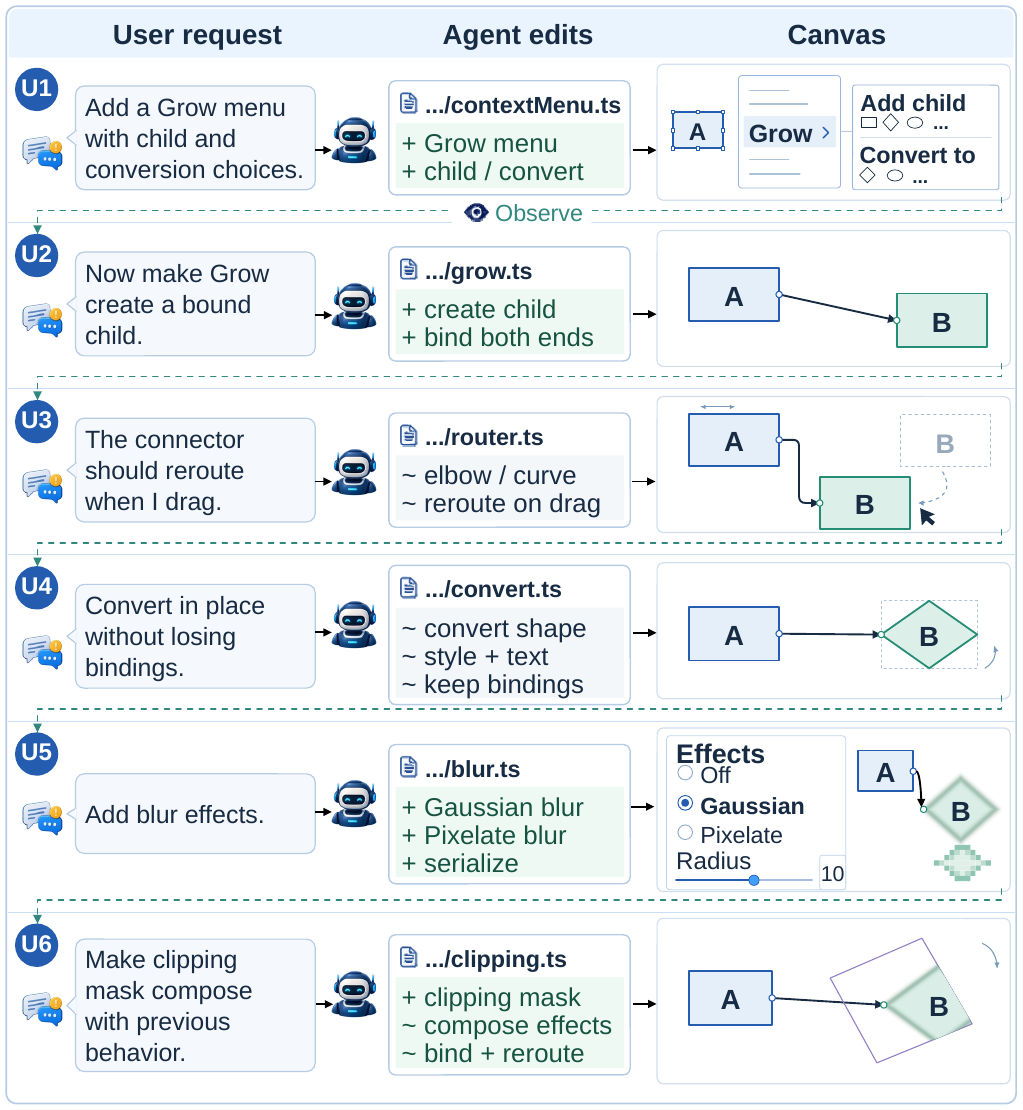}
  \caption{The complete six-stage Excalidraw case: successive user requests, illustrative code edits, and cumulative canvas behavior.}
  \label{fig:excalidraw-case-full}
\end{figure}

\subsubsection{Six Cumulative States}
\label{app:excalidraw-states}

The six states build on the same evolving implementation (Figure~\ref{fig:excalidraw-case-full}). Table~\ref{tab:excalidraw-stages} lists the new behavior and cumulative obligations at each state.

\begin{table}[htbp]
  \centering
  \footnotesize
  \setlength{\tabcolsep}{4pt}
  \renewcommand{\arraystretch}{1.12}
  \caption{The six cumulative states in the Excalidraw task.}
  \label{tab:excalidraw-stages}
  \begin{tabular}{@{}lL{0.25\linewidth}L{0.60\linewidth}@{}}
    \toprule
\rowcolor{FLHeader}
State & Behavior made explicit & What must hold in the cumulative workspace \\
    \midrule
    $Q_1$ & Grow menu & A selected fillable shape exposes one contextual Grow
      entry with child-creation and conversion choices, while the existing
      context menu retains its entries, predicates, and ordering. \\
\rowcolor{FLStripe}
    $Q_2$ & Bound child & A child choice creates exactly one non-overlapping
      shape of the requested type and one arrow bound at both ends; the new
      elements remain valid after serialization and restoration. \\
    $Q_3$ & Drag-aware routing & The grown connector visibly bends or curves,
      follows both bound shapes through dragging and a change in relative
      direction, and leaves manually drawn arrows unchanged. \\
\rowcolor{FLStripe}
    $Q_4$ & In-place conversion & Converting a shape preserves element count,
      geometry, style, bound text, and existing arrow bindings; dragging the
      converted shape still updates and reroutes its connectors. \\
    $Q_5$ & Blur effects & Gaussian and pixelate modes with an adjustable radius
      survive serialization and appear on the interactive canvas and in image
      exports; scenes with blur disabled retain their prior rendering. \\
\rowcolor{FLStripe}
    $Q_6$ & Clipping and composition & A fillable shape clips higher-layer
      content to its rendered outline, including rotated shapes, while clipping
      composes with blur, grown children, and bound rerouting connectors. \\
    \bottomrule
  \end{tabular}
\end{table}

\subsubsection{Feedback and Requirement Dependencies}
\label{app:excalidraw-feedback}

The sequence begins with the Grow menu ($Q_1$), followed by the parent--child relation created through its action ($Q_2$). Dragging either endpoint makes routing a continuing geometric obligation in $Q_3$. In-place conversion ($Q_4$) preserves labels, styles, and bindings. Blur ($Q_5$) and clipping ($Q_6$) add product capabilities drawn from the source requests, with serialization, export, and composition included in their acceptance requirements.

These transitions produce evolution feedback only when the newly revealed behavior remains missing. Completion feedback instead stays within the current state. For example, it may report that a child has only one bound arrow endpoint in $Q_2$, that routing remains at the old position after a drag in $Q_3$, or that blur appears on the canvas but not in the SVG export in $Q_5$. The message describes what the user can observe, without exposing a hidden test identifier, internal function, or Reference implementation.

In this case, each state introduces exactly one requirement group; writing $G_i$ for the group first introduced at state $i$, the semantic dependencies form the following directed acyclic graph, and the six user-visible stages follow a dependency-respecting disclosure order:
\begin{equation}
  G_1\rightarrow G_2\rightarrow G_3,\qquad
  G_1\rightarrow G_4,\qquad
  G_1\rightarrow G_5,\qquad
  \{G_2,G_5\}\rightarrow G_6.
  \label{eq:excalidraw-dependencies}
\end{equation}
Direct edges identify prerequisite capabilities. Preservation obligations carry every earlier active behavior forward even without a direct edge. Thus, although clipping directly composes with child creation and blur, success at $Q_6$ also requires retained rerouting, conversion, and baseline behavior.

\subsubsection{Executable Behavioral Evidence}
\label{app:case-evidence}

The executable checks cover six parts of the user-visible workflow:
\begin{enumerate}
  \item \textbf{UI and reachability:} Grow, child and conversion choices,
    Effects, and clipping controls appear under the appropriate selections
    while existing menu and property-panel behavior remains available.
  \item \textbf{Scene state:} element counts, types, finite geometry, layer
    order, bound text, arrow bindings, blur attributes, and clipping-mask state
    are inspected through public scene representations.
  \item \textbf{Interaction:} parent and child dragging, direction changes,
    conversion, and layer reordering exercise the subsequent connector,
    selection, and mask behavior.
  \item \textbf{Persistence:} serialize--restore checks preserve new
    attributes and relations while legacy scenes without them retain their
    previous semantics.
  \item \textbf{Rendering and export:} the interactive canvas, SVG export,
    and canvas export exercise routing, blur, outline-aware clipping, and their
    composition.
  \item \textbf{Baseline:} existing context-menu actions, manually created
    arrows, legacy scenes, frame clipping, and rendering with effects disabled
    remain valid.
\end{enumerate}
Together, these checks implement $Z_g(W)$ and $b(W)$ and distinguish a failure in the current behavior from a regression in an earlier requirement or the initial repository baseline.

\subsection{Construction Process}
\label{app:construction-process}

\begin{figure}[tbp]
\centering\includegraphics[width=\linewidth]{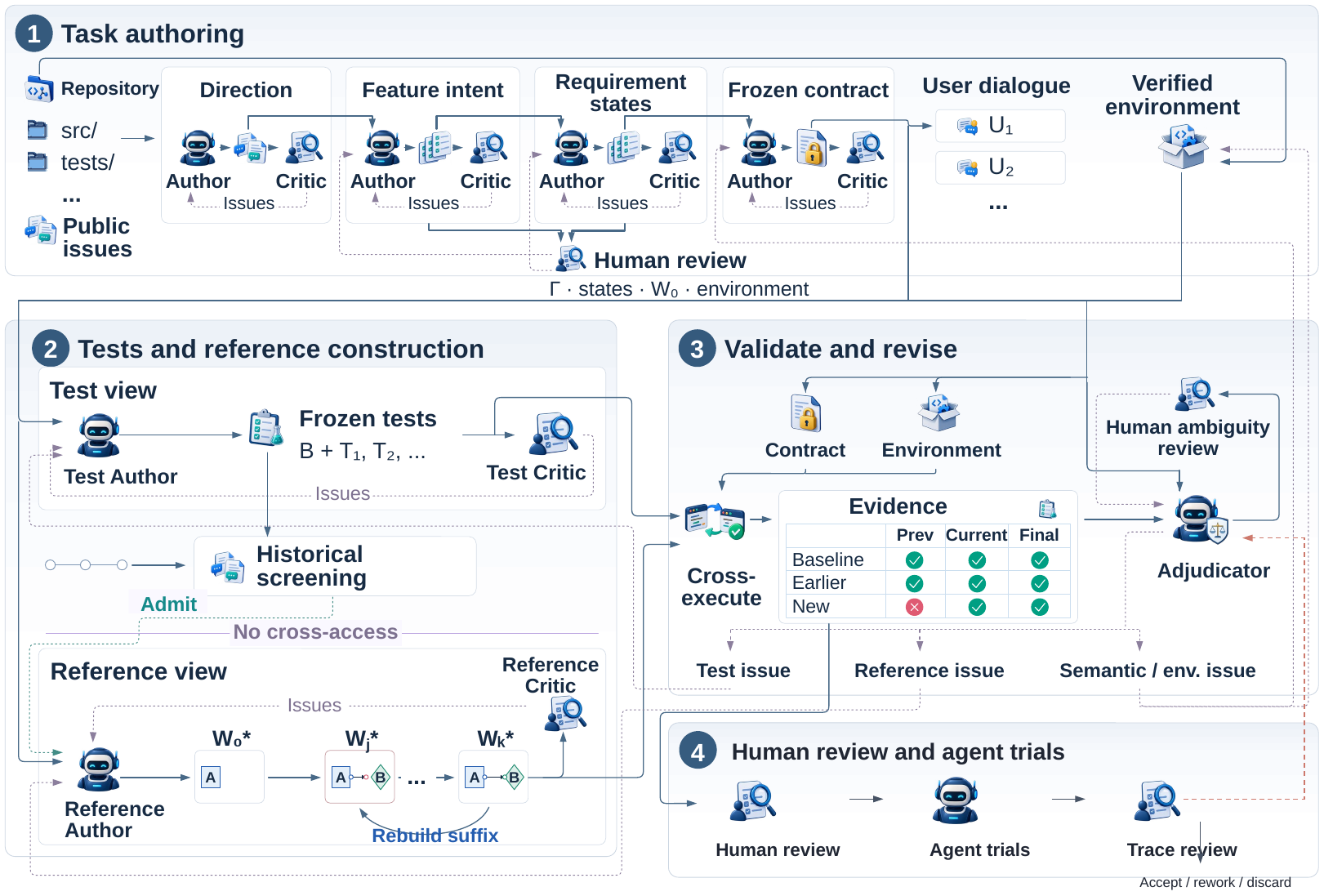}
\caption{Complete author--critic workflow. Solid paths link construction artifacts and execution stages; dashed paths return review findings for revision. The Test and Reference views mark their separate information access.}
\label{fig:construction-detail}
\end{figure}

\subsubsection{Repository Sources}
\label{app:source-grounding}

To anchor each task in real development needs, the Source Author selects at least two complete public issues that support a common user-facing direction at an exact repository checkpoint. The source record preserves the issue text, user motivation, relevant public interfaces, and supporting repository evidence. An independent reviewer checks that these sources support the proposed target and its scope. Later task components retain links to the accepted sources; the semantic analysis in Appendix~\ref{app:requirement-semantics} characterizes how closely each increment relates to earlier work.

\subsubsection{Feature Specification}
\label{app:feature-specification}

To give the construction roles a shared development target, the Feature Author states the complete user outcome, acceptance requirements, scope exclusions, and refinements supported by the accepted sources and repository. A requirement may come directly from a public request, follow the same direction through repository evidence, or describe how existing public capabilities must work together in the same workflow. This specification fixes the task's complete target; the dialogue determines when each part is revealed, and tests are authored separately.

\subsubsection{Requirement Sequence}
\label{app:requirement-trajectory}

To turn the development target into actionable stages, the Trajectory Author groups related behaviors into nodes of a directed acyclic dependency graph. An edge records a prerequisite relation between groups. The author then selects a stage sequence from $Q_1$ to $Q_K$ that respects these dependencies. Each state records the groups introduced there, all groups that remain active, direct dependencies, earlier behavior that must remain correct, and the use scenario that exposes the next gap. Source evidence informs a plausible development order; the frozen task record fixes the actual requirements and disclosure sequence used in evaluation.

\subsubsection{Executable Environment}
\label{app:executable-environment}

To make every stage reproducible, the Environment Author reconstructs an isolated environment from dependency locks, build configuration, and repository continuous-integration records. Validation includes independent clean runs, the original baseline, public entry points, offline execution, resource limits, and workspace integrity. The runs use the same environment definition but fresh writable workspaces. This definition determines how the task executes reproducibly, while the frozen sources and contract determine its requirements and behavior.

\subsubsection{Frozen Contract and Feedback Messages}
\label{app:frozen-contract}

The Contract Author combines the accepted sources, development target, requirement sequence, and environment into a versioned behavioral specification, called the contract $\Gamma$. This specification aligns the requests, tests, and reference implementations. Each group records the user action or input, expected and forbidden outcomes, earlier behavior that must remain correct, allowed implementation freedom, and a runnable check. An independent Contract Reviewer checks that the contract follows the sources, respects state dependencies, can be observed through public interfaces, leaves implementation choices open, and runs in the environment; deterministic validation checks the schema and every referenced file or check.

After $\Gamma$ is frozen, a Dialogue Author produces $u_0$ and the candidate evolution messages. Review verifies that each message reveals only its corresponding state, is consistent with its documented use scenario, and contains no group identifier, hidden-test wording, Reference information, or future requirement. Completion feedback does not create a new contract state; it is generated at runtime from failed user-visible checks for currently active groups.

\subsubsection{Isolated Test Construction}
\label{app:test-construction}

To evaluate the specified behavior independently of a solution, the Test Author works from the frozen contract, initial repository, environment, and complete state map in a separate workspace. It builds the group and cumulative-state checks, final suite, baseline checks, and transition checks. A Test Critic reviews their behavioral coverage, execution stability, and use of public interfaces. Reference code and execution results remain outside both roles' access.

\subsubsection{Historical Implementation Screening}
\label{app:history-screening}

To exclude targets with an already published complete solution, we screened every candidate against public repository history after its checkpoint and before a recorded cutoff. Screening used the frozen Test package and preceded Reference construction. The frozen behavioral checks were replayed on later commits or trees: inexpensive deterministic checks covered the relevant history exhaustively, while expensive checks such as browser execution used recorded targeted coverage. Candidates whose later public history passed the complete recorded behavior set were removed. The screening record preserves the checkpoint, cutoff, searched history, executed checks, coverage, and decision; reviewers checked partial matches and environment incompatibilities individually to determine which behaviors the historical version implemented.

\subsubsection{Isolated Reference Construction}
\label{app:reference-construction}

To establish a feasible implementation at every stage, the Reference Author receives the frozen contract, initial repository, environment, and state graph in an isolated workspace. Tests, fixtures, assertions, and Test results remain outside its access. Starting from the baseline $W_0^*=W_0$, it constructs $W_1^*,\ldots,W_K^*$ in one persistent workspace, with each checkpoint adding the current requirement while preserving earlier behavior. A Reference Critic with the same restricted information reviews contract coverage, repository consistency, retained behavior, and the boundary between checkpoints.

\subsubsection{Cross-State Validation}
\label{app:cross-validation}

To verify that each increment is implemented at its intended stage and preserved afterward, the controller runs the frozen Test suite against the Reference checkpoints. For every $g\in A_i$, it requires
\begin{equation}
  Z_g(W_{i-1}^*)=0,\qquad
  Z_g(W_j^*)=1\quad\forall j\geq i.
  \label{eq:cross-state-witness}
\end{equation}
Each checkpoint $W_i^*$ must also pass the complete active suite and repository baseline, and every later checkpoint must preserve that behavior. Selected candidates receive two additional probes: contract-valid alternatives test acceptance across different implementations, and targeted faulty variants test rejection of specific behavioral defects. The controller executed the constructed variants and recorded their outcomes alongside the required state-by-state checks. Additional checks can reveal defects missed by an original test suite \citep{liu2023evalplus}; running a test against both faulty and corrected versions also assesses whether it distinguishes the intended behavior \citep{mundler2024swtbench}. The two types of probe check complementary errors: rejecting valid behavior and accepting an implementation with a known defect.

\subsubsection{Human Review and Task Acceptance}
\label{app:review-admission}

For every accepted task, human review began with the task semantics. Reviewers checked whether the Contract was coherent, source-grounded, observable through the public interface, and permissive of different valid implementations, then verified that its Summary and Evidence records represented the same behavior. They next simulated the evaluation turn by turn. At each turn, they inspected the dialogue shown to the agent, ran the project at the corresponding Reference checkpoint, executed the current and retained tests, and followed the public observation points to confirm that the Reference behavior matched the intended user semantics. They also inspected Test assertions on adjacent checkpoints to verify that each assertion followed from the required public behavior.

The reviewer recorded a final accept, revise, or discard decision with findings tied to the responsible construction stage. A requested change returned to that stage and reran every affected downstream check. Each accepted task revision records the exact Source, Environment, Contract, Summary, Evidence, Dialogue, Test, historical-screening, Reference, Cross, and human-review artifacts, preserving earlier revisions for traceability.

\subsection{Roles and Information Boundaries}
\label{app:roles-boundaries}

Table~\ref{tab:construction-roles} separates the evidence available to each construction role. The key boundary is between Test and Reference: both receive the same approved task semantics and environment, but neither can inspect the other's private implementation or execution results.

\begin{table}[htbp]
  \centering
  \footnotesize
  \setlength{\tabcolsep}{3pt}
  \renewcommand{\arraystretch}{1.10}
  \caption{Construction roles and their information boundaries.}
  \label{tab:construction-roles}
  \begin{tabular}{@{}L{0.18\linewidth}L{0.35\linewidth}L{0.39\linewidth}@{}}
    \toprule
\rowcolor{FLHeader}
Role & May read & Must not read \\
    \midrule
    Source/Feature & Frozen source records, checkpoint, repository evidence &
      Private Test or Reference material, or evaluated-agent solution content \\
\rowcolor{FLStripe}
    Contract & Sources, feature specification, requirement sequence, environment report &
      Private Test or Reference material, or evaluated-agent solution content \\
    Dialogue & Frozen Contract and user-facing narrative context &
      Hidden tests, Reference, or future evaluated-agent behavior \\
\rowcolor{FLStripe}
    Test Author/Critic & Frozen Contract, $W_0$, environment, complete state map,
      and same-side Test files & Reference code, diffs, or execution results \\
    Reference Author/Critic & Frozen Contract, $W_0$, environment, complete
      state map, and same-side workspace & Tests, fixtures, assertions, or Test results \\
\rowcolor{FLStripe}
    Adjudicator & Contract, structured execution outcomes, suite summaries,
      and changed-file summaries & Joint write access or unrecorded transfer of
      one side's private implementation to the other \\
    Human reviewer & Sources, Contract, Summary and Evidence, Dialogue,
      Reference checkpoints, Test, and execution records & Unrecorded edits
      that bypass revision and revalidation \\
    \bottomrule
  \end{tabular}
\end{table}

Construction agents combine a foundation model with a workspace harness providing repository search, file editing, terminal execution, tests, and structured output. We use Claude Code, Codex, and CodeBuddy harnesses. Model--harness combinations follow each stage's role and compatibility requirements; repository views, tool permissions, and execution environments enforce the information boundaries.

Table~\ref{tab:construction-model-pool} lists the shared model pool. Routing uses the stage's output, task difficulty, review findings, and revision history to select an Author or Critic.

\begin{table}[htbp]
  \centering
  \footnotesize
  \setlength{\tabcolsep}{3pt}
  \renewcommand{\arraystretch}{1.15}
  \caption{Model pool used for agent-assisted benchmark construction.}
  \label{tab:construction-model-pool}
  \begin{tabular}{@{}L{0.15\linewidth}L{0.31\linewidth}L{0.46\linewidth}@{}}
    \toprule
\rowcolor{FLHeader}
Tier & Models & Routing role \\
    \midrule
    Low latency & Hy3; DeepSeek V4 Flash & High-throughput generation,
      organization, and low-risk revision \\
\rowcolor{FLStripe}
    Medium & DeepSeek V4 Pro; GPT-5.6 Terra & Routine Author work and
      workspace-based revision \\
    Strong & GLM 5.3; Kimi K3 & Difficult Author work, independent
      semantic checks, and root-cause analysis \\
\rowcolor{FLStripe}
    Flagship & GPT-5.6 Sol; Claude Opus 5 & High-complexity Author work,
      final independent review, and support for disputed decisions \\
    \bottomrule
  \end{tabular}
\end{table}

Strong and flagship models handle difficult semantic review, while low-latency models handle bounded generation and automated checks. Findings that require broader reasoning are escalated to a stronger model or a human reviewer. Deterministic checks govern schemas, artifact links, environment execution, state matrices, and result formatting.

Across semantic stages, Authors produce each stage's complete output from approved upstream evidence, while Critics work in independent contexts and return structured findings about source fidelity, observability, requirement completeness, implementation freedom, and execution evidence. The responsible Author revises that output, after which the stage and every affected later check are rerun. Test and Reference use separate authoring and review sessions and, by default, different model families. Separate repository views and tool permissions enforce this boundary for workspaces, conversation histories, and execution results.

Human reviewers remain responsible for feature direction, scope, interaction naturalness, and test fairness. They audit the Contract together with its Summary and Evidence, replay each dialogue turn at the corresponding Reference checkpoint, run the project and tests, compare the observed public behavior with the observation mapping, and verify that Test enforces the intended semantics. Human-requested changes return to the responsible stage and undergo every affected downstream check. The resulting audit supports the final decision to accept, revise, or discard the task.

\subsection{Quality Evidence}
\label{app:evidence-boundaries}

Task acceptance combines feasibility, behavioral discrimination, preservation, historical screening, and semantic review. Table~\ref{tab:evidence-boundaries} specifies the evidence supporting each component.
\begin{table}[htbp]
  \centering
  \footnotesize
  \setlength{\tabcolsep}{3pt}
  \renewcommand{\arraystretch}{1.15}
  \caption{Validation roles of the construction evidence.}
  \label{tab:evidence-boundaries}
  \begin{tabular}{@{}L{0.20\linewidth}L{0.33\linewidth}L{0.38\linewidth}@{}}
    \toprule
\rowcolor{FLHeader}
Evidence & Validation role & Use in task acceptance \\
    \midrule
    Historical screening & Checks later public repository history within a
      recorded cutoff and coverage & Required after Test is frozen and before
      Reference construction \\
\rowcolor{FLStripe}
    Reference pass & Establishes a feasible implementation under the frozen
      Contract and environment & Required at every state \\
    Adjacent-state check & Confirms that newly activated behavior distinguishes
      adjacent Reference checkpoints & Required for every state transition \\
\rowcolor{FLStripe}
    Earlier-state suite & Checks earlier active behavior on later Reference
      checkpoints & Required through the Reference suffix \\
    Baseline suite & Checks retained repository behavior at each Reference
      checkpoint & Required at every state \\
\rowcolor{FLStripe}
    Targeted mutation & Probes failure sensitivity for a selected behavioral
      defect & Executed for constructed targeted faulty versions \\
    Valid alternative & Probes acceptance of a different contract-valid
      implementation & Executed for constructed valid alternatives \\
\rowcolor{FLStripe}
    Human review & Audits Contract, Summary and Evidence, dialogue turns,
      checkpoint behavior, observations, and Test semantics & Required for the
      final accept, revise, or discard decision \\
    \bottomrule
  \end{tabular}
\end{table}

Acceptance evidence comes from executable observations such as unit or integration tests, CLI or PTY interaction, service calls, browser automation, lifecycle events, restart, and persistence. Some requirements are checked through relations between executions, such as export/import, save/restart/load, encode/decode, idempotence, conservation, or monotonicity. These are metamorphic checks \citep{segura2016metamorphic}. Comparisons with an independent reference provide additional evidence. Browser, TUI, and visual tasks remain programmatically checkable when geometry, DOM or public state, pixel relations, or protocol statistics can be recomputed from fixed inputs. Each accepted task records the evidence used for every requirement.

\subsection{Revision and Invalidation Rules}
\label{app:revision-rules}

A task revision reruns every check whose conclusion may have changed. Table~\ref{tab:invalidation} makes this dependency explicit: wording-only edits require dialogue review and an updated evaluator mapping, whereas changes to requirements, environments, tests, or Reference checkpoints invalidate the corresponding executable evidence.

\begin{table}[htbp]
  \centering
  \footnotesize
  \setlength{\tabcolsep}{4pt}
  \renewcommand{\arraystretch}{1.15}
  \caption{Task components that must be updated after a substantive revision.}
  \label{tab:invalidation}
  \begin{tabular}{@{}L{0.31\linewidth}L{0.61\linewidth}@{}}
    \toprule
\rowcolor{FLHeader}
Modified component & Required regeneration or re-execution \\
    \midrule
    Source or feature specification & Requirement sequence, Contract, Dialogue, Test,
      historical screening, Reference, Cross, and review \\
\rowcolor{FLStripe}
    Requirement semantics, dependency, or public observation & New Contract
      revision and all later semantic components and executable checks \\
    Wording only, with unchanged semantics & Presentation revision, evaluator
      mapping, and Dialogue review \\
\rowcolor{FLStripe}
    Environment, dependency, or runner & Baseline, historical screening,
      Reference execution, and all Cross evidence that uses the environment \\
    Test package & Historical screening and, for retained candidates,
      all Test--Reference executions \\
\rowcolor{FLStripe}
    Reference checkpoint $W_m^*$ & Reference suffix beginning at $m$ and its
      Cross evidence \\
    Feedback-observation mapping & Dialogue and feedback review, controller
      update, and affected feedback verification \\
    \bottomrule
  \end{tabular}
\end{table}

Before revising a task component, reviewers identify the source of a failed Test--Reference execution. A Test defect changes only Test-side files; a Reference defect rebuilds that checkpoint and the later Reference states; ambiguous requirements or states that cannot be separated create a new Contract revision; and environment failures are corrected before the evidence is rerun.

\subsection{Versioned Task Record}
\label{app:reproducibility-binding}

Every accepted task has a versioned manifest that records its repository and exact checkpoint together with the sources, Contract, Dialogue, environment, frozen Test package, historical-screening cutoff, coverage, and result, Reference checkpoints, final Cross evidence, and human-review decision. An update to requirements or executable conclusions creates a new revision and regenerates the affected downstream evidence; prior revisions remain unchanged. The \benchmark{} Excalidraw record links its task revision, environment, Test package, Reference sequence, and Cross evidence, allowing every state in Table~\ref{tab:excalidraw-stages} to be traced to the corresponding task record.

\section{Dataset Composition and Quality Evidence}
\label{app:dataset-quality}

This appendix reports the quality evidence and detailed composition behind Section~\ref{sec:dataset-analysis}. All dataset statistics use the final set of 61 \benchmark{} tasks.

\subsection{Trajectory Review and Final Selection}
\label{app:pre-freeze-validation}

Construction and initial human review produced 123 executable candidates with runnable task artifacts and state-by-state execution evidence. Hy3, DeepSeek V4 Flash, DeepSeek V4 Pro, and GLM 5.3 then ran every candidate under multiple protocols. Reviewers examined the complete messages, workspace changes, requirement-level scores, repeated runs, and execution errors, and compared failures with the frozen semantics and evidence.

Historical screening was completed after Test construction and before Reference construction (Appendix~\ref{app:history-screening}). Each admission decision then replayed the dialogue and every Reference checkpoint, including the project, tests, and public observation points (Appendix~\ref{app:review-admission}). Two independent automated checks by Hy3 and DeepSeek V4 Flash complemented this review. Confirmed task defects returned to the responsible stage, after which affected checks and model runs were repeated. The final 61 tasks passed this process and final human acceptance.

\subsection{Language and Software-Form Coverage}
\label{app:dataset-coverage}

\benchmark{} contains 19 Go, 15 Rust, 10 JavaScript, 9 TypeScript, 7 Python, and one Go/TypeScript task. Its software forms comprise 21 CLI, 13 service, ten TUI, seven browser, seven library, two framework, and one embedded task. These attributes describe overlapping repository settings: 14/15 Rust tasks are CLI tasks, 6/9 TypeScript tasks are browser tasks, and Go spans service, TUI, and CLI software (Figure~\ref{fig:dataset-overview}(a)).

\subsection{Requirement States and Acceptance Checks}
\label{app:requirement-horizon}

For a task with $K$ cumulative requirement states, adjacent states define $K-1$ possible evolution transitions. Whether the associated feedback is delivered depends on the behavior already present in the agent's current workspace.

\begin{figure}[H]
  \centering
  \includegraphics[width=\linewidth]{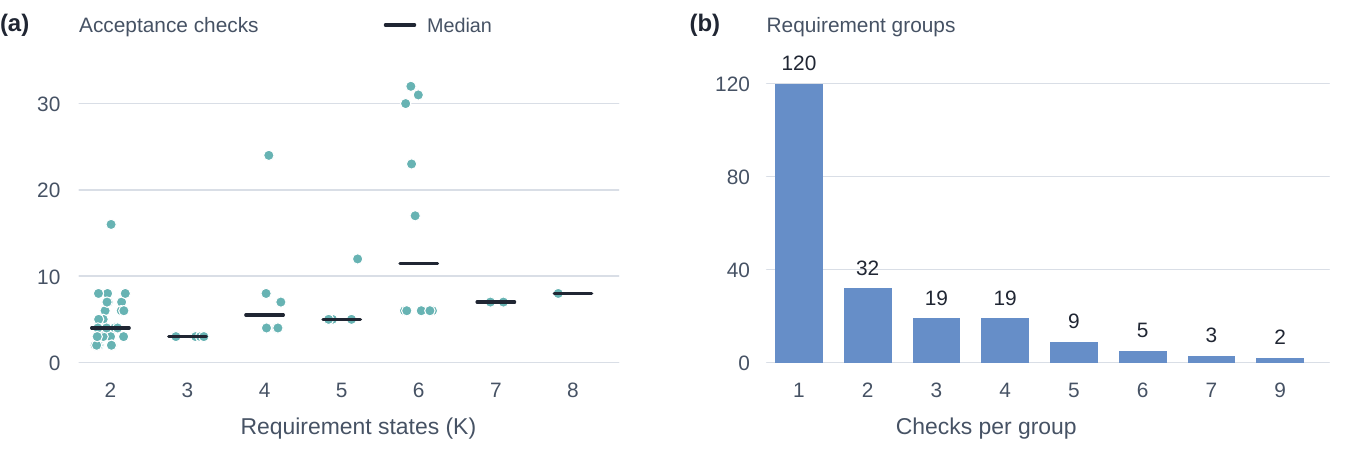}
  \caption{Acceptance-check distributions. (a) Checks per task by state count $K$; dots are tasks, ticks mark medians. (b) Checks per requirement group.}
  \label{fig:dataset-distributions}
\end{figure}

\begin{table}[t]
  \centering
  \small
  \setlength{\tabcolsep}{6pt}
  \caption{Distribution of the number of cumulative states per task.}
  \label{tab:dataset-depth}
  \begin{tabular}{@{}rrrrr@{}}
    \toprule
\rowcolor{FLHeader}
States $K$ & Tasks & States & Transitions & Acceptance checks \\
    \midrule
    2 & 33 & 66 & 33 & 153 \\
\rowcolor{FLStripe}
    3 & 5  & 15 & 10 & 15 \\
    4 & 6  & 24 & 18 & 51 \\
\rowcolor{FLStripe}
    5 & 4  & 20 & 16 & 27 \\
    6 & 10 & 60 & 50 & 163 \\
\rowcolor{FLStripe}
    7 & 2  & 14 & 12 & 14 \\
    8 & 1  & 8  & 7  & 8 \\
    \midrule
    \rowcolor{FLTotal}
    Total & 61 & 207 & 146 & 431 \\
    \bottomrule
  \end{tabular}
\end{table}

The 17 tasks with $K\geq5$ account for 102/207 states and 85/146 possible transitions (Table~\ref{tab:dataset-depth}). Ten six-state tasks contain 163 acceptance checks (37.8\% of all checks), as shown in Figure~\ref{fig:dataset-distributions}(a). The number of states and the number of checks describe different aspects of task size.

\begin{table}[t]
  \centering
  \small
  \setlength{\tabcolsep}{5pt}
  \caption{Software form, number of states, and executable checks.}
  \label{tab:form-depth}
  \begin{tabular}{@{}lrrrrrr@{}}
    \toprule
\rowcolor{FLHeader}
Form & Tasks & States & Mean $K$ & Median $K$ & $K\geq5$ & Checks \\
    \midrule
    Browser   & 7  & 27 & 3.86 & 4 & 3  & 121 \\
\rowcolor{FLStripe}
    CLI       & 21 & 86 & 4.10 & 4 & 10 & 132 \\
    Embedded  & 1  & 2  & 2.00 & 2 & 0  & 4 \\
\rowcolor{FLStripe}
    Framework & 2  & 12 & 6.00 & 6 & 2  & 12 \\
    Library   & 7  & 24 & 3.43 & 3 & 1  & 32 \\
\rowcolor{FLStripe}
    Service   & 13 & 29 & 2.23 & 2 & 0  & 52 \\
    TUI       & 10 & 27 & 2.70 & 2 & 1  & 78 \\
    \bottomrule
  \end{tabular}
\end{table}

The 207 states contain 209 requirement groups because a few states introduce more than one related group. The 431 acceptance checks correspond to a mean of 7.07 per task (median 5, range 2--32) and 2.06 per requirement group (median 1, range 1--9). As Figure~\ref{fig:dataset-distributions}(b) shows, 120/209 groups (57.4\%) have one check, while 57/209 (27.3\%) have at least three. The correlation between a task's state count and acceptance-check count is 0.49, so neither quantity alone captures how much behavior a task contains. Table~\ref{tab:form-depth} also shows that tasks with more states are concentrated in particular software forms.

\subsection{Executable Evidence across Requirement States}
\label{app:trajectory-metrics}

We measure three quantities at each requirement state. \emph{Test groups} counts groups in the cumulative suite. \emph{Test LOC} counts nonblank lines in the unique source files used by that suite. \emph{Reference $\Delta$LOC} is the sum of additions and deletions relative to the initial checkpoint, including replacements.

\begin{figure}[t]
  \centering
  \includegraphics[width=0.98\linewidth]{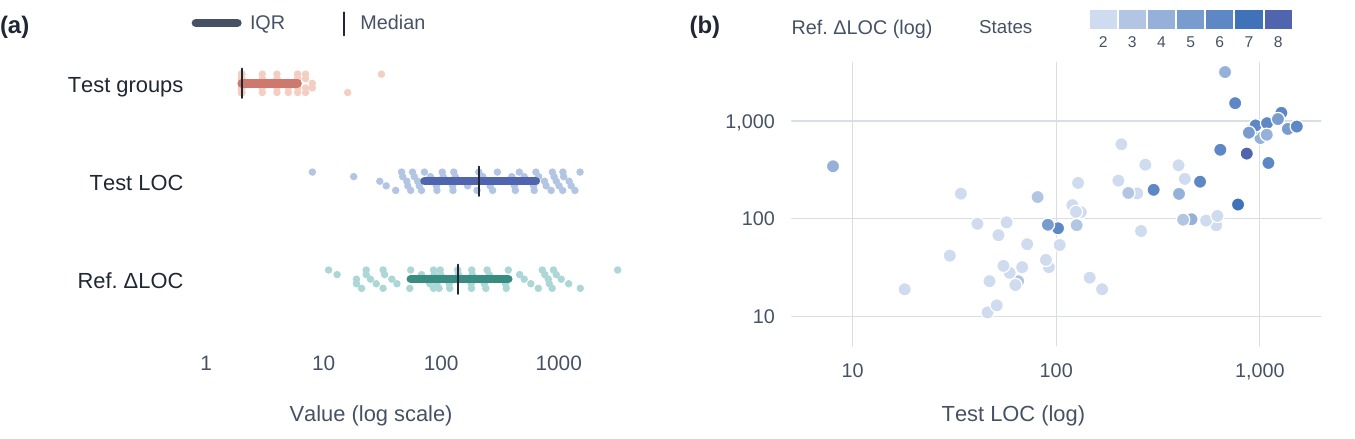}
  \caption{Final-state executable evidence across 61 tasks. (a) Task-level values on a log scale, with interquartile ranges and medians. (b) Final Test LOC versus Reference $\Delta$LOC, colored by state count $K$.}
  \label{fig:trajectory-metrics}
\end{figure}

A state-$i$ mean uses tasks with $K\geq i$. The final-state summary uses the last checkpoint of all 61 tasks. Table~\ref{tab:trajectory-metrics} reports the means, and Figure~\ref{fig:trajectory-metrics} shows their task-level distributions and the relation between Test and Reference edit volume.

\begin{table}[t]
  \centering
  \small
  \setlength{\tabcolsep}{5pt}
  \caption{Executable checks and Reference edits by requirement-state position. Each row averages tasks that
  reach the corresponding state. Test LOC counts nonblank lines in
  the cumulative Test suite; Reference $\Delta$LOC counts additions plus
  deletions from the base checkpoint.}
  \label{tab:trajectory-metrics}
  \begin{tabular}{@{}crrrrr@{}}
\toprule
\rowcolor{FLHeader}
State & Tasks & Test groups & Test LOC & Reference $\Delta$LOC \\
\midrule
1 & 61 & 1.15 & 101.8 & 82.9 \\
\rowcolor{FLStripe}
2 & 61 & 2.28 & 212.2 & 180.5 \\
3 & 28 & 4.04 & 408.1 & 394.0 \\
\rowcolor{FLStripe}
4 & 23 & 5.61 & 598.8 & 563.5 \\
5 & 17 & 6.47 & 732.8 & 557.2 \\
\rowcolor{FLStripe}
6 & 13 & 8.08 & 804.8 & 608.9 \\
7 & 3 & 7.00 & 827.7 & 457.7 \\
\rowcolor{FLStripe}
8 & 1 & 8.00 & 863.0 & 465.0 \\
\midrule
\rowcolor{FLTotal}
Final & 61 & 4.11 & 402.2 & 337.8 \\
\bottomrule
\end{tabular}

\end{table}

\subsection{Repository Coverage}
\label{app:repository-coverage}

\begin{figure}[t]
  \centering
  \includegraphics[width=0.98\linewidth]{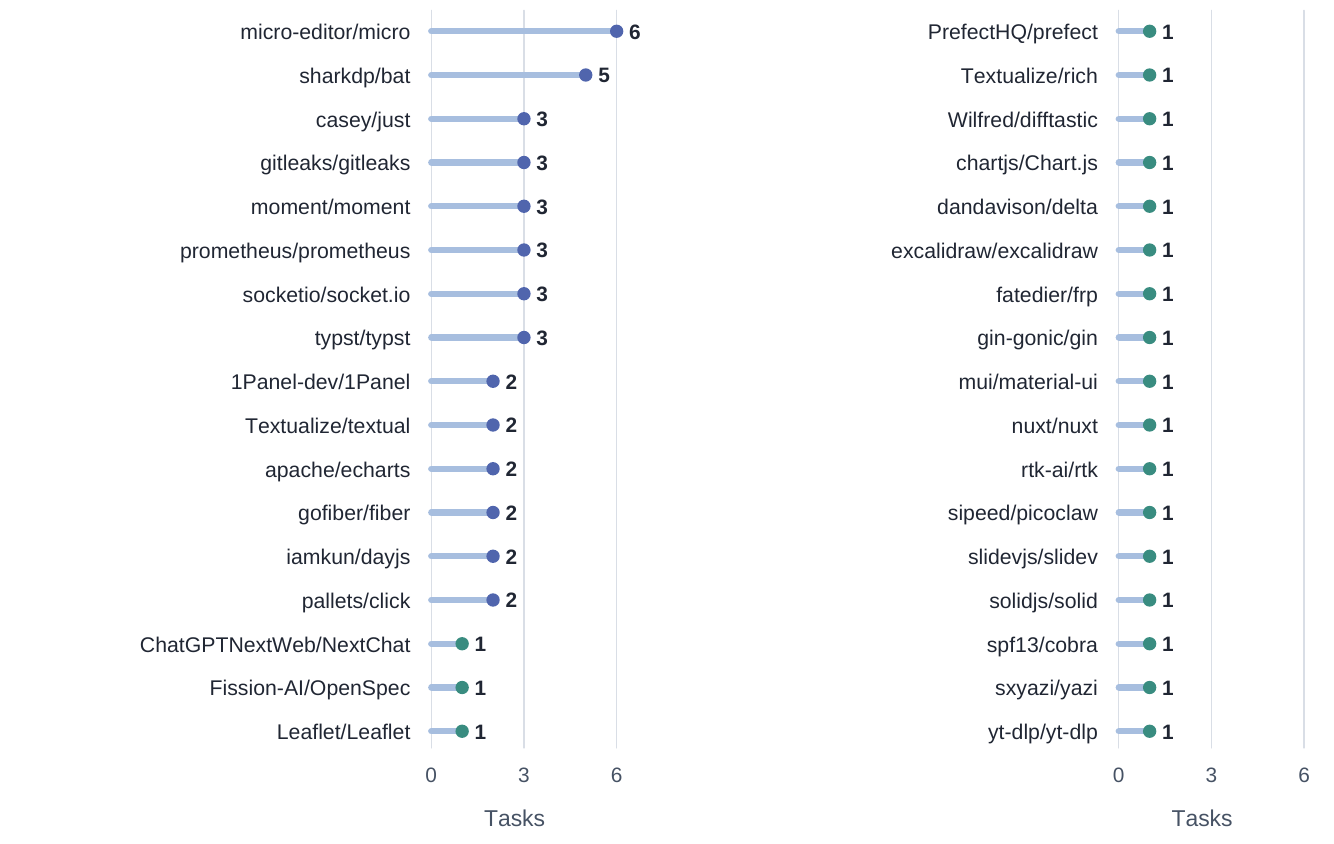}
  \caption{Tasks per repository. Both columns use the same count scale; blue marks indicate multiple-task repositories and teal marks indicate singletons.}
  \label{fig:repository-contribution}
\end{figure}

The 61 tasks come from 34 repositories. Twenty repositories contribute one task, the largest contributor contains six (9.8\%), and the three and five largest contributors account for 23.0\% and 32.8\% of the benchmark, respectively (Figure~\ref{fig:repository-contribution}). Table~\ref{tab:repository-inventory} gives the exact language, software form, state counts, and acceptance checks for every repository. Repository-clustered uncertainty estimates preserve these within-repository groups when resampling model outcomes.

\begin{table}[!htbp]
  \centering
  \footnotesize
  \setlength{\tabcolsep}{3pt}
  \renewcommand{\arraystretch}{1.00}
  \caption{Repository inventory for the 61 \benchmark{} tasks. ``States'' lists
  the number of cumulative requirement states for each task; ``Checks'' is the
  repository-level total of executable acceptance checks.}
  \label{tab:repository-inventory}
  \begin{tabular}{@{}L{0.28\linewidth}r L{0.13\linewidth}L{0.11\linewidth}C{0.19\linewidth}r@{}}
    \toprule
\rowcolor{FLHeader}
Repository & Tasks & Language & Form & States & Checks \\
    \midrule
    micro-editor/micro & 6 & Go & TUI & 6,2,2,2,2,2 & 62 \\
\rowcolor{FLStripe}
    sharkdp/bat & 5 & Rust & CLI & 5,6,2,2,2 & 28 \\
    casey/just & 3 & Rust & CLI & 2,3,2 & 9 \\
\rowcolor{FLStripe}
    gitleaks/gitleaks & 3 & Go & CLI & 2,7,6 & 19 \\
    moment/moment & 3 & JavaScript & Library & 4,4,2 & 14 \\
\rowcolor{FLStripe}
    prometheus/prometheus & 3 & Go & Service & 4,3,2 & 12 \\
    socketio/socket.io & 3 & JavaScript & Service & 2,2,2 & 11 \\
\rowcolor{FLStripe}
    typst/typst & 3 & Rust & CLI & 2,5,4 & 11 \\
    1Panel-dev/1Panel & 2 & Go; Go/TS & Service & 2,2 & 4 \\
\rowcolor{FLStripe}
    apache/echarts & 2 & TypeScript & Browser & 2,2 & 20 \\
    gofiber/fiber & 2 & Go & Service & 2,2 & 14 \\
\rowcolor{FLStripe}
    iamkun/dayjs & 2 & JavaScript & Library & 2,3 & 8 \\
    pallets/click & 2 & Python & CLI & 8,6 & 31 \\
\rowcolor{FLStripe}
    Textualize/textual & 2 & Python & TUI & 2,3 & 7 \\
    chartjs/Chart.js & 1 & TypeScript & Browser & 2 & 3 \\
\rowcolor{FLStripe}
    ChatGPTNextWeb/NextChat & 1 & TypeScript & Browser & 6 & 30 \\
    dandavison/delta & 1 & Rust & CLI & 5 & 5 \\
\rowcolor{FLStripe}
    excalidraw/excalidraw & 1 & TypeScript & Browser & 6 & 32 \\
    fatedier/frp & 1 & Go & Service & 2 & 4 \\
\rowcolor{FLStripe}
    Fission-AI/OpenSpec & 1 & TypeScript & CLI & 6 & 17 \\
    gin-gonic/gin & 1 & Go & Service & 2 & 5 \\
\rowcolor{FLStripe}
    Leaflet/Leaflet & 1 & JavaScript & Browser & 4 & 24 \\
    mui/material-ui & 1 & JavaScript & Library & 7 & 7 \\
\rowcolor{FLStripe}
    nuxt/nuxt & 1 & TypeScript & Framework & 6 & 6 \\
    PrefectHQ/prefect & 1 & Python & Service & 2 & 2 \\
\rowcolor{FLStripe}
    rtk-ai/rtk & 1 & Rust & CLI & 6 & 6 \\
    sipeed/picoclaw & 1 & Go & Embedded & 2 & 4 \\
\rowcolor{FLStripe}
    slidevjs/slidev & 1 & TypeScript & Browser & 5 & 12 \\
    solidjs/solid & 1 & TypeScript & Framework & 6 & 6 \\
\rowcolor{FLStripe}
    spf13/cobra & 1 & Go & Library & 2 & 3 \\
    sxyazi/yazi & 1 & Rust & TUI & 2 & 2 \\
\rowcolor{FLStripe}
    Textualize/rich & 1 & Python & TUI & 4 & 7 \\
    Wilfred/difftastic & 1 & Rust & CLI & 2 & 3 \\
\rowcolor{FLStripe}
    yt-dlp/yt-dlp & 1 & Python & CLI & 3 & 3 \\
    \bottomrule
  \end{tabular}
\end{table}

\subsection{Semantic Composition of Later Requirements}
\label{app:requirement-semantics}

We audited all 147 later requirement groups to characterize the information available before disclosure. The rubric separates support for a general direction from support for the exact tested behavior. It also records the effect of the missing behavior on prior delivery, distance from the preceding workflow, discovery depth, implementation burden, and premature-commitment risk. The resulting labels stratify the trajectory analysis; task definitions and scores are fixed independently.

\subsubsection{Two-stage annotation from the information available before each request}
\label{app:semantic-procedure}

The annotation uses two stages to assess prior information separately from the later requirement. First, the auditor reads the information available to the agent before the target group is disclosed, referred to as the \emph{prefix}. Assuming the previously requested behavior is implemented correctly, it records up to three plausible follow-up directions. Second, the target requirement and its behavioral contract are revealed, and the auditor assigns the eight labels. Reference-code changes are available at this stage to assess implementation burden.

Each judgment concerns the information available immediately before that group, including requests made during earlier turns. Implementation burden is assessed from the changes needed to satisfy the behavior. The evidence contains 138 reference diffs for individual increments and nine cumulative diffs used when an isolated increment is unavailable.

\subsubsection{Annotation provenance and review}
\label{app:semantic-provenance}

Three coding agents produced the initial annotations in batches of 49. In the first batch, the instructions enforced the two-stage reading order within a shared packet. The third batch used separate prefix and target packets and saved the first-stage output before revealing the target. Nine prefix packets in the second batch initially included the target because of an indexing error. We replaced those predictions: after correcting the generator, a fresh annotator read the corrected prefixes without seeing the targets. All 49 corrected packets were checked, and the nine replacement predictions were incorporated into the final labels.

One author reviewed all 147 items. A secondary model-assisted consistency audit then examined every item without solver identities, trajectories, pass/fail outcomes, scores, or rankings. It confirmed 134 items and recorded 13 changes: three substantive semantic corrections, one schema-delimiter correction, and nine provenance corrections for the replaced predictions. Six earlier adjudication overrides are also retained in the review record. The final version, \texttt{semantic-v2-final}, resolves all 147 items. Confidence is high for 65, medium for 81, and low for one; 11 items have mixed semantics.

Thus every item received an initial model annotation, author review, and a second consistency check that hid the evaluated agents' identities and item-level results.

\subsubsection{Rubric and cohort definitions}
\label{app:semantic-rubric}

Table~\ref{tab:semantic-axes} defines the eight dimensions. Levels are ordered within each dimension. Separating need foreseeability from exact-contract determinacy accommodates requirements whose direction is evident while their API, default, output format, or policy remains open.

\begin{table}[H]
\centering\small
\renewcommand{\arraystretch}{1.15}
\caption{Semantic annotation dimensions. The primary evolution type $T$ is categorical; the other axes use ordered levels.}
\label{tab:semantic-axes}
\begin{tabularx}{\linewidth}{@{}L{.06\linewidth}L{.26\linewidth}Y@{}}
\toprule
\rowcolor{FLHeader}
Axis & Question & Levels used in the audit \\
\midrule
$N$ & Is the need foreseeable? & 0: new direction; 1: weakly suggested; 2: a small set of natural continuations; 3: already entailed. \\
\rowcolor{FLStripe}
$X$ & Is exact behavior determined? & 0: new information; 1: several plausible contracts; 2: key behavior constrained; 3: essentially fixed. \\
$G$ & How incomplete is prior delivery? & 0: no gap in stated scope; 1: limited gap; 2: material gap; 3: blocked core use. \\
\rowcolor{FLStripe}
$W$ & How far is the new workflow? & 0: same behavior; 1: same feature; 2: adjacent workflow; 3: only repository/product-level adjacency. \\
$D$ & What reveals the need? & 0: static reasoning; 1: basic trial; 2: realistic workflow; 3: edge cases or lifecycle use; 4: new user information. \\
\rowcolor{FLStripe}
$T$ & What kind of evolution? & Ten categories covering completion, extension, composition, compatibility, robustness, lifecycle, observability, policy, and independent capability. \\
$B$ & How much implementation work? & 0: trivial; 1: bounded local change; 2: coupled implementation; 3: architectural change. \\
\rowcolor{FLStripe}
$C$ & What is the risk of guessing? & 0: low; 1: bounded or opt-in; 2: high-cost commitment to defaults, formats, policy, or shared behavior. \\
\bottomrule
\end{tabularx}
\end{table}

\begin{table}[H]
\centering\small
\renewcommand{\arraystretch}{1.12}
\caption{Requirement cohorts used to analyze prior information. The strict proactive cohort is a subset of the direction-foreseeable cohort. Later-information identifies a type of requirement and applies across development scenarios.}
\label{tab:semantic-cohorts}
\begin{tabularx}{\linewidth}{@{}L{.27\linewidth}L{.26\linewidth}rY@{}}
\toprule
\rowcolor{FLHeader}
Cohort & Rule & Groups & Interpretation \\
\midrule
Direction foreseeable & $N\geq2$ & 42 & Prior support for the need's direction \\
\rowcolor{FLStripe}
Strict proactive & $N\geq2,\ X\geq2,\ C\leq1$ & 16 & Enough support for exact, low-risk coverage \\
Directional, underdetermined & $N\geq2,\ X\leq1,\ C\leq1$ & 25 & Direction supported, exact contract unclear \\
\rowcolor{FLStripe}
Later-information & $N=0$ or $D=4$ & 85 & New direction or information unavailable earlier \\
\bottomrule
\end{tabularx}
\end{table}

Forty-two groups (28.6\%) have a strongly supported direction, but only 16 (10.9\%) meet the strict proactive criterion. The later-information cohort contains 85 groups (57.8\%). The remaining 20 fall outside both the 42-group direction-foreseeable cohort and the 85-group later-information cohort. Implementation burden is coupled or architectural for 93 groups (63.3\%); premature commitment is high-risk for 40 (27.2\%). These cohorts identify when earlier implementation is supported by the available information and when completing the task requires further input.

\paragraph{Workflow breadth.}
Thirty-two groups have workflow-distance label $W3$: they share the repository or product with the preceding request and introduce a distinct capability. For example, one 1Panel task combines OpenResty module validation with a duration-formatting helper. The corpus therefore spans both closely connected feature increments and broader cumulative repository development.

\begin{figure}[htbp]
\centering
\includegraphics[width=.78\linewidth]{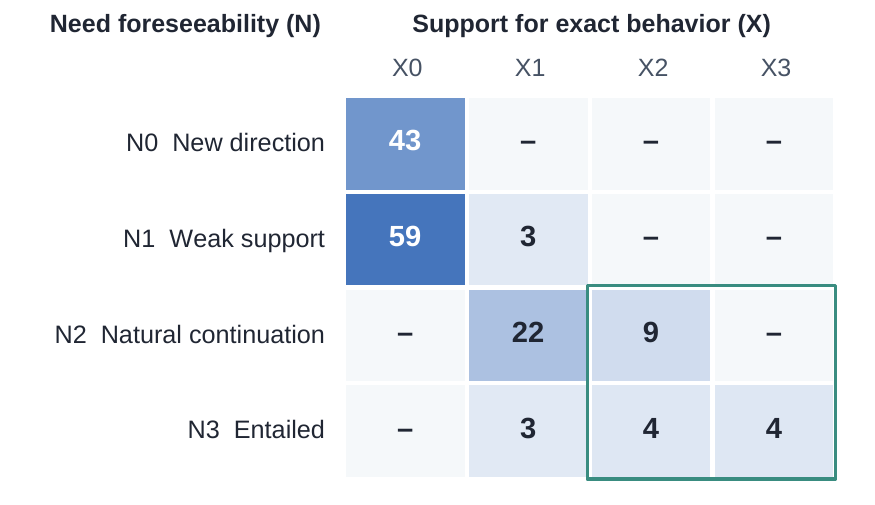}
\caption{Need foreseeability ($N$) and exact-contract determinacy ($X$) across 147 later requirement groups. Cells are group counts. The outlined region has $N\geq2$ and $X\geq2$; the additional commitment-risk filter $C\leq1$ selects the 16-group strict proactive cohort.}
\label{fig:semantic-foreseeability}
\end{figure}

\FloatBarrier

\subsubsection{Complete label inventory}
\label{app:complete-semantic-labels}

Table~\ref{tab:semantic-labels} lists all 147 final labels, identified by repository and the suffix of \texttt{REQEVO-LATER-}. $Q/G$ gives the state and group number. Numeric columns give the level on each axis; $+$ marks mixed semantics, and H/M/L indicates confidence. Evolution types are CC (contract completion), IE (interface extension), WE (workflow extension), CO (composition), CP (compatibility), RO (robustness), SL (state lifecycle), EO (error observability), PP (policy preference), and OC (orthogonal capability).

\begingroup
\footnotesize
\setlength{\tabcolsep}{4pt}
\renewcommand{\arraystretch}{1.12}
\begin{longtable}{@{}rlccccccccccc@{}}
\caption{Complete semantic labels for all 147 later requirement groups. See Table~\ref{tab:semantic-axes} for axis meanings.}\label{tab:semantic-labels}\\
\toprule
\rowcolor{FLHeader}
ID & Repository & $Q/G$ & $N$ & $X$ & $G$ & $W$ & $D$ & $B$ & $C$ & $T$ & Mix & Conf. \\\midrule
\endfirsthead
\multicolumn{13}{l}{\textit{Table \thetable\ continued}}\\
\toprule
\rowcolor{FLHeader}
ID & Repository & $Q/G$ & $N$ & $X$ & $G$ & $W$ & $D$ & $B$ & $C$ & $T$ & Mix & Conf. \\\midrule
\endhead
\bottomrule\endfoot
1 & 1Panel & 2/2 & 0 & 0 & 0 & 3 & 4 & 1 & 1 & OC & -- & H \\
\rowcolor{FLStripe}
2 & 1Panel & 2/2 & 0 & 0 & 0 & 3 & 4 & 1 & 1 & OC & -- & H \\
3 & NextChat & 2/2 & 1 & 0 & 0 & 1 & 4 & 2 & 1 & IE & -- & M \\
\rowcolor{FLStripe}
4 & NextChat & 3/3 & 1 & 0 & 0 & 1 & 4 & 3 & 1 & IE & -- & M \\
5 & NextChat & 4/4 & 1 & 0 & 0 & 2 & 4 & 3 & 1 & WE & -- & M \\
\rowcolor{FLStripe}
6 & NextChat & 5/5 & 3 & 1 & 3 & 0 & 3 & 2 & 0 & EO & + & M \\
7 & NextChat & 6/6 & 0 & 0 & 0 & 2 & 4 & 2 & 1 & PP & -- & H \\
\rowcolor{FLStripe}
8 & OpenSpec & 2/2 & 0 & 0 & 0 & 2 & 4 & 2 & 1 & WE & -- & H \\
9 & OpenSpec & 3/3 & 2 & 1 & 2 & 0 & 2 & 3 & 2 & SL & -- & M \\
\rowcolor{FLStripe}
10 & OpenSpec & 4/4 & 1 & 0 & 1 & 1 & 4 & 2 & 2 & WE & -- & H \\
11 & OpenSpec & 5/5 & 0 & 0 & 0 & 2 & 4 & 2 & 1 & WE & -- & H \\
\rowcolor{FLStripe}
12 & OpenSpec & 6/6 & 0 & 0 & 0 & 2 & 4 & 2 & 1 & WE & -- & H \\
13 & Leaflet & 2/2 & 0 & 0 & 0 & 3 & 4 & 1 & 1 & OC & -- & H \\
\rowcolor{FLStripe}
14 & Leaflet & 3/3 & 0 & 0 & 0 & 3 & 4 & 3 & 1 & OC & -- & H \\
15 & Leaflet & 4/4 & 1 & 0 & 1 & 1 & 4 & 2 & 1 & SL & -- & M \\
\rowcolor{FLStripe}
16 & prefect & 2/2 & 2 & 1 & 3 & 0 & 0 & 1 & 0 & RO & -- & M \\
17 & rich & 2/3 & 0 & 0 & 0 & 2 & 4 & 1 & 1 & IE & -- & H \\
\rowcolor{FLStripe}
18 & rich & 3/4 & 1 & 0 & 1 & 2 & 2 & 2 & 1 & CO & -- & M \\
19 & rich & 4/5 & 1 & 0 & 0 & 1 & 4 & 2 & 1 & WE & + & M \\
\rowcolor{FLStripe}
20 & textual & 2/2 & 1 & 0 & 0 & 2 & 3 & 2 & 1 & SL & -- & H \\
21 & textual & 2/2 & 0 & 0 & 0 & 3 & 3 & 1 & 0 & RO & -- & H \\
\rowcolor{FLStripe}
22 & textual & 3/3 & 0 & 0 & 0 & 3 & 3 & 1 & 0 & RO & -- & H \\
23 & difftastic & 2/2 & 1 & 0 & 0 & 2 & 4 & 2 & 1 & WE & -- & M \\
\rowcolor{FLStripe}
24 & echarts & 2/2 & 3 & 2 & 3 & 0 & 2 & 2 & 0 & CC & + & M \\
25 & echarts & 2/2 & 3 & 2 & 3 & 0 & 2 & 3 & 0 & CC & -- & H \\
\rowcolor{FLStripe}
26 & just & 2/2 & 1 & 0 & 0 & 1 & 4 & 2 & 1 & IE & -- & M \\
27 & just & 2/2 & 1 & 0 & 0 & 1 & 4 & 2 & 1 & PP & -- & M \\
\rowcolor{FLStripe}
28 & just & 3/3 & 0 & 0 & 0 & 3 & 4 & 2 & 1 & OC & -- & H \\
29 & just & 2/2 & 0 & 0 & 0 & 3 & 4 & 1 & 1 & OC & -- & H \\
\rowcolor{FLStripe}
30 & Chart.js & 2/2 & 0 & 0 & 0 & 3 & 4 & 1 & 0 & OC & -- & H \\
31 & delta & 2/2 & 1 & 0 & 1 & 1 & 2 & 3 & 1 & PP & -- & M \\
\rowcolor{FLStripe}
32 & delta & 3/3 & 1 & 0 & 1 & 1 & 3 & 3 & 1 & RO & -- & M \\
33 & delta & 4/4 & 3 & 3 & 3 & 0 & 1 & 3 & 0 & CP & -- & H \\
\rowcolor{FLStripe}
34 & delta & 5/5 & 2 & 1 & 2 & 1 & 0 & 1 & 0 & CC & -- & M \\
35 & excalidraw & 2/2 & 3 & 2 & 3 & 0 & 1 & 3 & 1 & CC & + & H \\
\rowcolor{FLStripe}
36 & excalidraw & 3/3 & 1 & 0 & 2 & 0 & 1 & 2 & 1 & CC & + & L \\
37 & excalidraw & 4/4 & 3 & 2 & 3 & 0 & 1 & 3 & 1 & CC & + & H \\
\rowcolor{FLStripe}
38 & excalidraw & 5/5 & 0 & 0 & 0 & 3 & 4 & 3 & 2 & OC & -- & H \\
39 & excalidraw & 6/6 & 1 & 0 & 0 & 1 & 4 & 3 & 2 & IE & -- & M \\
\rowcolor{FLStripe}
40 & frp & 2/2 & 1 & 0 & 1 & 2 & 2 & 2 & 2 & WE & -- & M \\
41 & gin & 2/2 & 0 & 0 & 0 & 3 & 3 & 1 & 0 & RO & -- & H \\
\rowcolor{FLStripe}
42 & gitleaks & 2/2 & 1 & 0 & 0 & 0 & 2 & 2 & 2 & CO & -- & M \\
43 & gitleaks & 2/2 & 1 & 0 & 0 & 2 & 4 & 2 & 2 & WE & -- & M \\
\rowcolor{FLStripe}
44 & gitleaks & 3/3 & 0 & 0 & 0 & 2 & 4 & 1 & 2 & WE & -- & H \\
45 & gitleaks & 4/4 & 1 & 0 & 1 & 2 & 4 & 2 & 1 & IE & -- & M \\
\rowcolor{FLStripe}
46 & gitleaks & 5/5 & 1 & 0 & 1 & 2 & 3 & 2 & 1 & RO & -- & M \\
47 & gitleaks & 6/6 & 2 & 1 & 2 & 0 & 2 & 2 & 2 & CP & + & M \\
\rowcolor{FLStripe}
48 & gitleaks & 7/7 & 0 & 0 & 0 & 2 & 4 & 1 & 0 & IE & -- & H \\
49 & gitleaks & 2/2 & 0 & 0 & 0 & 2 & 4 & 2 & 2 & WE & -- & H \\
\rowcolor{FLStripe}
50 & gitleaks & 3/3 & 2 & 1 & 1 & 0 & 3 & 2 & 2 & CO & -- & M \\
51 & gitleaks & 4/4 & 0 & 0 & 0 & 2 & 4 & 1 & 1 & IE & -- & H \\
\rowcolor{FLStripe}
52 & gitleaks & 5/5 & 2 & 1 & 2 & 0 & 3 & 1 & 0 & EO & -- & M \\
53 & gitleaks & 6/6 & 1 & 0 & 0 & 1 & 4 & 2 & 1 & PP & -- & M \\
\rowcolor{FLStripe}
54 & fiber & 2/2 & 0 & 0 & 0 & 3 & 4 & 1 & 1 & OC & -- & H \\
55 & fiber & 2/2 & 1 & 0 & 0 & 3 & 4 & 2 & 0 & SL & -- & H \\
\rowcolor{FLStripe}
56 & dayjs & 2/2 & 1 & 0 & 0 & 1 & 4 & 1 & 1 & IE & -- & H \\
57 & dayjs & 2/2 & 2 & 2 & 2 & 0 & 1 & 2 & 0 & CC & -- & M \\
\rowcolor{FLStripe}
58 & dayjs & 3/3 & 2 & 2 & 2 & 0 & 2 & 2 & 0 & CO & -- & M \\
59 & micro & 2/2 & 1 & 0 & 0 & 1 & 4 & 2 & 1 & WE & -- & M \\
\rowcolor{FLStripe}
60 & micro & 3/3 & 1 & 0 & 0 & 1 & 4 & 2 & 1 & WE & -- & M \\
61 & micro & 4/4 & 1 & 0 & 0 & 2 & 4 & 2 & 2 & WE & -- & M \\
\rowcolor{FLStripe}
62 & micro & 5/5 & 2 & 2 & 2 & 0 & 2 & 1 & 1 & CP & -- & M \\
63 & micro & 6/6 & 1 & 0 & 1 & 2 & 2 & 2 & 2 & SL & -- & M \\
\rowcolor{FLStripe}
64 & micro & 2/2 & 0 & 0 & 0 & 3 & 4 & 0 & 0 & OC & -- & H \\
65 & micro & 2/2 & 0 & 0 & 0 & 3 & 4 & 1 & 1 & PP & -- & H \\
\rowcolor{FLStripe}
66 & micro & 2/2 & 0 & 0 & 0 & 3 & 4 & 1 & 1 & OC & -- & H \\
67 & micro & 2/2 & 0 & 0 & 0 & 3 & 4 & 1 & 2 & OC & -- & H \\
\rowcolor{FLStripe}
68 & micro & 2/2 & 0 & 0 & 0 & 3 & 4 & 2 & 1 & OC & -- & H \\
69 & moment & 2/2 & 1 & 0 & 0 & 1 & 3 & 2 & 1 & PP & -- & M \\
\rowcolor{FLStripe}
70 & moment & 3/3 & 3 & 1 & 2 & 0 & 3 & 2 & 1 & CO & + & M \\
71 & moment & 4/4 & 3 & 3 & 3 & 2 & 2 & 2 & 0 & CC & -- & H \\
\rowcolor{FLStripe}
72 & moment & 2/2 & 2 & 1 & 1 & 0 & 3 & 1 & 0 & RO & -- & M \\
73 & moment & 3/3 & 1 & 0 & 0 & 1 & 4 & 2 & 1 & IE & -- & M \\
\rowcolor{FLStripe}
74 & moment & 4/4 & 1 & 0 & 0 & 1 & 4 & 2 & 1 & EO & -- & H \\
75 & moment & 2/2 & 0 & 0 & 0 & 3 & 4 & 1 & 1 & OC & -- & H \\
\rowcolor{FLStripe}
76 & material-ui & 2/2 & 2 & 1 & 2 & 1 & 1 & 2 & 1 & CC & -- & M \\
77 & material-ui & 3/3 & 1 & 0 & 0 & 1 & 4 & 2 & 1 & PP & -- & M \\
\rowcolor{FLStripe}
78 & material-ui & 4/4 & 1 & 0 & 0 & 1 & 4 & 2 & 1 & PP & -- & H \\
79 & material-ui & 5/5 & 1 & 1 & 0 & 2 & 2 & 1 & 0 & CC & -- & M \\
\rowcolor{FLStripe}
80 & material-ui & 6/6 & 1 & 1 & 0 & 2 & 2 & 1 & 2 & CC & -- & M \\
81 & material-ui & 7/7 & 1 & 0 & 0 & 1 & 4 & 2 & 1 & IE & -- & H \\
\rowcolor{FLStripe}
82 & nuxt & 2/2 & 3 & 3 & 3 & 0 & 1 & 2 & 0 & CC & -- & H \\
83 & nuxt & 3/3 & 2 & 1 & 1 & 0 & 3 & 1 & 2 & CP & -- & M \\
\rowcolor{FLStripe}
84 & nuxt & 4/4 & 1 & 0 & 0 & 2 & 4 & 2 & 2 & PP & -- & H \\
85 & nuxt & 5/5 & 1 & 0 & 1 & 2 & 2 & 2 & 2 & IE & -- & M \\
\rowcolor{FLStripe}
86 & nuxt & 6/6 & 0 & 0 & 0 & 2 & 4 & 2 & 2 & IE & -- & H \\
87 & click & 2/2 & 1 & 0 & 0 & 2 & 4 & 2 & 1 & WE & -- & M \\
\rowcolor{FLStripe}
88 & click & 3/3 & 2 & 1 & 1 & 2 & 2 & 2 & 2 & CO & -- & M \\
89 & click & 4/4 & 1 & 0 & 0 & 2 & 4 & 2 & 2 & IE & -- & H \\
\rowcolor{FLStripe}
90 & click & 5/5 & 2 & 1 & 2 & 1 & 3 & 1 & 1 & CO & + & M \\
91 & click & 6/6 & 1 & 0 & 0 & 2 & 4 & 1 & 2 & RO & -- & M \\
\rowcolor{FLStripe}
92 & click & 7/7 & 0 & 0 & 0 & 2 & 4 & 3 & 2 & IE & -- & H \\
93 & click & 8/8 & 1 & 0 & 0 & 2 & 4 & 2 & 1 & IE & -- & H \\
\rowcolor{FLStripe}
94 & click & 2/2 & 1 & 0 & 0 & 2 & 4 & 2 & 1 & WE & -- & M \\
95 & click & 3/3 & 1 & 0 & 0 & 2 & 4 & 2 & 2 & IE & -- & H \\
\rowcolor{FLStripe}
96 & click & 4/4 & 2 & 1 & 1 & 1 & 2 & 2 & 2 & EO & -- & M \\
97 & click & 5/5 & 0 & 0 & 0 & 3 & 4 & 1 & 2 & OC & -- & H \\
\rowcolor{FLStripe}
98 & click & 6/6 & 1 & 0 & 0 & 2 & 4 & 3 & 2 & IE & -- & H \\
99 & prometheus & 2/2 & 1 & 0 & 0 & 1 & 4 & 1 & 1 & WE & -- & M \\
\rowcolor{FLStripe}
100 & prometheus & 3/3 & 2 & 1 & 0 & 1 & 3 & 1 & 0 & RO & -- & M \\
101 & prometheus & 4/4 & 1 & 0 & 0 & 1 & 4 & 1 & 1 & IE & -- & M \\
\rowcolor{FLStripe}
102 & prometheus & 2/2 & 2 & 2 & 2 & 0 & 2 & 1 & 0 & EO & -- & M \\
103 & prometheus & 3/3 & 0 & 0 & 0 & 3 & 4 & 3 & 1 & OC & -- & H \\
\rowcolor{FLStripe}
104 & prometheus & 2/2 & 0 & 0 & 0 & 2 & 3 & 1 & 0 & RO & -- & H \\
105 & rtk & 2/2 & 2 & 2 & 0 & 2 & 2 & 0 & 1 & WE & -- & H \\
\rowcolor{FLStripe}
106 & rtk & 3/3 & 1 & 0 & 0 & 2 & 4 & 2 & 1 & IE & -- & M \\
107 & rtk & 4/4 & 1 & 1 & 0 & 1 & 2 & 1 & 1 & IE & -- & M \\
\rowcolor{FLStripe}
108 & rtk & 5/5 & 1 & 0 & 0 & 2 & 4 & 2 & 1 & IE & -- & M \\
109 & rtk & 6/6 & 1 & 0 & 0 & 2 & 4 & 2 & 2 & IE & -- & M \\
\rowcolor{FLStripe}
110 & bat & 2/2 & 2 & 1 & 1 & 0 & 2 & 2 & 1 & CC & -- & M \\
111 & bat & 3/3 & 0 & 0 & 0 & 3 & 4 & 1 & 1 & OC & -- & H \\
\rowcolor{FLStripe}
112 & bat & 4/4 & 0 & 0 & 0 & 3 & 3 & 1 & 0 & RO & -- & H \\
113 & bat & 5/5 & 0 & 0 & 0 & 3 & 4 & 2 & 2 & PP & -- & H \\
\rowcolor{FLStripe}
114 & bat & 2/2 & 0 & 0 & 0 & 3 & 4 & 2 & 2 & OC & -- & H \\
115 & bat & 3/3 & 0 & 0 & 0 & 3 & 2 & 1 & 0 & RO & -- & H \\
\rowcolor{FLStripe}
116 & bat & 4/4 & 0 & 0 & 0 & 3 & 4 & 2 & 2 & PP & -- & H \\
117 & bat & 5/5 & 1 & 0 & 0 & 1 & 4 & 1 & 2 & CP & -- & M \\
\rowcolor{FLStripe}
118 & bat & 6/6 & 2 & 2 & 1 & 2 & 0 & 1 & 0 & CC & -- & H \\
119 & bat & 2/2 & 0 & 0 & 0 & 3 & 4 & 2 & 1 & PP & -- & H \\
\rowcolor{FLStripe}
120 & bat & 2/2 & 1 & 0 & 0 & 2 & 2 & 2 & 1 & PP & -- & M \\
121 & bat & 2/2 & 0 & 0 & 0 & 3 & 4 & 1 & 1 & OC & -- & H \\
\rowcolor{FLStripe}
122 & picoclaw & 2/2 & 1 & 0 & 0 & 2 & 3 & 1 & 0 & RO & -- & M \\
123 & slidev & 2/2 & 2 & 1 & 2 & 1 & 1 & 2 & 2 & CC & -- & M \\
\rowcolor{FLStripe}
124 & slidev & 2/3 & 2 & 1 & 1 & 1 & 1 & 2 & 1 & IE & -- & M \\
125 & slidev & 3/4 & 1 & 0 & 0 & 1 & 2 & 0 & 1 & IE & -- & M \\
\rowcolor{FLStripe}
126 & slidev & 4/5 & 1 & 0 & 0 & 2 & 3 & 1 & 0 & SL & -- & M \\
127 & slidev & 5/6 & 2 & 2 & 2 & 2 & 2 & 1 & 1 & CO & -- & H \\
\rowcolor{FLStripe}
128 & socket.io & 2/2 & 2 & 1 & 1 & 1 & 2 & 2 & 1 & WE & -- & M \\
129 & socket.io & 2/2 & 0 & 0 & 0 & 3 & 3 & 1 & 1 & EO & -- & H \\
\rowcolor{FLStripe}
130 & socket.io & 2/2 & 0 & 0 & 0 & 3 & 3 & 1 & 2 & IE & -- & H \\
131 & solid & 2/2 & 2 & 1 & 2 & 1 & 2 & 2 & 1 & WE & -- & M \\
\rowcolor{FLStripe}
132 & solid & 3/3 & 3 & 1 & 3 & 0 & 2 & 2 & 0 & CC & -- & M \\
133 & solid & 4/4 & 2 & 1 & 2 & 2 & 3 & 3 & 1 & CO & -- & M \\
\rowcolor{FLStripe}
134 & solid & 5/5 & 2 & 1 & 1 & 1 & 2 & 2 & 1 & SL & -- & M \\
135 & solid & 6/6 & 2 & 2 & 2 & 0 & 0 & 1 & 0 & CC & + & M \\
\rowcolor{FLStripe}
136 & cobra & 2/2 & 0 & 0 & 0 & 3 & 3 & 1 & 0 & RO & -- & H \\
137 & yazi & 2/2 & 1 & 0 & 0 & 1 & 4 & 1 & 1 & IE & -- & M \\
\rowcolor{FLStripe}
138 & typst & 2/2 & 2 & 2 & 2 & 0 & 2 & 2 & 2 & CC & -- & M \\
139 & typst & 2/2 & 2 & 1 & 1 & 0 & 2 & 2 & 1 & RO & -- & M \\
\rowcolor{FLStripe}
140 & typst & 3/3 & 1 & 0 & 0 & 2 & 4 & 2 & 2 & WE & -- & M \\
141 & typst & 4/4 & 1 & 0 & 0 & 2 & 3 & 2 & 2 & RO & -- & M \\
\rowcolor{FLStripe}
142 & typst & 5/5 & 1 & 0 & 0 & 1 & 4 & 2 & 1 & IE & -- & M \\
143 & typst & 2/2 & 1 & 0 & 0 & 2 & 3 & 3 & 2 & IE & -- & M \\
\rowcolor{FLStripe}
144 & typst & 3/3 & 1 & 0 & 0 & 2 & 4 & 3 & 2 & IE & -- & M \\
145 & typst & 4/4 & 1 & 0 & 0 & 1 & 4 & 3 & 2 & WE & + & M \\
\rowcolor{FLStripe}
146 & yt-dlp & 2/2 & 2 & 1 & 2 & 0 & 3 & 1 & 0 & RO & -- & M \\
147 & yt-dlp & 3/3 & 3 & 3 & 3 & 0 & 3 & 0 & 0 & CC & -- & H \\
\end{longtable}
\endgroup

\FloatBarrier

\section{Experimental Details and Protocol Implementation}
\label{app:experimental-details}

\subsection{Evaluated Models}
\label{app:experimental-models}

We evaluate 12 models, with one result for each of the 61 tasks under each of the three workflows: 2,196 model--task--workflow runs in total. Table~\ref{tab:experimental-models} records the releases or major updates used to define this July--September 2026 cohort and links the corresponding provider documentation.

\begin{table}[htbp]
\centering\small
\setlength{\tabcolsep}{5pt}
\renewcommand{\arraystretch}{1.12}
\caption{Evaluated models and their release or update dates. Dates use year--month--day or year--month. 
The linked pages identify the model. DeepSeek V4 dates refer to updates after the earlier preview.}
\label{tab:experimental-models}
\begin{tabular}{@{}llcl@{}}
\toprule
\rowcolor{FLHeader}
Model & Provider & Release date & Source \\
\midrule
DeepSeek V4 Flash & DeepSeek & 2026-07-31 & \href{https://huggingface.co/deepseek-ai/DeepSeek-V4-Flash}{Model record} \\
\rowcolor{FLStripe}
DeepSeek V4 Pro & DeepSeek & 2026-08-13 & \href{https://huggingface.co/deepseek-ai/DeepSeek-V4-Pro}{Model record} \\
DeepSeek V4.1 Flash & DeepSeek & 2026-09-10 & \href{https://huggingface.co/deepseek-ai/DeepSeek-V4.1-Flash}{Model record} \\
\rowcolor{FLStripe}
GLM 5.3 & Zhipu AI & 2026-08-19 & \href{https://huggingface.co/zai-org/GLM-5.3/commits/main/README.md}{Model record} \\
GLM 5.3 Flash & Zhipu AI & 2026-08-26 & \href{https://z.ai/blog/glm-5.3-flash}{Release note} \\
\rowcolor{FLStripe}
Hy3 & Tencent Hunyuan & 2026-07-06 & \href{https://www.tencent.com/tencent-hy3-now-available-globally-extending-practical-ai-across-products-workflows-and-cloud-services/}{Release note} \\
Kimi K3 & Moonshot AI & 2026-07-19 & \href{https://www.kimi.ai/blog/kimi-k3}{Release note} \\
\rowcolor{FLStripe}
Qwen 3.8 Flash & Alibaba Qwen & 2026-08-26 & \href{https://huggingface.co/Qwen/Qwen3.8-Flash-Next}{Model record} \\
GPT-5.6 Sol & OpenAI & 2026-07-09 & \href{https://openai.com/index/gpt-5-6/}{Release note} \\
\rowcolor{FLStripe}
GPT-6 Astra & OpenAI & 2026-09-03 & \href{https://openai.com/index/safety-overview-gpt-6-astra/}{Release note} \\
Grok 4.6 & xAI & 2026-08-12 & \href{https://docs.x.ai/developers/release-notes}{Release note} \\
\rowcolor{FLStripe}
Hy4-Preview & Tencent Hunyuan & 2026-08-28 & \href{https://hy.tencent.ai/research/hy4-preview}{Release note} \\
\bottomrule
\end{tabular}
\end{table}

\subsection{Harness and Runtime Configuration}
\label{app:harness-configuration}

Agent interfaces affect repository-task performance \citep{yang2024sweagent}. We therefore use Claude Code 2.1.74 as the shared coding harness for all 2,196 model--task--scenario runs. Workspaces are isolated across model--task--protocol runs. Within each progressive run, the writable repository and conversation persist throughout development. The active context window is 200,000 tokens, with high reasoning effort where supported by the model route. Sampling uses each provider's default temperature and top-$p$. Table~\ref{tab:runtime-configuration} records the shared configuration.

\begin{table}[htbp]
\centering\small
\renewcommand{\arraystretch}{1.12}
\caption{Execution settings. A response can contain multiple model calls and tool interactions.}
\label{tab:runtime-configuration}
\begin{tabularx}{\linewidth}{@{}L{.30\linewidth}Y@{}}
\toprule
\rowcolor{FLHeader}
Setting & Value \\
\midrule
Coding harness & Claude Code 2.1.74 \\
\rowcolor{FLStripe}
Context window & 200,000 tokens per active context \\
Reasoning effort & High in the recorded configuration \\
\rowcolor{FLStripe}
Temperature / top-$p$ & Provider defaults \\
Full budget & One complete coding-agent session \\
\rowcolor{FLStripe}
Evo budget & One coding-agent response per disclosed state \\
Evo+Comp budget & Initial response and at most two completion revisions per state \\
\rowcolor{FLStripe}
Feedback generator & Hy4-Preview, high reasoning effort \\
Feedback-generation limit & 20,000 tokens per generation \\
\rowcolor{FLStripe}
Observation selection & At most two predefined public observations per message \\
Evaluation environment & Frozen task environment; network-disabled execution \\
\bottomrule
\end{tabularx}
\end{table}

A complete run contains repeated model invocations and context reuse. The context limit applies to the active context of an invocation; Agent Tokens accumulates reported input and output tokens over every call in the run.

Feedback quality and revision compute affect code-repair outcomes \citep{olausson2024selfrepair}. To standardize these aspects, completion feedback is generated by a fixed Hy4-Preview model with high reasoning effort and a 20,000-token generation limit. Its input contains the current public check status and the task's predefined mapping from failed checks to observable problems. It selects up to two observations and expresses them as a natural-language report. This limit keeps the amount of information in each report comparable across models and focuses the next revision on a small set of concrete problems. The two-revision limit separately bounds repeated intervention at each stage. Executable checks determine correctness; the generator expresses the resulting evidence to the agent.

\subsection{Protocol Execution}
\label{app:protocol-implementation}

\begin{figure}[htbp]
\centering
\includegraphics[width=\linewidth]{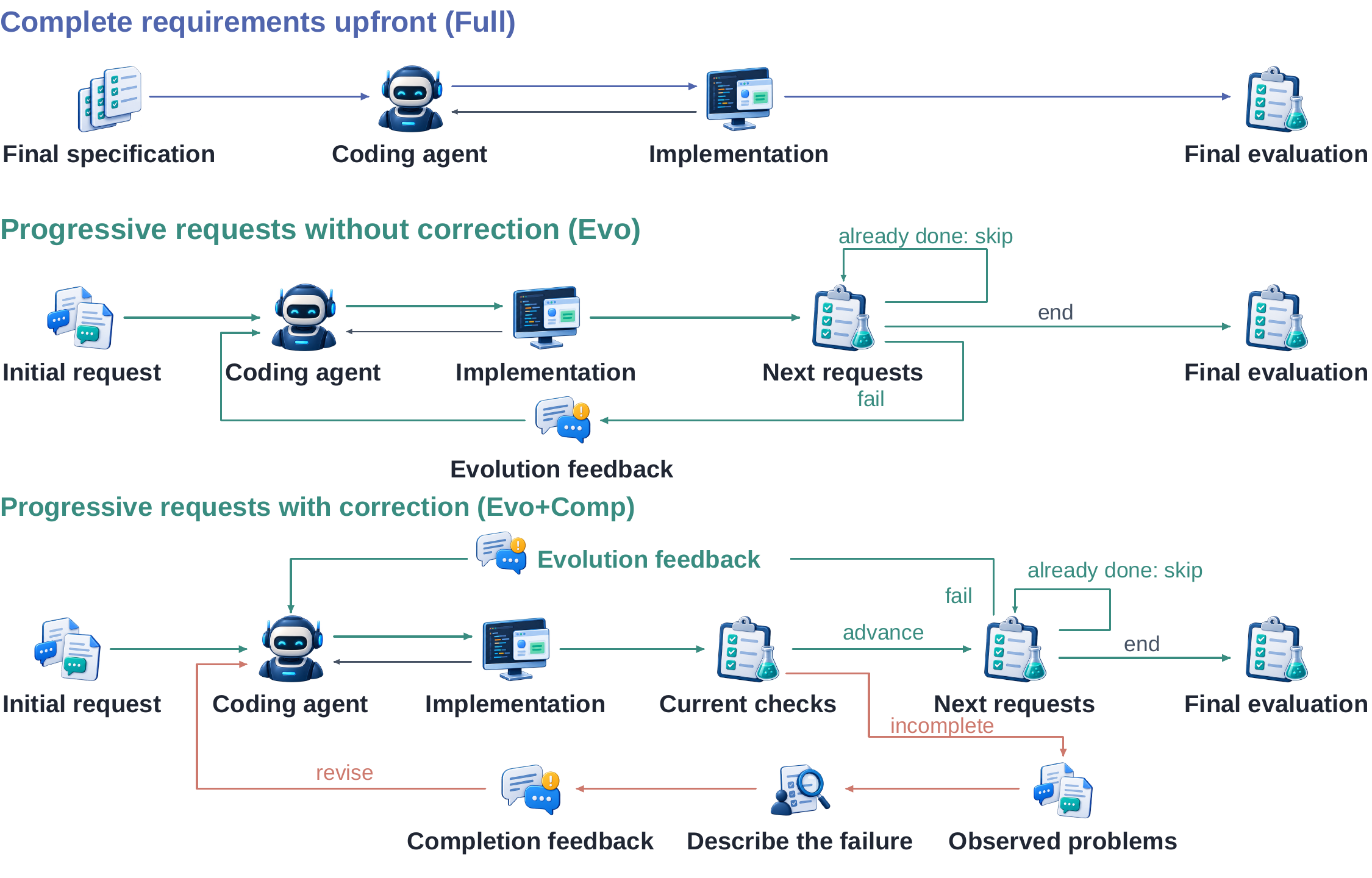}
\caption{Full protocol information flow. Green paths carry requirement disclosures; orange paths carry reports of observed failures. The skip branches bypass behavior already implemented.}
\label{fig:protocols-full}
\end{figure}

All three workflows target the same final requirement set $Q_K$ and start from the same task checkpoint, but each is an independent execution. Their information schedules and opportunities for action differ.

\paragraph{Full Specification (Full).}
The controller supplies the complete final specification in the opening request. The agent explores, edits, and tests the repository within one coding session. The evaluator then checks the final workspace. This workflow measures implementation from complete upfront information.

\paragraph{Evolution-Only Feedback (Evo).}
The controller first supplies $u_0$. After the agent's response, it evaluates the persistent workspace against the next requirement state. Requirements already satisfied need not be disclosed. Otherwise, it supplies the corresponding evolution message and lets the agent modify the same workspace. The process continues through the fixed state sequence. Failures of previously stated behavior do not produce completion messages, but those requirements remain active in all subsequent cumulative checks.

\paragraph{Evolution + Completion Feedback (Evo+Comp).}
This protocol adds within-state correction. After a response, the evaluator checks the active cumulative requirements. When it obtains attributable failure evidence, the feedback generator expresses up to two predefined observations as a user-facing message. The agent can revise the workspace at most twice at that state. The controller then advances even if requirements remain unmet. Earlier failures remain active obligations in the agent's workspace.

\paragraph{Normal execution path.}
The controller performs the following steps:
\begin{enumerate}
\setlength{\itemsep}{2pt}
\item Deliver the initial request and obtain a coding-agent response.
\item Execute the current cumulative checks. In Evo+Comp, an observed failure can trigger a completion message and another response, up to the two-revision limit.
\item Before each later disclosure, check the current workspace. Skip already satisfied behavior; otherwise disclose the missing requirement information for that state.
\item Obtain the agent's response, apply the protocol's within-state checking and correction rule, and advance to the next state.
\item After the final state and its permitted revisions, evaluate every active requirement group and the repository baseline.
\end{enumerate}

The generator uses the predefined observations to describe public behavior, such as a command returning the wrong result or an interface failing after a specified action. Hidden test identifiers, assertions, reference code, and future requirements stay outside the message context. After two corrective revisions, the controller proceeds to the next requirement stage and keeps all earlier requirements in the cumulative checks.

\subsection{Information and Response Opportunities}
\label{app:protocol-comparison}

\begin{table}[htbp]
\centering\small
\renewcommand{\arraystretch}{1.12}
\caption{Information and response opportunities. Earlier requirements remain final acceptance obligations in all protocols.}
\label{tab:protocol-comparison}
\begin{tabularx}{\linewidth}{@{}lYYY@{}}
\toprule
\rowcolor{FLHeader}
Protocol & Initial information & Subsequent information & Agent responses \\
\midrule
Full & Complete final specification & None & One session \\
\rowcolor{FLStripe}
Evo & Initial request & Missing later requirements & One per disclosed state \\
Evo+Comp & Initial request & Missing later requirements and observed failures of active requirements & First response plus at most two revisions per state \\
\bottomrule
\end{tabularx}
\end{table}

Each comparison matches the same model and task across independently executed workflows. The paired change summarizes the complete development process; direct recovery measures a failed state repaired within one execution.

\subsection{Observation Availability and Feedback-Event Accounting}
\label{app:event-accounting}

The interaction ledger and the recovery denominator describe different units. The 732 Evo+Comp runs contain 2,484 state records and 1,723 delivered completion messages. Of these states, 733 have an observed first-response failure and qualify as recovery opportunities. One opportunity can receive either one or two messages. Table~\ref{tab:feedback-event-ledger} reconciles these counts.

\begin{table}[htbp]
\centering\small
\caption{Aggregate feedback-event accounting for Evo+Comp. Recovery eligibility requires an observed first-response failure; message counts record delivery regardless of eligibility.}
\label{tab:feedback-event-ledger}
\begin{tabular}{@{}ccrr@{}}
\toprule
\rowcolor{FLHeader}
Messages per state & Recovery-eligible & States & Delivered messages \\
\midrule
0 & No & 1,564 & 0 \\
\rowcolor{FLStripe}
1 & Yes & 95 & 95 \\
2 & Yes & 638 & 1,276 \\
\rowcolor{FLStripe}
1 & No & 22 & 22 \\
2 & No & 165 & 330 \\
\midrule
\rowcolor{FLTotal}
\multicolumn{2}{l}{Total} & 2,484 & 1,723 \\
\bottomrule
\end{tabular}
\end{table}

The 187 feedback-bearing states with unavailable first-response observations contribute 352 messages. The controller records typed execution or suite-availability outcomes for these states. Message accounting retains the delivered interaction, while recovery analysis uses the 733 observed failures. Terminal scoring uses each run's final execution result.

Metric-specific projections select the observations required by each definition. FDAC uses 2,390 states with usable scoring observations. The state-outcome summary contains 151 unobserved records and 140 delayed solutions; recovery additionally requires an observed first-response failure, selecting 113 repairs. Appendix~\ref{app:feedback-metrics} defines the recovery unit, and Appendix~\ref{app:fdac} gives FDAC's coverage rule.

\section{Metrics and Statistical Analysis}
\label{app:experiment-metrics}

All metrics use the frozen task definitions and recorded trajectories. A model--task--protocol run is the unit of terminal evaluation. A requirement state is the unit of controller interaction and direct recovery. Requirement groups are the units of local correctness and the semantic analysis. The 207 states contain 209 groups, and the 146 later states contain 147 later groups.

\subsection{Final Task Completion}
\label{app:terminal-metrics}

For model $m$, task $\tau$, and protocol $p$, let $W_T$ be the final workspace, $b(W_T)$ its baseline result, and $Z_g(W_T)$ the result for requirement group $g$. Then
\begin{equation}
\operatorname{StrictPass}_{m,p}(\tau)
=\mathbf{1}\!\left[b(W_T)=1\ \land\ \bigwedge_{g\in Q_K}Z_g(W_T)=1\right].
\label{eq:strict-pass}
\end{equation}
The model's StrictPass rate is the mean over 61 tasks. Partial completion is
\begin{equation}
\operatorname{NodeCredit}_{m,p}(\tau)
=\frac{1}{|Q_K|}\sum_{g\in Q_K}Z_g(W_T).
\label{eq:node-credit}
\end{equation}
NodeCredit gives each task equal weight and retains partial requirement completion. StrictPass additionally requires every group and the repository baseline to pass.

\subsection{Initial Delivery}
\label{app:initial-delivery-metrics}

Initial-delivery results use runs with progressive requests and no corrective feedback. Let $\operatorname{precheck}_i$ indicate that a state passes at the pre-disclosure check, and let $\operatorname{firstpass}_i$ indicate that it passes after the first agent response at that state. Initial Usable Delivery (IUD) is
\begin{equation}
\operatorname{IUD}(\tau)=\mathbf{1}[\operatorname{precheck}_1\lor\operatorname{firstpass}_1].
\label{eq:iud}
\end{equation}
It measures usability within the initial evaluation window, including pre-existing behavior. Initial Complete Delivery (ICD) additionally requires every later state to pass before its disclosure:
\begin{equation}
\operatorname{ICD}(\tau)=\operatorname{StrictPass}(\tau)\operatorname{IUD}(\tau)
\prod_{i=2}^{K}\mathbf{1}[\operatorname{precheck}_i].
\label{eq:icd}
\end{equation}
ICD measures whether the initial delivery already satisfies the task's frozen final requirements. Proactive coverage is analyzed separately using the prior-information cohorts in Appendix~\ref{app:requirement-semantics}.

\subsection{Feedback-Discounted Agent Completion}
\label{app:fdac}

FDAC measures complete delivery while assigning less credit to requirements completed after additional follow-up. Let $I_\tau$ contain states whose required observations are usable. For $i\in I_\tau$, define
\begin{equation}
q_i=\begin{cases}
1, & \text{pre-disclosure pass, or initial-state first-response pass},\\
0.5, & \text{later-state pass immediately after disclosure},\\
0.5\times0.4^r, & \text{pass after completion revision }r\in\{1,2\},\\
0, & \text{no pass within the state's allowed responses}.
\end{cases}
\label{eq:fdac-state-factor}
\end{equation}
The task score is
\begin{equation}
\operatorname{FDAC}(\tau)=\operatorname{StrictPass}(\tau)
\left(\prod_{i\in I_\tau}q_i\right)^{1/|I_\tau|}.
\label{eq:fdac}
\end{equation}
The geometric mean prevents a long task from being discounted merely by multiplying more factors, but still makes any unresolved usable state decisive. Model-level FDAC averages task scores. The state partition is fixed before evaluation.

\paragraph{Interpretation and edge cases.}
The state weights encode the benchmark's preference for earlier completion with fewer follow-ups: 1 before a later disclosure or after the initial response, 0.5 after a later disclosure, and an additional factor of 0.4 for each corrective revision. The same revision rule gives an initial state repaired on the first revision a weight of 0.2. A state unresolved within its local budget keeps weight zero even when later work repairs it. Thus FDAC incorporates timely intermediate delivery as well as final correctness.

For a successful three-state trajectory, $q=(1,0.5,0.5)$ gives FDAC $0.25^{1/3}\approx0.630$. One correction at the second state changes this to $q=(1,0.2,0.5)$ and FDAC $0.1^{1/3}\approx0.464$. Missing that state's budget yields $q=(1,0,0.5)$ and score zero.

\paragraph{Observation coverage.}
Across the 732 runs with both feedback types, 2,390 of 2,484 states have the observations required for FDAC. The geometric mean uses these states, with at least one usable state in every run. Table~\ref{tab:fdac-coverage} reports per-model coverage.

\begin{table}[htbp]
\centering\small
\caption{FDAC state coverage. Every model has 207 possible requirement states.}
\label{tab:fdac-coverage}
\begin{tabular}{@{}lrr@{}}
\toprule
\rowcolor{FLHeader}
Model & Usable & Omitted \\
\midrule
DeepSeek V4 Flash & 205 & 2 \\
\rowcolor{FLStripe}
DeepSeek V4 Pro & 201 & 6 \\
DeepSeek V4.1 Flash & 207 & 0 \\
\rowcolor{FLStripe}
GLM 5.3 & 202 & 5 \\
GLM 5.3 Flash & 197 & 10 \\
\rowcolor{FLStripe}
Hy3 & 184 & 23 \\
Kimi K3 & 204 & 3 \\
\rowcolor{FLStripe}
Qwen 3.8 Flash & 194 & 13 \\
GPT-5.6 Sol & 202 & 5 \\
\rowcolor{FLStripe}
GPT-6 Astra & 203 & 4 \\
Grok 4.6 & 192 & 15 \\
\rowcolor{FLStripe}
Hy4-Preview & 199 & 8 \\
\midrule
\rowcolor{FLTotal}
Pooled & 2,390 & 94 \\
\bottomrule
\end{tabular}
\end{table}

\paragraph{Partial-credit variant.}
FDAC-soft replaces strict final success with NodeCredit and uses an arithmetic mean:
\begin{equation}
\operatorname{FDAC\text{-}soft}(\tau)=\operatorname{NodeCredit}(\tau)
\frac{1}{|I_\tau|}\sum_{i\in I_\tau}q_i.
\label{eq:fdac-soft}
\end{equation}
This partial-credit diagnostic combines NodeCredit with the arithmetic mean of the state weights.

\subsection{Paired Protocol Transitions}
\label{app:paired-transition-metric}

For two protocols, match runs by model and task. A pair is \emph{both pass}, a \emph{rescue} (only the second protocol passes), a \emph{loss} (only the first passes), or \emph{both fail}. The net rate difference is
\begin{equation}
\Delta_{b-a}=\frac{n_{\mathrm{rescue}}-n_{\mathrm{loss}}}{N_{\mathrm{paired}}}.
\label{eq:paired-transition}
\end{equation}
The contrast compares independently executed workflows on matched model--task pairs. Direct repair and code regression are measured separately from within-run execution records.

\subsection{Individual Requirement Completion and Cumulative Correctness}
\label{app:group-metrics}

For a later group $g$ activated at state $i$, local success is $Z_g(W)$ and cumulative success is $P_i(W)=\mathbf{1}[\forall h\in Q_i,Z_h(W)=1]$ on the same workspace. The event $Z_g(W)=1, P_i(W)=0$ identifies a completed current requirement with at least one other stated requirement unfinished. Complete delivery additionally requires baseline preservation: $C_i(W)=b(W)P_i(W)$.

Pre-disclosure coverage checks whether a group already passes before its request. First-response completion evaluates disclosed groups after the agent's first response. Local Repair@2 and Prefix Closure@2 use the same 455 failed later-group instances: the group fails after the first response and its stage receives completion feedback. Local Repair@2 counts a subsequent pass of that group within two revisions. Prefix Closure@2 requires every group introduced so far to pass within the same window. The selected feedback observations concern the stage; an individual failed group may remain unmentioned.

The 147 later groups map to 146 states. Slidev's Q002 introduces two groups, so its execution result contributes one observation to each group. Message counts, token totals, and state recovery count this shared state once. Each metric uses records with the observations required by its definition.

\subsection{Feedback Recovery and Requirement Preservation}
\label{app:feedback-metrics}

A state-level recovery opportunity requires an observed first-response failure and completion feedback at that state. Recover@$r$ is the fraction of such opportunities whose recorded cumulative state check passes within $r$ revisions. Recover@1 and Recover@2 share the denominator of 733, including opportunities not recovered within the budget. The complete set of delivered messages is larger and includes fallback events without an attributable first-response observation (Appendix~\ref{app:event-accounting}).

Regression testing checks whether changes preserve existing behavior \citep{yoo2012regression}. Observed retention uses groups that pass at least once and have a check at a later requirement stage. A group is retained when it remains passing at every such observation and at the final checkpoint.

State-Evolution Reliability (SER@$r$) measures cumulative-state completion along the 685 trajectories with usable results at every later state. A run counts when every cumulative state passes by its recorded checkpoint after at most $r$ revisions. The three checkpoints reuse the same trajectory. Stalled responses add no progress under the cumulative checks; regression events record an earlier passing behavior failing after a later response.

\subsection{Conditional Proactive Coverage}
\label{app:pcc-metric}

Proactive Contract Coverage (PCC) restricts pre-disclosure exact coverage to groups with $N\geq2$, $X\geq2$, and $C\leq1$: sufficient prior support for the need and exact behavior, with limited premature-commitment risk. Labels use the information available immediately before each group is disclosed. Micro PCC pools usable model--group observations; macro PCC averages the within-group rates. The strict cohort comprises 16 groups in 14 tasks.

\subsection{Interaction and Token Cost}
\label{app:cost-metrics}

$N_E$ and $N_C$ count delivered evolution and completion messages after the initial request. Agent Tokens sum the evaluated agent's reported input and output tokens across model calls. Cached-input usage is recorded separately to avoid double counting. These measurements characterize the interaction and solver-side token cost of the development path.

Let $\mathcal R_p$ be runs with reported telemetry under protocol $p$, and $T_\tau$ their token totals. We report
\begin{equation}
\operatorname{MeanTokens}_p=\frac{\sum_{\tau\in\mathcal R_p}T_\tau}{|\mathcal R_p|},\qquad
\operatorname{TokensPerSuccess}_p=\frac{\sum_{\tau\in\mathcal R_p}T_\tau}
{\sum_{\tau\in\mathcal R_p}\operatorname{StrictPass}(\tau)}.
\label{eq:token-cost}
\end{equation}
Tokens per success includes the cost of failed runs. The reported-telemetry subsets contain 726/732 runs for the upfront specification, 732/732 for progressive requirements, and 731/732 for progressive requirements with correction; their success denominators are 295, 378, and 468. The mean-token comparison uses these reported subsets. Thus the 47.2\% increase measures the change in mean reported agent tokens per run.

\subsection{Uncertainty and Rounding}
\label{app:experiment-statistics}

Per-model task-completion intervals use the 95\% Wilson interval \citep{wilson1927probable}. Pooled rates and workflow differences use 10,000 repository-clustered bootstrap samples with seed 20260923. Resampling retains all tasks and model outcomes belonging to a selected repository, measuring sensitivity to the repository sample. Semantic subgroup sensitivity is reported separately. Percentages and differences are calculated from integer counts or unrounded scores and then rounded for display.

\setcounter{topnumber}{4}
\setcounter{totalnumber}{6}
\renewcommand{\topfraction}{0.95}
\renewcommand{\textfraction}{0.05}
\renewcommand{\floatpagefraction}{0.98}
\section{Complete Results and Model-Level Analysis}
\label{app:full-results}

The main text groups the results by implementation difficulty, model differences, and resource use. This appendix retains the corresponding event counts, full distributions, conditional populations, and uncertainty estimates. Table~\ref{tab:finding-denominators} provides a direct map from the main findings to their counting units.

\begin{table}[!htbp]
\centering\small
\setlength{\tabcolsep}{4pt}\renewcommand{\arraystretch}{1.13}
\caption{Observation populations behind the main findings. A requirement group can appear in several model runs; a state may contain several groups. Each row uses its own explicitly defined observation set.}
\label{tab:finding-denominators}
\begin{tabularx}{\linewidth}{@{}L{.23\linewidth}Yrr@{}}
\toprule\rowcolor{FLHeader}
Observation & Population and window & Count & Rate (\%)\\
\midrule
New requirement implemented & First responses to later-information groups after disclosure & 717/927 & 77.3\\
\rowcolor{FLStripe}
Other requirements still unmet & Locally successful first responses above that do not satisfy all stated requirements & 76/717 & 10.6\\
Complete state repaired & Observed failed states that receive feedback and pass within two revisions & 113/733 & 15.4\\
\rowcolor{FLStripe}
Individual requirement repaired & Failed later groups at stages receiving correction, within two revisions & 91/455 & 20.0\\
Other requirements still unmet after repair & Successful individual repairs above without cumulative completion within the same window & 19/91 & 20.9\\
\rowcolor{FLStripe}
Previously correct behavior lost & Passing groups with later-stage observations that subsequently fail & 73/642 & 11.4\\
User-path verification missing or misdirected & Final failed runs with the corresponding secondary forensic label & 198/263 & 75.3\\
\bottomrule
\end{tabularx}
\end{table}

\subsection{Final Task Completion}
\label{app:complete-terminal-results}

Final task completion and the path to that result reveal different model profiles (Table~\ref{tab:complete-results}). GPT-6 Astra records the highest completion rate and feedback-discounted score. GPT-5.6 Sol, Kimi K3, DeepSeek V4 Flash, and DeepSeek V4.1 Flash each complete 43 tasks, while their path scores distinguish the guidance required along the way.

\begin{table*}[!htbp]
  \centering
  \small
  \setlength{\tabcolsep}{4pt}
  \renewcommand{\arraystretch}{1.10}
  \caption{Complete results, ordered by FDAC. Model StrictPass cells give successes/61 and percentages (the pooled denominator is 732); NodeCredit, FDAC, and FDAC-soft are percentages. Best observed model values in each column are bold.}
  \label{tab:complete-results}
  \resizebox{\textwidth}{!}{%
  \begin{tabular}{@{}lrrrrrr@{}}
    \toprule
\rowcolor{FLHeader}
& \multicolumn{3}{c}{StrictPass $\uparrow$} & \multicolumn{3}{c}{Evo+Comp scores $\uparrow$} \\
    \cmidrule(lr){2-4}\cmidrule(lr){5-7}
    \rowcolor{FLHeader}
    Model & Full & Evo & Evo+Comp & NodeCredit & FDAC & FDAC-soft \\
    \midrule
    GPT-6 Astra & \textbf{36/61 (59.0)} & \textbf{42/61 (68.9)} & \textbf{44/61 (72.1)} & 82.2 & \textbf{46.5} & \textbf{52.6} \\
\rowcolor{FLStripe}
    GPT-5.6 Sol & 32/61 (52.5) & 38/61 (62.3) & 43/61 (70.5) & 79.3 & 45.4 & 51.9 \\
    Kimi K3 & 24/61 (39.3) & 33/61 (54.1) & 43/61 (70.5) & 84.6 & 44.2 & 51.6 \\
\rowcolor{FLStripe}
    Hy4-Preview & 29/61 (47.5) & 32/61 (52.5) & 42/61 (68.9) & 81.2 & 42.7 & 49.6 \\
    DeepSeek V4 Flash & 23/61 (37.7) & 35/61 (57.4) & 43/61 (70.5) & 82.0 & 41.2 & 49.3 \\
\rowcolor{FLStripe}
    DeepSeek V4.1 Flash & 26/61 (42.6) & 35/61 (57.4) & 43/61 (70.5) & \textbf{86.5} & 40.8 & 50.3 \\
    GLM 5.3 & 27/61 (44.3) & 31/61 (50.8) & 40/61 (65.6) & 79.0 & 40.5 & 47.8 \\
\rowcolor{FLStripe}
    Qwen 3.8 Flash & 24/61 (39.3) & 29/61 (47.5) & 35/61 (57.4) & 70.2 & 37.3 & 43.8 \\
    GLM 5.3 Flash & 22/61 (36.1) & 30/61 (49.2) & 38/61 (62.3) & 71.0 & 37.2 & 44.1 \\
\rowcolor{FLStripe}
    Grok 4.6 & 17/61 (27.9) & 23/61 (37.7) & 39/61 (63.9) & 70.3 & 37.1 & 41.8 \\
    DeepSeek V4 Pro & 22/61 (36.1) & 30/61 (49.2) & 37/61 (60.7) & 76.3 & 36.7 & 44.8 \\
\rowcolor{FLStripe}
    Hy3 & 13/61 (21.3) & 20/61 (32.8) & 22/61 (36.1) & 50.1 & 21.2 & 27.9 \\
    \midrule
    \rowcolor{FLTotal}
    \textbf{Pooled} & \textbf{295/732 (40.3)} & \textbf{378/732 (51.6)} & \textbf{469/732 (64.1)} & \textbf{76.1} & \textbf{39.2} & \textbf{46.3} \\
    \bottomrule
  \end{tabular}}
\end{table*}

Table~\ref{tab:strictpass-intervals} gives the per-model 95\% Wilson intervals. The repository-clustered intervals for pooled final completion are $[29.9,50.1]$ with an upfront specification, $[41.0,61.5]$ with progressive requirements, and $[53.8,72.8]$ with progressive requirements and correction.

\begin{table*}[!htbp]
  \centering
  \small
  \caption{Per-model 95\% Wilson intervals for StrictPass (\%).}
  \label{tab:strictpass-intervals}
  \begin{tabular}{@{}lrrr@{}}
    \toprule
\rowcolor{FLHeader}
Model & Full & Evo & Evo+Comp \\
    \midrule
    DeepSeek V4 Flash   & 26.6--50.3 & 44.9--69.0 & 58.1--80.4 \\
\rowcolor{FLStripe}
    DeepSeek V4 Pro     & 25.2--48.6 & 37.1--61.4 & 48.1--71.9 \\
    DeepSeek V4.1 Flash & 31.0--55.1 & 44.9--69.0 & 58.1--80.4 \\
\rowcolor{FLStripe}
    GLM 5.3             & 32.5--56.7 & 38.6--62.9 & 53.0--76.3 \\
    GLM 5.3 Flash       & 25.2--48.6 & 37.1--61.4 & 49.7--73.4 \\
\rowcolor{FLStripe}
    Hy3                 & 12.9--33.1 & 22.3--45.3 & 25.2--48.6 \\
    Kimi K3             & 28.1--51.9 & 41.7--66.0 & 58.1--80.4 \\
\rowcolor{FLStripe}
    Qwen 3.8 Flash      & 28.1--51.9 & 35.5--59.8 & 44.9--69.0 \\
    GPT-5.6 Sol         & 40.2--64.5 & 49.7--73.4 & 58.1--80.4 \\
\rowcolor{FLStripe}
    GPT-6 Astra         & 46.5--70.5 & 56.4--79.1 & 59.8--81.8 \\
    Grok 4.6            & 18.2--40.2 & 26.6--50.3 & 51.4--74.8 \\
\rowcolor{FLStripe}
    Hy4-Preview         & 35.5--59.8 & 40.2--64.5 & 56.4--79.1 \\
    \bottomrule
  \end{tabular}
\end{table*}

Partial completion also separates models tied on final success. Averaged across all tasks, DeepSeek V4.1 Flash satisfies 86.5\% of final requirement groups and Kimi K3 satisfies 84.6\%, compared with 82.2\% for GPT-6 Astra. Hy3 reaches 50.1\%. These values retain progress on runs that fall short of complete delivery.


\subsection{Failure Attribution Protocol}
\label{app:failure-attribution}

We reviewed all 263 final failures in the workflow with both feedback types. The review used the request, feedback, visible agent actions, workspace changes, response-level scores, and final execution evidence. In one exhaustive forensic pass, each failure received a primary label for the earliest supported mistake explaining the later outcome. Independently supported validation and repair problems received secondary labels. Table~\ref{tab:failure-taxonomy} gives the criteria; the evidence record links each decision to its trajectory.

\begin{table*}[!htbp]
  \centering
  \small
  \caption{Primary-cause taxonomy for final failure attribution.}
  \label{tab:failure-taxonomy}
  \begin{tabular}{@{}L{0.05\textwidth}L{0.24\textwidth}L{0.63\textwidth}@{}}
    \toprule
\rowcolor{FLHeader}
Code & Cause & Decision criterion \\
    \midrule
    G1 & Relevant code not found & The workspace exists, but the agent does not locate or use the relevant implementation area. \\
\rowcolor{FLStripe}
    G2 & User-path connection & Logic exists, but the real API, CLI, UI, registry, lifecycle, or consumer does not use it. \\
    G3 & Disclosed branch omitted & A behavior or compatibility branch already expressed to the agent remains unimplemented. \\
\rowcolor{FLStripe}
    I1 & Behavior or API mismatch & The implementation targets the right area but violates the required type, ordering, state, protocol, representation, or timing. \\
    I2 & Build, type, or runtime break & The agent's patch prevents compilation, startup, test entry, or execution through the user-facing path. \\
\rowcolor{FLStripe}
    I3 & Inconsistent shared state & Parallel registries, shims, generated interfaces, or state sources become inconsistent. \\
    R3 & Earlier behavior regressed & A previously passing behavior is broken by a later change and is not restored. \\
\rowcolor{FLStripe}
    V3 & No relevant implementation & The necessary information and workspace are available, but the run stops without implementing the requested behavior. \\
    \bottomrule
  \end{tabular}
\end{table*}

Secondary labels capture validation aimed at a substitute for the public path, known failures carried forward, and further patching on an already broken workspace. Assignment uses the implementation and execution evidence, with tool calls and agent responses locating the relevant decisions in the trace.


\subsection{What Causes Final Failures}
\label{app:failure-structure}

Table~\ref{tab:failure-primary-causes} gives the full primary-cause distribution. Behavior or API mismatches, build or runtime failures, and code that does not reach the user-facing path together account for 196/263 failures (74.5\%). Build, type, or runtime failures are particularly severe: 65/69 satisfy none of the final requirement groups, whereas 58/70 behavior or API mismatches and 44/57 user-path failures retain partial credit.

\begin{table}[!htbp]
  \centering
  \small
  \caption{Primary causes across all final failures under Evo+Comp.}
  \label{tab:failure-primary-causes}
  \begin{tabular}{lrr}
    \toprule
\rowcolor{FLHeader}
Primary cause & Failures & Share (\%) \\
    \midrule
    Behavior or API mismatch (I1) & 70 & 26.6 \\
\rowcolor{FLStripe}
    Build, type, or runtime break (I2) & 69 & 26.2 \\
    User-path connection (G2) & 57 & 21.7 \\
\rowcolor{FLStripe}
    Inconsistent shared state (I3) & 39 & 14.8 \\
    Disclosed branch omitted (G3) & 13 & 4.9 \\
\rowcolor{FLStripe}
    Earlier behavior regressed (R3) & 13 & 4.9 \\
    Relevant code not found (G1) & 1 & 0.4 \\
\rowcolor{FLStripe}
    No relevant implementation (V3) & 1 & 0.4 \\
    \midrule
    \rowcolor{FLTotal}
    Total & 263 & 100.0 \\
    \bottomrule
  \end{tabular}
\end{table}

The secondary labels locate recurring obstacles to repair: 198/263 failures omit a relevant public-path check or validate a substitute, 140 carry a known problem forward, and 59 continue patching before restoring a working version. These traces expose how a run can continue responding to feedback while leaving the user's failing behavior unchanged.

Separating tasks by whether any model solves them distinguishes common limitations from model-specific ones (Table~\ref{tab:failure-task-strata}). The 38 tasks solved by some models contribute 155 failures, of which 147 have direct execution evidence for the assigned cause. The nine tasks solved by none contribute 108 failures; because no successful peer run is available, their attribution relies on the request, implementation, and observed failing path. Table~\ref{tab:failure-causes-by-stratum} shows why the distinction matters: inconsistent shared state is concentrated in tasks solved by no model, while behavior mismatches and build failures more often separate models.

\begin{table}[!htbp]
  \centering
  \small
  \caption{Task groups under Evo+Comp.}
  \label{tab:failure-task-strata}
  \begin{tabular}{lrr}
    \toprule
\rowcolor{FLHeader}
Outcome across models & Tasks & Failed runs \\
    \midrule
    All models pass & 14 & 0 \\
\rowcolor{FLStripe}
    Some models pass (1--11) & 38 & 155 \\
    No model passes & 9 & 108 \\
    \bottomrule
  \end{tabular}
\end{table}

\begin{table*}[!htbp]
  \centering
  \small
  \caption{Detailed primary-cause counts by task group.}
  \label{tab:failure-causes-by-stratum}
  \begin{tabular}{lrrrrrrrr}
    \toprule
\rowcolor{FLHeader}
Task group & G1 & G2 & G3 & I1 & I2 & I3 & R3 & V3 \\
    \midrule
    Some models pass ($n=155$) & 1 & 33 & 8 & 46 & 54 & 4 & 8 & 1 \\
\rowcolor{FLStripe}
    No model passes ($n=108$) & 0 & 24 & 5 & 24 & 15 & 35 & 5 & 0 \\
    \bottomrule
  \end{tabular}
\end{table*}


\subsection{Model-Level Failure Profiles}
\label{app:failure-model-profiles}

\paragraph{Task coverage among the tied models.}
GPT-5.6 Sol, Kimi K3, DeepSeek V4 Flash, and DeepSeek V4.1 Flash each succeed on 43 of 61 tasks. Intersecting their final failed-task sets leaves 10 tasks failed by all four, so the union of their successful-task sets contains $61-10=51$ tasks. Pairwise disagreement in task outcome ranges from 6 to 14 tasks. These counts are recomputed from the 263-row final-failure annotation table, with one model--task entry per failed run. They describe complementary coverage across the existing, independently recorded runs.

Equal final success rates conceal different failure sets. Among the four models completing 43/61 tasks, pairwise failure overlap on tasks solved by at least one model ranges from 12.5\% to 50.0\%, with a median of 20.0\%. Figure~\ref{fig:failure-models} and Table~\ref{tab:model-failure-profile} show which implementation problems account for each model's remaining failures.

\begin{table*}[!htbp]
  \centering
  \small
  \caption{Grouped primary-cause counts by model. ``None / some'' separates failures on tasks solved by no model from failures on tasks solved by at least one model.}
  \label{tab:model-failure-profile}
  \resizebox{\textwidth}{!}{%
  \begin{tabular}{lrrrrrrr}
    \toprule
\rowcolor{FLHeader}
Model & Failures & None / some & Behavior/API & Build/runtime & User path & Shared state & Other \\
    \midrule
    GPT-6 Astra         & 17 & 9 / 8  & 4  & 3  & 5  & 4 & 1 \\
\rowcolor{FLStripe}
    GPT-5.6 Sol         & 18 & 9 / 9  & 4  & 6  & 4  & 2 & 2 \\
    Kimi K3             & 18 & 9 / 9  & 7  & 1  & 4  & 4 & 2 \\
\rowcolor{FLStripe}
    Hy4-Preview         & 19 & 9 / 10 & 6  & 4  & 6  & 2 & 1 \\
    DeepSeek V4 Flash   & 18 & 9 / 9  & 5  & 4  & 3  & 4 & 2 \\
\rowcolor{FLStripe}
    DeepSeek V4.1 Flash & 18 & 9 / 9  & 6  & 0  & 3  & 4 & 5 \\
    GLM 5.3             & 21 & 9 / 12 & 10 & 4  & 2  & 3 & 2 \\
\rowcolor{FLStripe}
    Qwen 3.8 Flash      & 26 & 9 / 17 & 5  & 8  & 8  & 3 & 2 \\
    GLM 5.3 Flash       & 23 & 9 / 14 & 6  & 8  & 5  & 1 & 3 \\
\rowcolor{FLStripe}
    Grok 4.6            & 22 & 9 / 13 & 3  & 9  & 3  & 3 & 4 \\
    DeepSeek V4 Pro     & 24 & 9 / 15 & 8  & 5  & 4  & 4 & 3 \\
\rowcolor{FLStripe}
    Hy3                 & 39 & 9 / 30 & 6  & 17 & 10 & 5 & 1 \\
    \bottomrule
  \end{tabular}%
  }
\end{table*}


The following profiles connect the numerical comparisons in Tables~\ref{tab:model-failure-profile} and~\ref{tab:feedback-recovery} to the implementation paths found in the failed runs.

\paragraph{GPT-6 Astra: strong delivery with remaining integration gaps.}
Remaining failures involve public entry-point integration, exact input representations, and inconsistent shared state. The model has the highest final completion and preserves every requirement in its retention-eligible subset. The broader event record still contains two regressions, outside that narrower retention population. Its remaining failures therefore concern specific end-to-end paths rather than a general inability to implement the requested behavior.

\paragraph{GPT-5.6 Sol: strong terminal results, limited direct recovery.}
Several unsuccessful traces investigate relevant code without producing a compilable implementation connected to the public interface. The strong final score coexists with a low observed repair fraction once a stage has failed. This profile calls attention to the transition from locating relevant code to validating a working repair, rather than repository exploration alone.

\paragraph{Kimi K3: partial implementations with public-interface gaps.}
The remaining gap often lies between functioning internal logic and the chart, output representation, or earlier format exposed to the user. Fifteen of its 18 final failures retain partial requirement credit. These traces distinguish incomplete public behavior from a wholly unsuccessful implementation.

\paragraph{Hy4-Preview: repairs coexist with incorrect validation targets.}
Failed traces include checks of an internal merge component while the live endpoint still uses the old behavior, and configuration changes that never reach rendering. Such checks leave the public failure intact despite progress elsewhere in the code. The issue is which execution path the agent validates, not simply whether it calls a test tool.

\paragraph{DeepSeek V4 Flash: substitute checks miss the default path.}
Failed traces often check a custom example, an explicitly registered component, or a helper instead of the default renderer, command, or required output. Frequent tool use consequently produces evidence about a substitute execution path. This pattern coexists with a relatively high recovery fraction among the model's observed failed stages.

\paragraph{DeepSeek V4.1 Flash: partial success with compatibility omissions.}
Legacy calling forms, timing bounds, output representations, and changes to earlier behavior account for the remaining compatibility gaps. Seventeen of its 18 final failures retain partial credit, and none has a build or runtime failure as its primary label. Its profile differs from models whose changes frequently leave the repository unable to execute.

\paragraph{GLM 5.3: exact behavior remains unresolved.}
Ordering, update timing, error text, naming, and boundary calculations remain incorrect in representative traces after feedback. Behavior or API mismatches account for ten of its 21 final failures. The unresolved work lies in exact public requirements within otherwise relevant implementations.

\paragraph{Qwen 3.8 Flash: build failures and disconnected implementations.}
Some runs continue using unverified interfaces or adding features after the workspace has stopped working. Build/runtime failures and disconnected public-path implementations each account for eight final failures. Continued activity in those runs does not resolve the problem blocking the requested behavior.

\paragraph{GLM 5.3 Flash: incomplete migrations affect later work.}
Traces include partial data-structure migrations and guessed signatures followed by unresolved compiler errors. Fourteen of the 23 final failures satisfy no final requirement group. An incomplete interface change can therefore affect both the current requirement and subsequent development, rather than remaining a local defect.

\paragraph{Grok 4.6: frequent recovery with subsequent regression.}
The model's relatively high repair fraction coexists with weak preservation of already working requirements. The NextChat trace in Appendix~\ref{app:failure-cases} illustrates the contrast: an early repair is followed by a runtime error that persists while later capabilities are added. Responding effectively to some corrections does not ensure that the repaired implementation remains usable through later edits.

\paragraph{DeepSeek V4 Pro: unresolved obligations persist across stages.}
Some runs add new functionality while earlier command, interface, count, or build problems remain unresolved. Custom checks miss the actual execution path, allowing progress on a later requirement to coexist with an unfinished earlier obligation. The accumulated task remains incomplete even as the latest change appears plausible.

\paragraph{Hy3: basic execution failures dominate.}
The remaining failures concentrate on using existing repository interfaces, maintaining an executable workspace, and exposing the implementation through the intended entry point. Build/runtime errors and public-path integration failures account for 27 of its 39 failed runs. Repeated correction often encounters these execution-level obstacles before finer behavioral requirements can be validated.

Across models, the traces distinguish four obstacles to low-intervention development: exact behavior left unfinished, functionality disconnected from the public interface, repairs lost during later edits, and broken execution paths carried into subsequent work. These obstacles explain why terminal completion, direct repair, and retention provide complementary views of the same development process.


\subsection{Failure Cases over Time}
\label{app:failure-cases}

\paragraph{A concrete failure report leads to a lasting repair.} In Qwen 3.8 Flash's 1Panel run, the initial endpoint depended on a global validator initialized only by server bootstrap and therefore failed under direct-router use. One completion-feedback revision moved binding and validation into the handler, recovered the first state, and left the second state correct on its first response; the repair remained valid at the final checkpoint.

\paragraph{One exact interface mismatch can stop an otherwise strong run.} GPT-6 Astra treats a Socket.IO worker object as a process identifier. The implementation therefore returns and compares the wrong public data model, finishes only one of two requirements, and remains unresolved after two completion-feedback messages. Eight of the other 11 models pass this task, showing that this task is solvable by peers under the same evaluation. The execution trace localizes the failure to the public representation mismatch.

\begin{figure}[!htbp]
  \centering
  \includegraphics[width=0.94\textwidth]{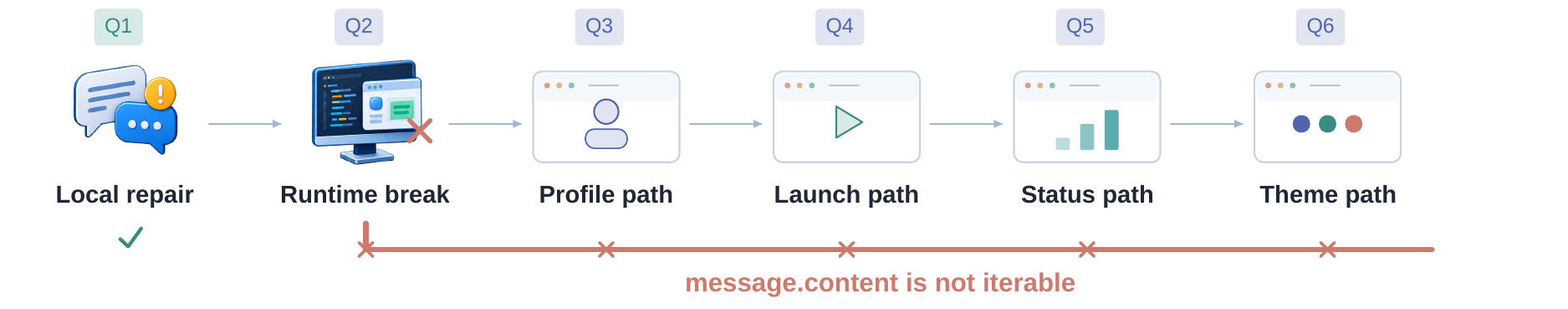}
  \caption{Grok 4.6's NextChat trajectory. Columns follow successive requirement stages; the lower track records the runtime status.}
  \label{fig:failure-case-nextchat}
\end{figure}

\paragraph{A local repair can still be lost during later work.} Grok 4.6 recovers the first NextChat state after feedback, then introduces \texttt{message.content is not iterable} while revising the second state. The error persists through the remaining four states as the run continues to add profile, launch, status, and theme paths, producing 15 regressions and a final score of 1/6 (Figure~\ref{fig:failure-case-nextchat}). The trajectory shows how continued feature work can obscure the fact that the user-facing path stopped working several states earlier.

\subsection{Initial Delivery}
\label{app:initial-delivery-results}

Initial usability and compatibility with the frozen final requirements are different outcomes. These initial-delivery results use the progressive-requirements workflow. Table~\ref{tab:iud-icd} gives exact counts: the initial state is usable within the first evaluation window for 63.9--80.3\% of runs by model, while no model already satisfies the eventual feature in more than 3.3\%. All 14 ICD successes occur in two tasks (nine model successes on ECharts and five on Typst). Tasks containing at least one later-information group account for 0/540 ICD successes. Appendix~\ref{app:requirement-semantics} characterizes the prior information available for each later requirement.

\begin{table*}[!htbp]
  \centering
  \small
  \caption{Initial Usable Delivery (IUD) and Initial Complete Delivery (ICD) under Evo. Each model has 61 runs.}
  \label{tab:iud-icd}
  \begin{tabular}{@{}lrr@{}}
    \toprule
\rowcolor{FLHeader}
Model & IUD & ICD \\
    \midrule
    DeepSeek V4 Flash   & 47/61 (77.0) & 2/61 (3.3) \\
\rowcolor{FLStripe}
    DeepSeek V4 Pro     & 45/61 (73.8) & 1/61 (1.6) \\
    DeepSeek V4.1 Flash & 49/61 (80.3) & 1/61 (1.6) \\
\rowcolor{FLStripe}
    GLM 5.3             & 48/61 (78.7) & 0/61 (0.0) \\
    GLM 5.3 Flash       & 45/61 (73.8) & 1/61 (1.6) \\
\rowcolor{FLStripe}
    Hy3                 & 39/61 (63.9) & 1/61 (1.6) \\
    Kimi K3             & 47/61 (77.0) & 2/61 (3.3) \\
\rowcolor{FLStripe}
    Qwen 3.8 Flash      & 46/61 (75.4) & 0/61 (0.0) \\
    GPT-5.6 Sol         & 47/61 (77.0) & 2/61 (3.3) \\
\rowcolor{FLStripe}
    GPT-6 Astra         & 49/61 (80.3) & 2/61 (3.3) \\
    Grok 4.6            & 41/61 (67.2) & 1/61 (1.6) \\
\rowcolor{FLStripe}
    Hy4-Preview         & 48/61 (78.7) & 1/61 (1.6) \\
    \midrule
    \rowcolor{FLTotal}
    Pooled              & 551/732 (75.3) & 14/732 (1.9) \\
    \bottomrule
  \end{tabular}
\end{table*}


\subsection{Local Uptake and Cumulative Closure}
\label{app:group-prefix-results}

\begin{figure}[!htbp]
\centering\includegraphics[width=.75\linewidth]{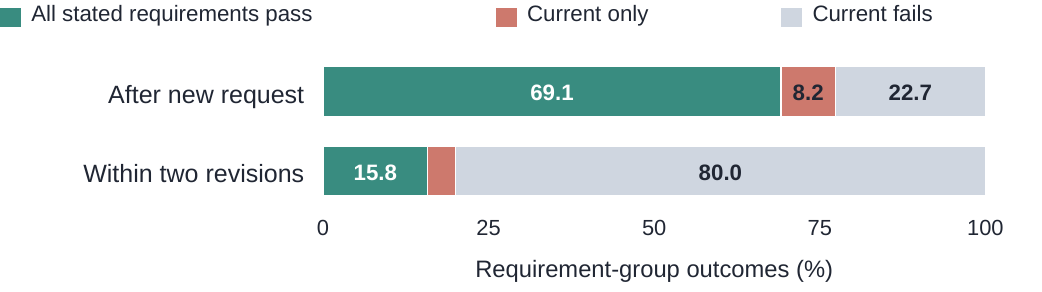}
\caption{Local success and complete requirement-state success. The new-request population contains 927 first responses to later-information groups; the corrective population contains 455 failed later-group opportunities. The populations differ, while each row partitions its own observations into mutually exclusive outcomes.}
\label{fig:local-cumulative-gap}
\end{figure}

Table~\ref{tab:group-prefix-results} combines the requirement labels with execution results. Pre-disclosure pass checks whether a later group already works before its request. First-response local success checks that group after the response; cumulative success checks all groups introduced so far. Local Repair@2 counts a failed group repaired within two revisions at a stage receiving completion feedback. Prefix Closure@2 additionally requires all previously introduced groups to pass. Each rate uses observations available for its specific check.

The 147 groups map to 146 disclosure states: Slidev Q002 contains two related groups in one state. Each Slidev row refers to a different requirement group within the same observed state.

\begin{table}[!htbp]
\centering\footnotesize
\setlength{\tabcolsep}{4pt}\renewcommand{\arraystretch}{1.12}
\caption{Early coverage and first-response outcomes by semantic cohort under Evo+Comp. Entries are successes/opportunities (percent). }
\label{tab:group-prefix-results}
\begin{tabular}{@{}lrrr@{}}
\toprule
\rowcolor{FLHeader}
Cohort & Pre-disclosure & First-response local & First-response cumulative \\
\midrule
All groups & 75/1660 (4.5) & 1078/1533 (70.3) & 942/1533 (61.4) \\
\rowcolor{FLStripe}
Strict proactive & 44/179 (24.6) & 67/134 (50.0) & 44/134 (32.8) \\
Directional, unclear & 5/291 (1.7) & 170/276 (61.6) & 146/276 (52.9) \\
\rowcolor{FLStripe}
Later-information & 1/956 (0.1) & 717/927 (77.3) & 641/927 (69.1) \\
\bottomrule
\end{tabular}
\end{table}
\begin{table}[!htbp]
\centering\small
\caption{Local repair and cumulative closure within two revisions. The two columns use the same observed failed-group opportunities within each cohort.}
\label{tab:group-repair-results}
\begin{tabular}{@{}lrr@{}}
\toprule
\rowcolor{FLHeader}
Cohort & Local Repair@2 & Prefix Closure@2 \\
\midrule
All groups & 91/455 (20.0) & 72/455 (15.8) \\
\rowcolor{FLStripe}
Strict proactive & 6/67 (9.0) & 4/67 (6.0) \\
Directional, unclear & 15/106 (14.2) & 13/106 (12.3) \\
\rowcolor{FLStripe}
Later-information & 50/210 (23.8) & 38/210 (18.1) \\
\bottomrule
\end{tabular}
\end{table}

Prior information separates two kinds of early coverage: 44/179 evaluations pass in the strict proactive cohort, compared with 1/956 in the later-information cohort. Once a requirement is disclosed, implementing it alone can still leave earlier work unfinished. This occurs in 76 of 717 locally successful first responses in the later-information cohort and 136 of 1,078 across all groups. Similarly, 19 of 91 successful individual repairs leave another stated requirement unmet within the same revision window. These incomplete implementations include both requirements that never passed and behavior broken by later edits.

\subsection{Matched Protocol Transitions}
\label{app:paired-transitions}

\begin{figure}[!htbp]
\centering\includegraphics[width=.72\linewidth]{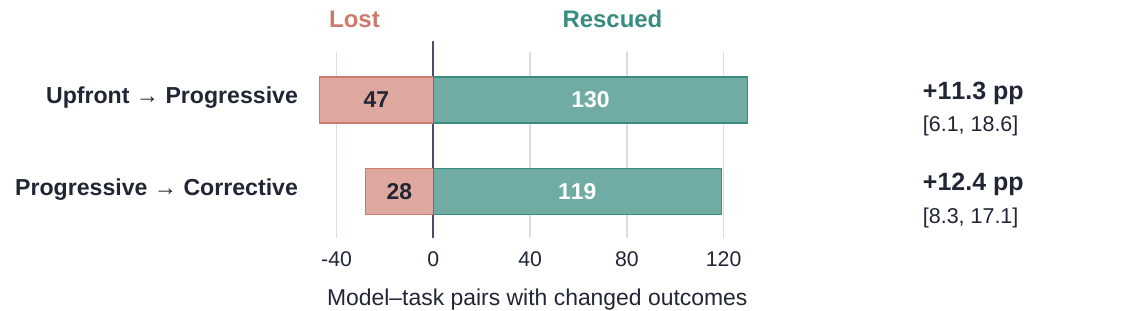}
\caption{Final-outcome changes over 732 matched model--task pairs per comparison. Bars distinguish newly successful and newly unsuccessful runs; intervals resample repositories. Each comparison uses independent executions of the complete workflows.}
\label{fig:paired-transitions}
\end{figure}

\begin{figure}[!htbp]
  \centering
  \includegraphics[width=0.94\textwidth]{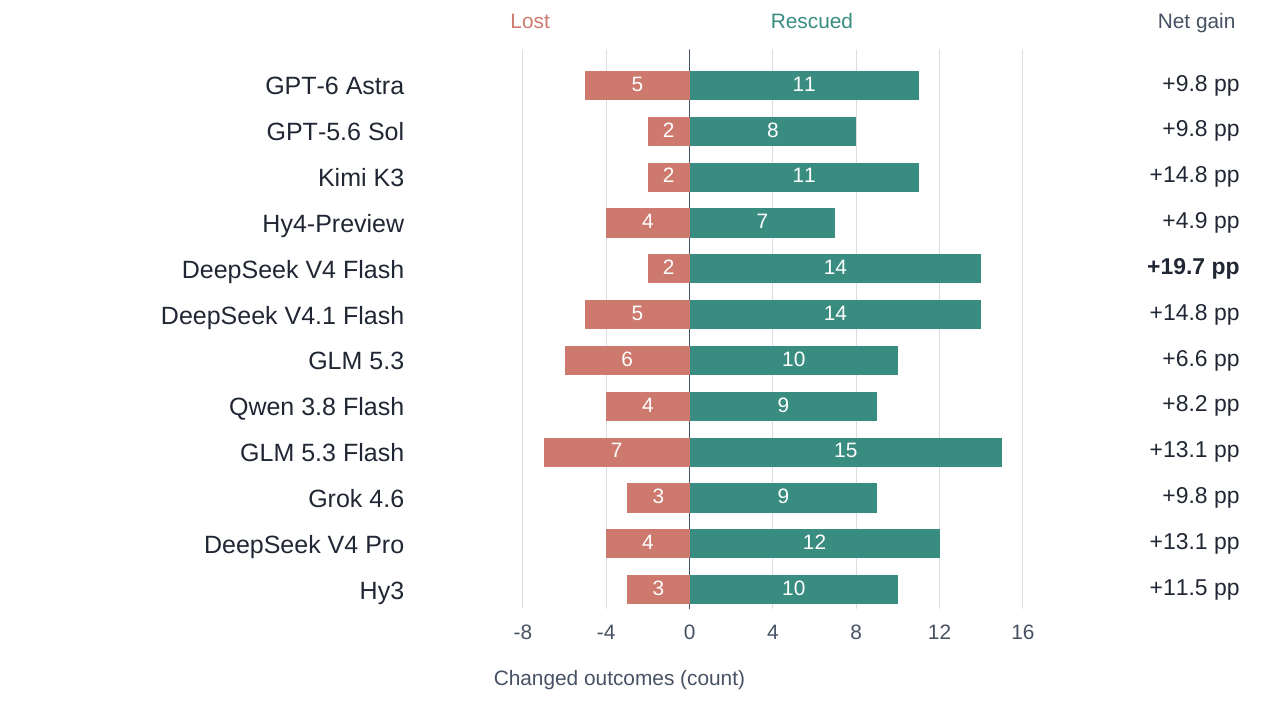}
  \caption{Paired Full$\rightarrow$Evo transitions over 61 tasks per model. Bars count lost successes and rescued failures; the right column gives net percentage-point gain.}
  \label{fig:transition-full-evo-model}
\end{figure}

\begin{figure}[!htbp]
  \centering
  \includegraphics[width=0.94\textwidth]{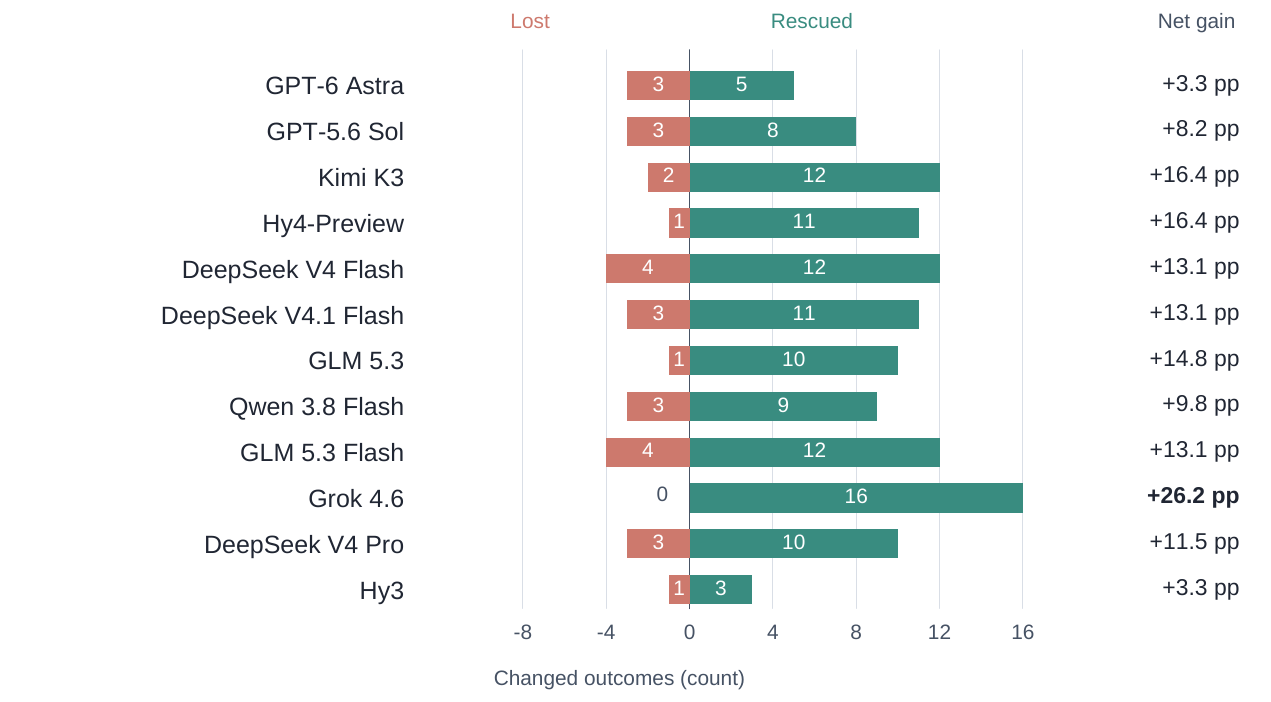}
  \caption{Paired Evo$\rightarrow$Evo+Comp transitions over 61 tasks per model. Bars count lost successes and rescued failures; the right column gives net percentage-point gain.}
  \label{fig:transition-evo-comp-model}
\end{figure}

The Full$\rightarrow$Evo workflow contrast gives the largest net change for DeepSeek V4 Flash and the smallest for Hy4-Preview. The Evo$\rightarrow$Evo+Comp contrast reverses that pattern for Hy4-Preview and gives its largest change for Grok 4.6, while changing little for GPT-6 Astra and Hy3. Table~\ref{tab:paired-transitions} gives the matched counts; Figures~\ref{fig:transition-full-evo-model} and~\ref{fig:transition-evo-comp-model} show the complete model-level breakdown.

\begin{table*}[!htbp]
  \centering
  \small
  \caption{Matched protocol transitions by model. Rescue/loss gives counts on the same 61 tasks; net change is in percentage points.}
  \label{tab:paired-transitions}
  \begin{tabular}{lrrrr}
    \toprule
\rowcolor{FLHeader}
Model & \multicolumn{2}{c}{Full $\rightarrow$ Evo} & \multicolumn{2}{c}{Evo $\rightarrow$ Evo+Comp} \\
    \cmidrule(lr){2-3}\cmidrule(lr){4-5}
          & Rescue / loss & Net change & Rescue / loss & Net change \\
    \midrule
    DeepSeek V4 Flash   & 14 / 2 & +19.7 & 12 / 4 & +13.1 \\
\rowcolor{FLStripe}
    DeepSeek V4 Pro     & 12 / 4 & +13.1 & 10 / 3 & +11.5 \\
    DeepSeek V4.1 Flash & 14 / 5 & +14.8 & 11 / 3 & +13.1 \\
\rowcolor{FLStripe}
    GLM 5.3             & 10 / 6 & +6.6  & 10 / 1 & +14.8 \\
    GLM 5.3 Flash       & 15 / 7 & +13.1 & 12 / 4 & +13.1 \\
\rowcolor{FLStripe}
    Hy3                 & 10 / 3 & +11.5 & 3 / 1  & +3.3 \\
    Kimi K3             & 11 / 2 & +14.8 & 12 / 2 & +16.4 \\
\rowcolor{FLStripe}
    Qwen 3.8 Flash      & 9 / 4  & +8.2  & 9 / 3  & +9.8 \\
    GPT-5.6 Sol         & 8 / 2  & +9.8  & 8 / 3  & +8.2 \\
\rowcolor{FLStripe}
    GPT-6 Astra         & 11 / 5 & +9.8  & 5 / 3  & +3.3 \\
    Grok 4.6            & 9 / 3  & +9.8  & 16 / 0 & +26.2 \\
\rowcolor{FLStripe}
    Hy4-Preview         & 7 / 4  & +4.9  & 11 / 1 & +16.4 \\
    \bottomrule
  \end{tabular}
\end{table*}

At the pooled level, replacing the one-shot complete-contract workflow with progressive feature development gives a $+11.3$ point net change, with a repository-clustered 95\% bootstrap interval of $[+6.1,+18.6]$. Enabling the completion-feedback workflow, including failure evidence and limited revisions, gives a $+12.4$ point net change, with interval $[+8.3,+17.1]$ (Figure~\ref{fig:paired-transitions}).

Of the 119 model--task pairs rescued across the two independently executed progressive workflows, 57 receive no completion message in the successful run and 54 contain an observed same-state recovery. The paired table records changes in final outcomes; the within-run recovery counts identify repairs following failure feedback.

\subsection{Outcomes by Implementation Language and Software Form}
\label{app:dimension-results}
\label{sec:dimension-results}

\begin{figure}[!htbp]
  \centering
  \includegraphics[width=0.94\textwidth]{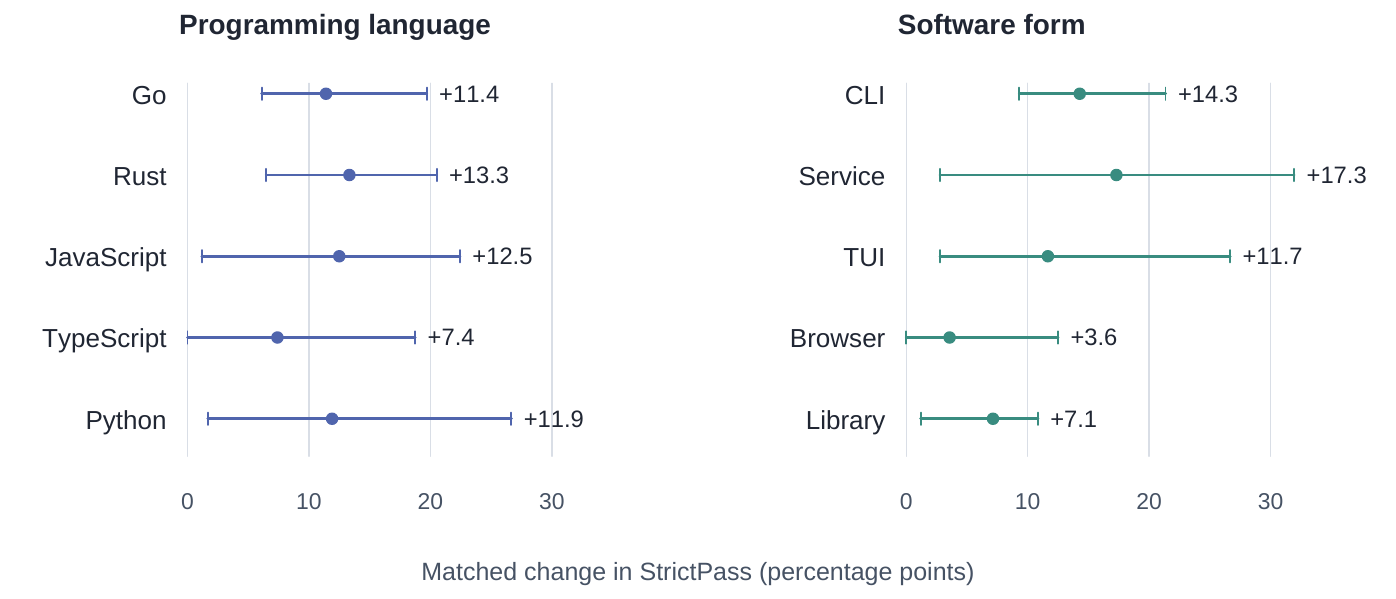}
  \caption{Matched Evo$\rightarrow$Evo+Comp change by implementation language and software form. Intervals use repository-clustered bootstrap samples; sparse descriptive-only strata are separated from the primary slices.}
  \label{fig:dimension-paired-gains}
\end{figure}

Table~\ref{tab:dimension-results} reports the primary descriptive strata: five single-language slices and five software forms with at least seven tasks from at least three repositories. The point estimate of every matched Evo$\rightarrow$Evo+Comp change is positive, but the intervals for TypeScript and Browser reach zero. Recovery, stability, and token use vary substantially (Table~\ref{tab:dimension-diagnostics}). Figure~\ref{fig:dimension-paired-gains} shows that Service has the largest primary matched gain, whereas Browser has the smallest. Repository-clustered intervals describe sensitivity to the sampled repository mixture.

\begin{table}[!htbp]
\centering\footnotesize\setlength{\tabcolsep}{4pt}\renewcommand{\arraystretch}{1.12}
\caption{Final completion by primary language and software form. $n$ is tasks, $r$ repositories, and the denominator for each protocol is $12n$ runs. $\Delta$ is the paired Evo-to-Evo+Comp change in percentage points with a repository-clustered 95\% interval.}
\label{tab:dimension-results}
\begin{tabular}{@{}lrrcccc@{}}
\toprule
\rowcolor{FLHeader}
Stratum & $n$ & $r$ & Full & Evo & Evo+Comp & $\Delta$ [95\% CI] \\
\midrule
Go & 19 & 9 & 57.0 & 70.2 & 81.6 & +11.4 [6.1, 19.7] \\
\rowcolor{FLStripe}
Rust & 15 & 7 & 24.4 & 35.0 & 48.3 & +13.3 [6.5, 20.5] \\
JavaScript & 10 & 5 & 45.8 & 73.3 & 85.8 & +12.5 [1.2, 22.4] \\
\rowcolor{FLStripe}
TypeScript & 9 & 8 & 12.0 & 10.2 & 17.6 & +7.4 [0.0, 18.8] \\
Python & 7 & 5 & 58.3 & 66.7 & 78.6 & +11.9 [1.7, 26.7] \\
\midrule
\rowcolor{FLStripe}
CLI & 21 & 10 & 36.5 & 43.3 & 57.5 & +14.3 [9.3, 21.4] \\
Service & 13 & 7 & 53.2 & 66.7 & 84.0 & +17.3 [2.8, 31.9] \\
\rowcolor{FLStripe}
TUI & 10 & 4 & 46.7 & 59.2 & 70.8 & +11.7 [2.8, 26.7] \\
Browser & 7 & 6 & 26.2 & 27.4 & 31.0 & +3.6 [0.0, 12.5] \\
\rowcolor{FLStripe}
Library & 7 & 4 & 50.0 & 83.3 & 90.5 & +7.1 [1.2, 10.8] \\
\bottomrule\end{tabular}\end{table}
\begin{table}[!htbp]
\centering\footnotesize\setlength{\tabcolsep}{4pt}\renewcommand{\arraystretch}{1.12}
\caption{Evo+Comp diagnostics for the same primary strata. Recovery and SER cells give successes/eligible opportunities. Tokens are millions per reported run; telemetry reports available/total runs. Each stratum groups the sampled tasks by one attribute.}
\label{tab:dimension-diagnostics}
\begin{tabular}{@{}lrrrrr@{}}
\toprule
\rowcolor{FLHeader}
Stratum & Recover@2 & SER@2 & FDAC (\%) & Tokens & Telemetry \\
\midrule
Go & 36/100 & 186/224 & 52.2 & 8.64 & 227/228 \\
\rowcolor{FLStripe}
Rust & 30/159 & 85/139 & 28.2 & 27.11 & 180/180 \\
JavaScript & 15/42 & 103/120 & 53.7 & 7.28 & 120/120 \\
\rowcolor{FLStripe}
TypeScript & 14/354 & 14/106 & 9.8 & 37.86 & 108/108 \\
Python & 11/63 & 65/84 & 47.1 & 13.47 & 84/84 \\
\midrule
\rowcolor{FLStripe}
CLI & 42/224 & 137/209 & 32.4 & 25.56 & 252/252 \\
Service & 35/70 & 131/154 & 50.9 & 6.53 & 156/156 \\
\rowcolor{FLStripe}
TUI & 20/77 & 85/118 & 45.5 & 9.01 & 119/120 \\
Browser & 7/188 & 26/84 & 21.1 & 27.84 & 84/84 \\
\rowcolor{FLStripe}
Library & 5/23 & 76/84 & 59.5 & 8.99 & 84/84 \\
\bottomrule\end{tabular}\end{table}

The mixed Go/TypeScript task, two Framework tasks, and one Embedded task are reported descriptively. Their matched changes are $+66.7$, $0.0$, and $+41.7$ points, respectively, without inferential intervals. Language--form composition is sparse: 25 of 42 possible cells are empty, Rust is almost entirely CLI, and TypeScript is concentrated in Browser and Framework tasks. These strata describe the observed combinations of repository, language, software form, and task horizon.

\subsection{Recovery after Completion Feedback}
\label{app:feedback-recovery-results}

\paragraph{Recovery and retention have different populations.}
Recovery starts from a failed first response and asks whether the current requirement state passes within its two corrective revisions. Retention starts only after an individual requirement has passed and asks whether it remains correct at later observed stages and the final check. Across models, 642 requirement-group instances have a passing observation and a subsequent-stage observation. Of these, 569 retain success and 73 subsequently fail. Thus $569/642=88.6\%$ is observed retention and $73/642=11.4\%$ is its complementary loss rate. These instances are requirements evaluated in model runs, not 642 distinct benchmark requirements.

\begin{figure}[!htbp]
\centering\includegraphics[width=.64\linewidth]{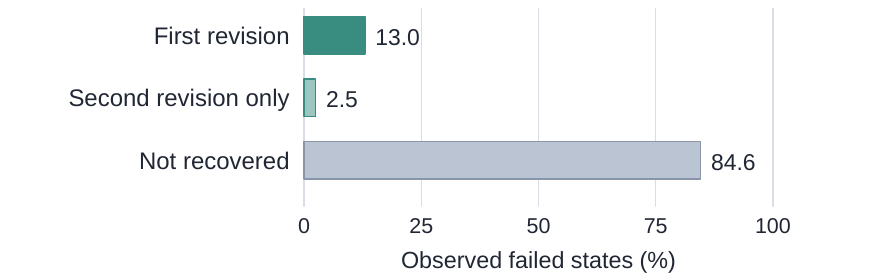}
\caption{Recovery after a failed first response. Segments partition the 733 observed failed states by their first successful revision, with unresolved states shown separately.}
\label{fig:feedback-conversion}
\end{figure}

\begin{figure}[!htbp]
  \centering
  \includegraphics[width=0.94\textwidth]{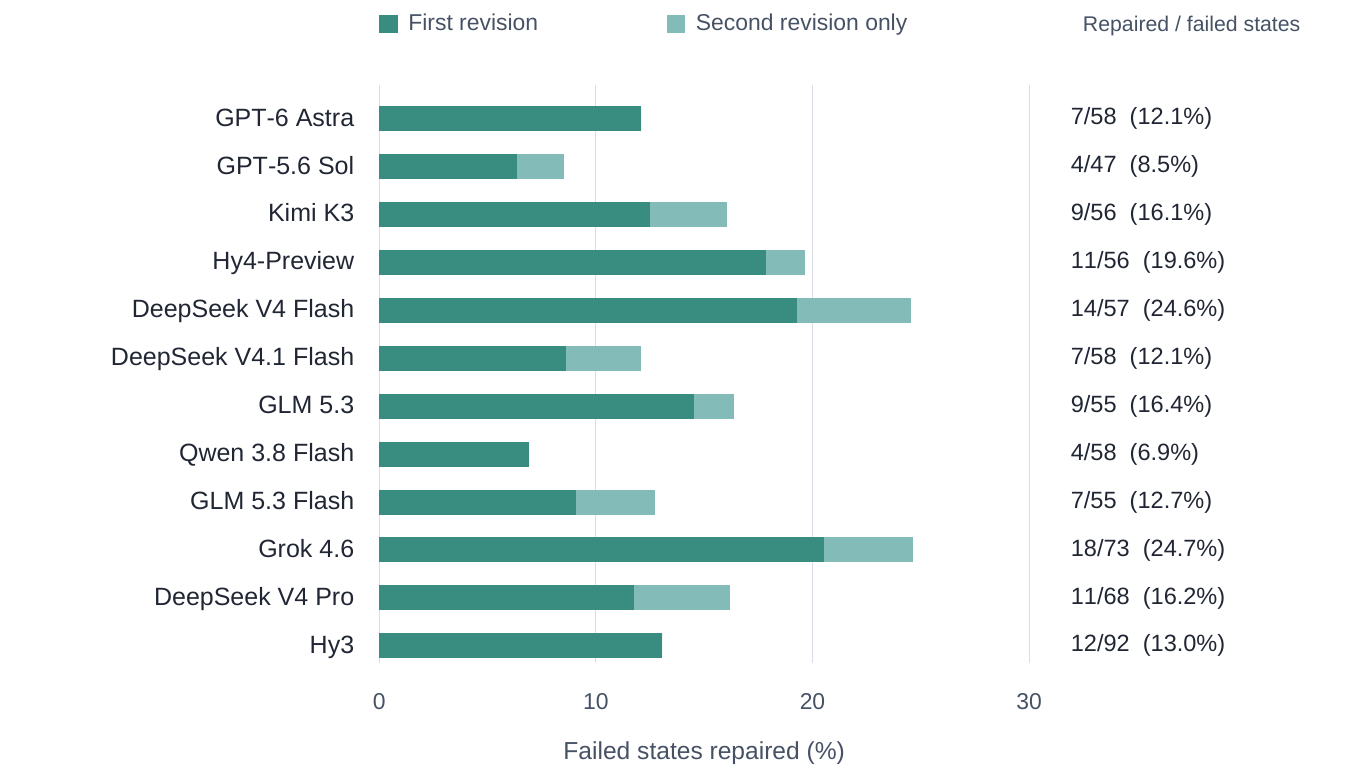}
  \caption{Completion-feedback recovery by model. Bars show fixes obtained after the first revision and those added by a second; the right columns give Recover@2 and opportunity counts.}
  \label{fig:feedback-conversion-by-model}
\end{figure}

Figure~\ref{fig:feedback-conversion-by-model} shows that direct recovery varies substantially across models; Table~\ref{tab:feedback-recovery} reports both feedback conversion and the separate group-level retention measure. Grok 4.6 and DeepSeek V4 Flash have the highest Recover@2, while GPT-5.6 Sol has a lower observed recovery fraction despite strong final completion. Each recovery rate uses the failed states encountered by that model. The second revision adds three recoveries for Grok 4.6, DeepSeek V4 Flash, and DeepSeek V4 Pro, but none for Hy3, Qwen 3.8 Flash, or GPT-6 Astra.

\begin{table*}[!htbp]
  \centering
  \small
  \caption{Completion-feedback conversion and observed requirement retention by model. Recover@1 and Recover@2 use feedback opportunities as their denominator. Retention uses only previously satisfied requirement groups with at least one later-state observation.}
  \label{tab:feedback-recovery}
  \begin{tabular}{lrrrr}
    \toprule
\rowcolor{FLHeader}
Model & Opportunities & Recover@1 & Recover@2 & Observed retention \\
    \midrule
    DeepSeek V4 Flash   & 57 & 11 (19.3) & 14 (24.6) & 52/62 (83.9) \\
\rowcolor{FLStripe}
    DeepSeek V4 Pro     & 68 & 8 (11.8)  & 11 (16.2) & 46/53 (86.8) \\
    DeepSeek V4.1 Flash & 58 & 5 (8.6)   & 7 (12.1)  & 61/62 (98.4) \\
\rowcolor{FLStripe}
    GLM 5.3             & 55 & 8 (14.5)  & 9 (16.4)  & 51/59 (86.4) \\
    GLM 5.3 Flash       & 55 & 5 (9.1)   & 7 (12.7)  & 46/51 (90.2) \\
\rowcolor{FLStripe}
    Hy3                 & 92 & 12 (13.0) & 12 (13.0) & 27/33 (81.8) \\
    Kimi K3             & 56 & 7 (12.5)  & 9 (16.1)  & 56/59 (94.9) \\
\rowcolor{FLStripe}
    Qwen 3.8 Flash      & 58 & 4 (6.9)   & 4 (6.9)   & 46/50 (92.0) \\
    GPT-5.6 Sol         & 47 & 3 (6.4)   & 4 (8.5)   & 47/55 (85.5) \\
\rowcolor{FLStripe}
    GPT-6 Astra         & 58 & 7 (12.1)  & 7 (12.1)  & 57/57 (100.0) \\
    Grok 4.6            & 73 & 15 (20.5) & 18 (24.7) & 33/46 (71.7) \\
\rowcolor{FLStripe}
    Hy4-Preview         & 56 & 10 (17.9) & 11 (19.6) & 47/55 (85.5) \\
    \midrule
    \rowcolor{FLTotal}
    Pooled              & 733 & 95 (13.0) & 113 (15.4) & 569/642 (88.6) \\
    \bottomrule
  \end{tabular}
\end{table*}

Repository-clustered 95\% bootstrap intervals are $[7.5,20.0]$ for Recover@1 and $[9.2,23.5]$ for Recover@2.

\subsection{State Outcomes and User Feedback Burden}
\label{app:state-outcome-results}

\begin{figure}[!htbp]
  \centering
  \includegraphics[width=0.94\textwidth]{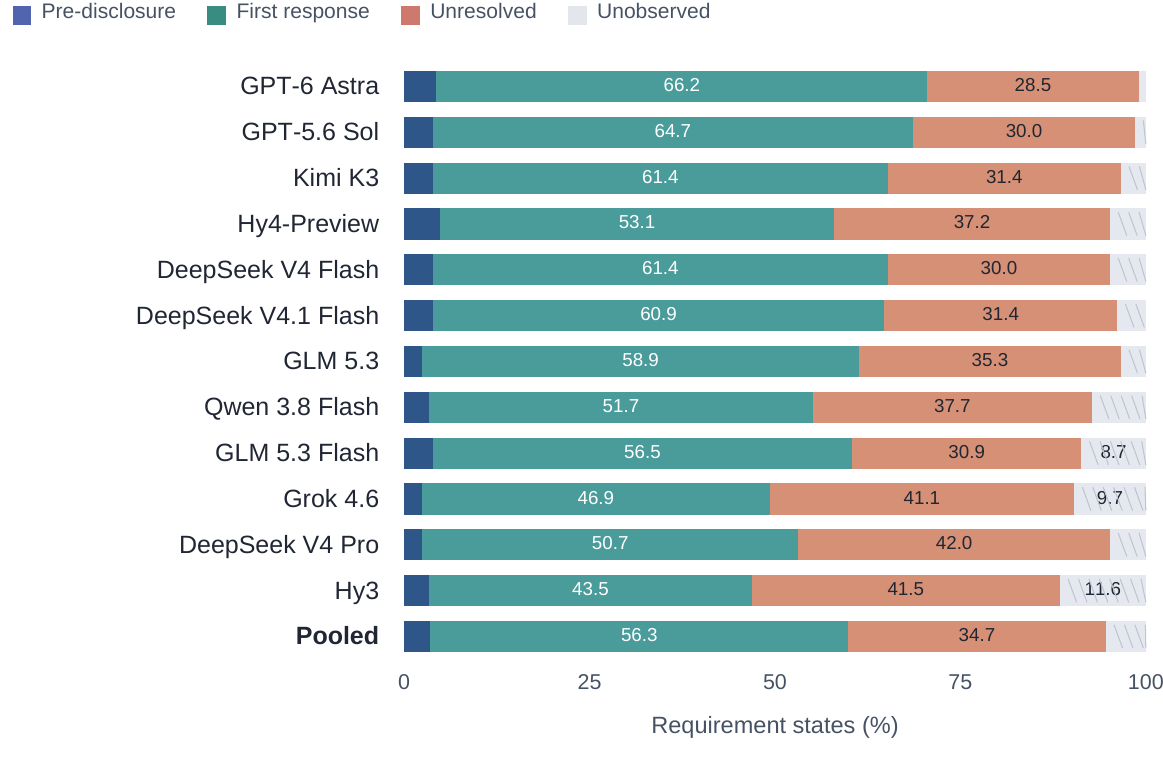}
  \caption{State outcomes under Evo, over 207 possible states per model; Pooled aggregates all 12 models. Hatching denotes unobserved records.}
  \label{fig:evo-state-outcomes}
\end{figure}

\begin{figure}[!htbp]
  \centering
  \includegraphics[width=0.94\textwidth]{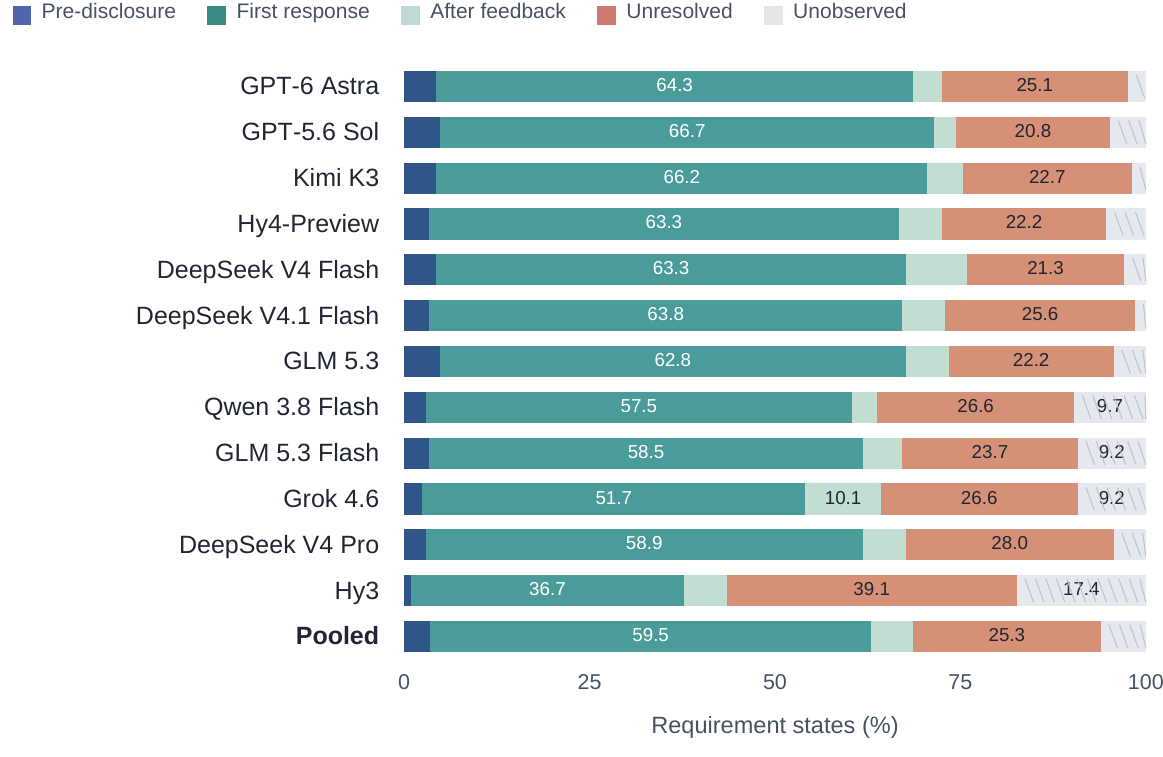}
  \caption{State outcomes under Evo+Comp, over 207 possible states per model; Pooled aggregates all 12 models. After-feedback denotes delayed resolution; hatching denotes unobserved records.}
  \label{fig:evo-comp-state-outcomes}
\end{figure}

With progressive requests and no correction, 56.3\% of 2,484 requirement states pass after the first response, 3.5\% are already satisfied before disclosure, 34.7\% remain unresolved, and 5.4\% are unobserved in the recorded trace. With corrective feedback, 59.5\% pass on the first response, 5.6\% pass after a revision, 3.5\% are already satisfied, 25.3\% remain unresolved, and 6.1\% are unobserved. Figures~\ref{fig:evo-state-outcomes} and~\ref{fig:evo-comp-state-outcomes} show how these outcomes vary by model.

\begin{table}[!htbp]
  \centering
  \small
  \caption{Feedback and path diagnostics under Evo+Comp. $N_E$ and $N_C$ are mean delivered messages per task. Triggered runs receive completion feedback; Success is their final StrictPass rate. Stalled is the percentage of responses with no observed cumulative progress; regressions are events, not distinct tasks.}
  \label{tab:feedback-burden}
  \resizebox{\textwidth}{!}{%
  \begin{tabular}{lrrrrrr}
    \toprule
\rowcolor{FLHeader}
Model & Mean $N_E$ & Mean $N_C$ & Triggered runs & Success (\%) & Stalled (\%) & Regressions \\
    \midrule
    DeepSeek V4 Flash   & 2.25 & 1.97 & 31 & 41.9 & 43.9 & 9 \\
\rowcolor{FLStripe}
    DeepSeek V4 Pro     & 2.30 & 2.44 & 32 & 25.0 & 52.6 & 7 \\
    DeepSeek V4.1 Flash & 2.28 & 2.07 & 29 & 37.9 & 44.3 & 6 \\
\rowcolor{FLStripe}
    GLM 5.3             & 2.23 & 2.02 & 28 & 25.0 & 45.2 & 8 \\
    GLM 5.3 Flash       & 2.28 & 2.48 & 32 & 28.1 & 49.8 & 6 \\
\rowcolor{FLStripe}
    Hy3                 & 2.36 & 4.03 & 46 & 15.2 & 67.1 & 4 \\
    Kimi K3             & 2.25 & 1.87 & 26 & 30.8 & 45.8 & 5 \\
\rowcolor{FLStripe}
    Qwen 3.8 Flash      & 2.30 & 2.59 & 31 & 16.1 & 52.4 & 5 \\
    GPT-5.6 Sol         & 2.23 & 1.89 & 23 & 21.7 & 41.0 & 9 \\
\rowcolor{FLStripe}
    GPT-6 Astra         & 2.25 & 2.00 & 23 & 26.1 & 47.5 & 2 \\
    Grok 4.6            & 2.31 & 2.82 & 33 & 33.3 & 52.1 & 35 \\
\rowcolor{FLStripe}
    Hy4-Preview         & 2.28 & 2.08 & 29 & 34.5 & 44.8 & 8 \\
    \midrule
    \rowcolor{FLTotal}
    Pooled              & 2.27 & 2.35 & 363 & 27.5 & 49.2 & 104 \\
    \bottomrule
  \end{tabular}}
\end{table}

Evolution-feedback counts vary little because models follow the same task states, whereas completion-feedback burden depends on the current workspace (Table~\ref{tab:feedback-burden}). Hy3 receives 4.03 completion-feedback messages per task and makes no observable progress on 67.1\% of responses. GPT-5.6 Sol and Kimi K3 receive fewer than 1.9 completion messages per task, while DeepSeek V4 Flash has the highest final success rate among tasks that trigger completion feedback. Grok 4.6 combines a large between-protocol net gain with 35 within-run regression events. 


\subsection{Tasks That Separate Models}
\label{app:task-discrimination}

Across the three independently executed workflows, the number of tasks failed by all models decreases, while the number solved by some but not all models increases (Table~\ref{tab:task-discrimination}). When both feedback types are available, 38 tasks are solved by some but not all models, and nine remain unsolved by all 12.

\begin{table}[!htbp]
  \centering
  \small
  \caption{How many tasks are solved by all, some, or none of the models under each protocol.}
  \label{tab:task-discrimination}
  \begin{tabular}{lrrr}
    \toprule
\rowcolor{FLHeader}
Protocol & All pass & Some pass & None pass \\
    \midrule
    Full     & 6  & 35 & 20 \\
\rowcolor{FLStripe}
    Evo      & 10 & 37 & 14 \\
    Evo+Comp & 14 & 38 & 9 \\
    \bottomrule
  \end{tabular}
\end{table}


\subsection{Preserving Behavior across Later States}
\label{app:stability-results}

\begin{figure}[p]
  \centering
  \includegraphics[width=0.88\textwidth]{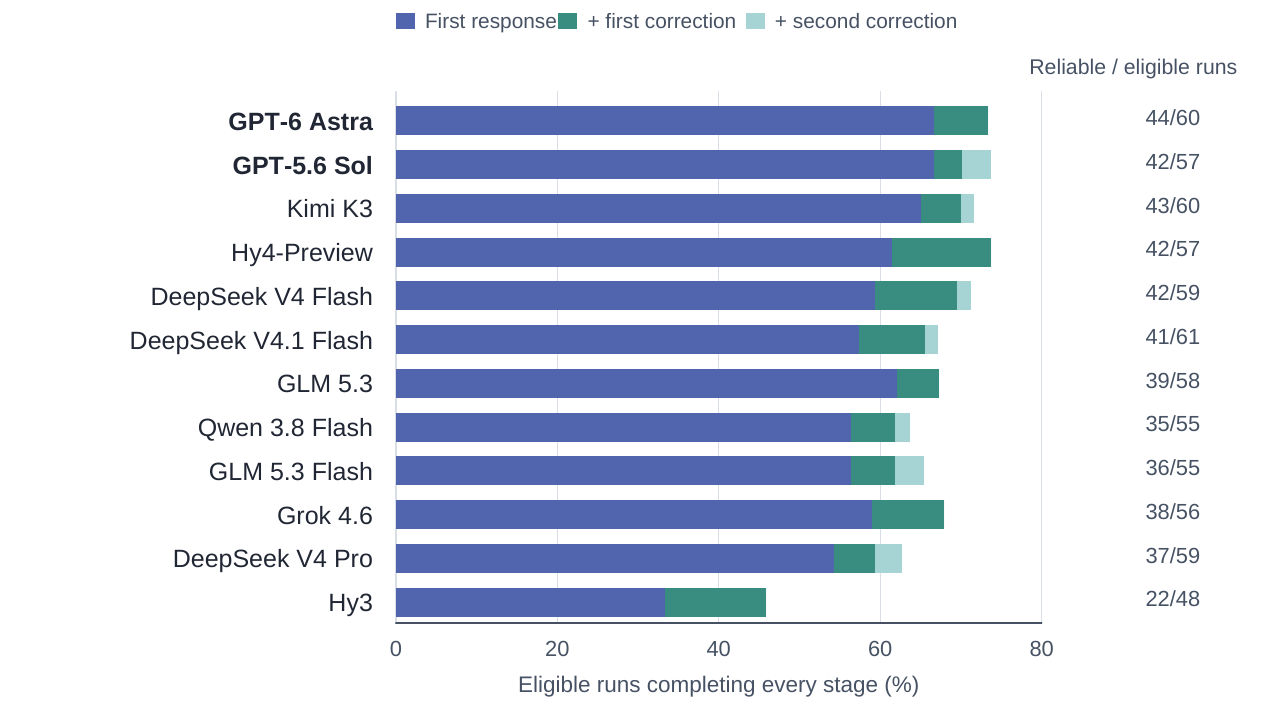}
  \caption{Trajectory reliability by model when both feedback types are available. Segments give the fraction of eligible runs that complete every cumulative stage by the first response, first corrective revision, or second corrective revision within that stage.}
  \label{fig:ser-by-model}
\end{figure}

Table~\ref{tab:ser-results} checks the same 685 eligible runs at three points within each stage: after the first response, after the first correction, and after the second. At each checkpoint, trajectory reliability requires all cumulative stages to have passed by that point in their respective revision sequences. GPT-5.6 Sol and GPT-6 Astra have the highest observed SER@0. Hy4-Preview obtains the largest first-revision gain, while DeepSeek V4 Pro, GLM 5.3 Flash, and GPT-5.6 Sol gain the most from the second revision. Figure~\ref{fig:ser-by-model} makes the small and model-dependent contribution of the second revision visible.

\begin{table}[!htbp]
  \centering
  \small
  \caption{Trajectory reliability at the first-response, first-revision, and second-revision checkpoints of the same runs. ``Runs'' counts trajectories with a usable result at every later state.}
  \label{tab:ser-results}
  \begin{tabular}{lrrrr}
    \toprule
\rowcolor{FLHeader}
Model & SER@0 & SER@1 & SER@2 & Runs \\
    \midrule
    DeepSeek V4 Flash   & 59.3 & 69.5 & 71.2 & 59 \\
\rowcolor{FLStripe}
    DeepSeek V4 Pro     & 54.2 & 59.3 & 62.7 & 59 \\
    DeepSeek V4.1 Flash & 57.4 & 65.6 & 67.2 & 61 \\
\rowcolor{FLStripe}
    GLM 5.3             & 62.1 & 67.2 & 67.2 & 58 \\
    GLM 5.3 Flash       & 56.4 & 61.8 & 65.5 & 55 \\
\rowcolor{FLStripe}
    Hy3                 & 33.3 & 45.8 & 45.8 & 48 \\
    Kimi K3             & 65.0 & 70.0 & 71.7 & 60 \\
\rowcolor{FLStripe}
    Qwen 3.8 Flash      & 56.4 & 61.8 & 63.6 & 55 \\
    GPT-5.6 Sol         & 66.7 & 70.2 & 73.7 & 57 \\
\rowcolor{FLStripe}
    GPT-6 Astra         & 66.7 & 73.3 & 73.3 & 60 \\
    Grok 4.6            & 58.9 & 67.9 & 67.9 & 56 \\
\rowcolor{FLStripe}
    Hy4-Preview         & 61.4 & 73.7 & 73.7 & 57 \\
    \midrule
    \rowcolor{FLTotal}
    Pooled              & 58.5 & 65.8 & 67.3 & 685 \\
    \bottomrule
  \end{tabular}
\end{table}

\subsection{Token Cost}
\label{app:cost-results}

\begin{figure}[!htbp]
  \centering
  \includegraphics[width=0.88\textwidth]{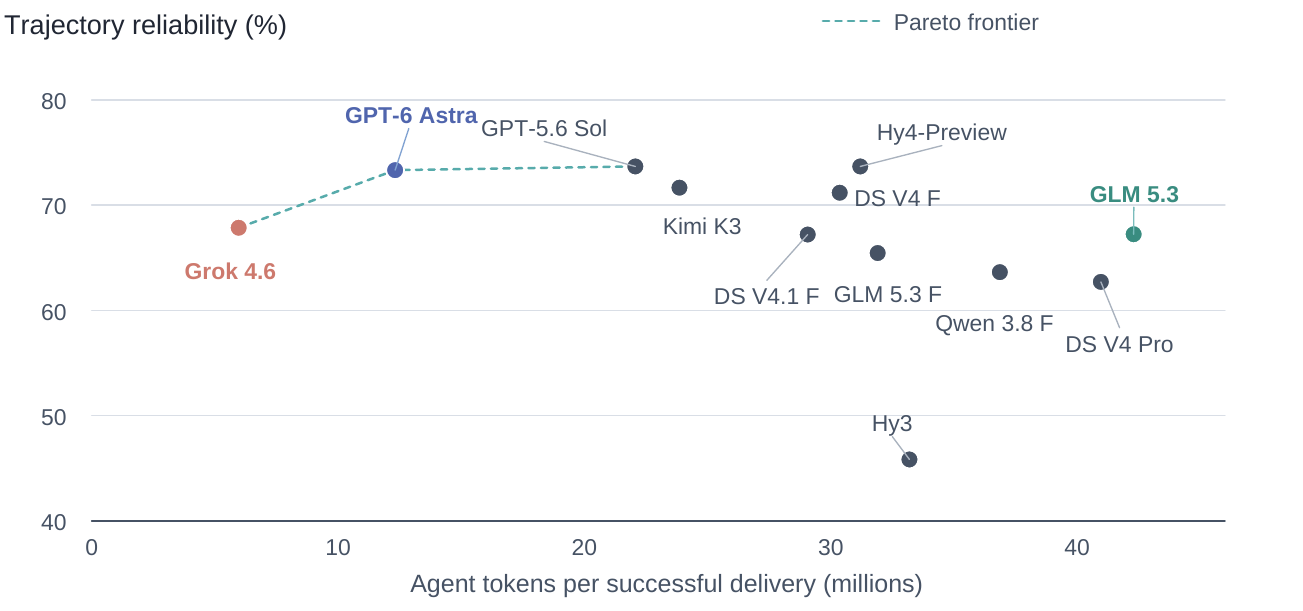}
  \caption{SER@2 versus reported tokens per successful final delivery across all models. Cost includes tokens from reported failed runs; the dashed line marks the observed Pareto frontier.}
  \label{fig:efficiency-by-model}
\end{figure}

Table~\ref{tab:token-cost} reports mean tokens per reported run and tokens per successful final delivery. GPT-6 Astra and Grok 4.6 use substantially fewer tokens per successful delivery than the remaining models. GLM 5.3 has the highest token cost when both feedback types are available, while Hy3's low mean run cost is offset by its low completion rate. Grok 4.6 and GLM 5.3 have similar SER@2 (67.9\% and 67.2\%), yet GLM 5.3 consumes about seven times as many tokens per successful delivery (Figure~\ref{fig:efficiency-by-model}).

\begin{table}[!htbp]
  \centering\small
  \setlength{\tabcolsep}{4pt}
  \caption{Token consumption in millions using only runs with reported telemetry. Each protocol separates mean tokens per reported run from total reported tokens per reported StrictPass success, including tokens spent on failures. Coverage is 726/732 for Full, 732/732 for Evo, and 731/732 for Evo+Comp. Minimum reported token costs within each column are bold.}
  \label{tab:token-cost}
  \begin{tabular}{@{}lrrrrrr@{}}
    \toprule
\rowcolor{FLHeader}
& \multicolumn{2}{c}{Full} & \multicolumn{2}{c}{Evo} & \multicolumn{2}{c}{Evo+Comp} \\
    \cmidrule(lr){2-3}\cmidrule(lr){4-5}\cmidrule(lr){6-7}
    \rowcolor{FLHeader}
    Model & Mean & / success & Mean & / success & Mean & / success \\
    \midrule
    GPT-6 Astra & \textbf{1.55} & \textbf{2.63} & 8.22 & 11.94 & 8.88 & 12.32 \\
\rowcolor{FLStripe}
    GPT-5.6 Sol & 5.19 & 9.89 & 14.18 & 22.76 & 15.56 & 22.07 \\
    Kimi K3 & 6.63 & 16.85 & 13.31 & 24.60 & 16.82 & 23.86 \\
\rowcolor{FLStripe}
    Hy4-Preview & 7.34 & 15.44 & 13.19 & 25.14 & 21.48 & 31.20 \\
    DeepSeek V4 Flash & 11.47 & 29.91 & 14.18 & 24.71 & 21.40 & 30.36 \\
\rowcolor{FLStripe}
    DeepSeek V4.1 Flash & 8.76 & 20.55 & 13.13 & 22.89 & 20.49 & 29.06 \\
    GLM 5.3 & 11.30 & 25.53 & 20.11 & 39.56 & 27.73 & 42.29 \\
\rowcolor{FLStripe}
    Qwen 3.8 Flash & 7.34 & 18.35 & 13.04 & 27.44 & 21.15 & 36.86 \\
    GLM 5.3 Flash & 6.64 & 18.42 & 12.42 & 25.26 & 19.87 & 31.90 \\
\rowcolor{FLStripe}
    Grok 4.6 & 1.93 & 6.92 & \textbf{2.29} & \textbf{6.09} & \textbf{3.82} & \textbf{5.97} \\
    DeepSeek V4 Pro & 10.09 & 27.06 & 13.83 & 28.12 & 24.58 & 40.96 \\
\rowcolor{FLStripe}
    Hy3 & 4.77 & 21.67 & 7.23 & 22.05 & 11.97 & 33.19 \\
    \midrule
    \rowcolor{FLTotal}
    \textbf{Pooled} & \textbf{6.91} & \textbf{17.00} & \textbf{12.09} & \textbf{23.42} & \textbf{17.80} & \textbf{27.81} \\
    \bottomrule
  \end{tabular}
\end{table}

\makeatletter
\setlength{\@fpsep}{8pt}
\makeatother
\subsection{Conditional Proactive Coverage and Model Profiles}
\label{app:proactive-results}
\label{sec:proactive-results}

On the strict eligible cohort, micro PCC is 44/179 (24.6\%). Averaging first within each of the 16 groups gives macro PCC of 22.9\%. Excluding mixed-semantics items changes the micro result to 24/131 (18.3\%) and macro PCC to 16.7\%. Thus this diagnostic is sensitive to cohort composition. The heavy-implementation subset covers a small, distinct mixture of requirements and repositories.

Table~\ref{tab:semantic-model-scorecard} reports the complete model profile with its actual denominators. Every model has only 14--16 usable strict proactive opportunities. The profiles distinguish early implementation from completion after new information, individual repair, and successful completion of all stated requirements. Repair denominators consist of the groups that each model failed.

\begin{table}[!htbp]
\centering\footnotesize\setlength{\tabcolsep}{4pt}\renewcommand{\arraystretch}{1.12}
\caption{Conditional model profiles under Evo+Comp. Each entry is successes/opportunities. Uptake uses the later-information cohort; repair and closure use all eligible failed later groups. Percentages can be recovered directly from the displayed counts.}
\label{tab:semantic-model-scorecard}
\begin{tabular}{@{}lrrrrr@{}}
\toprule
\rowcolor{FLHeader}
Model & PCC & Uptake@1 & Repair@2 & Closure@2 & Retention \\
\midrule
DeepSeek V4 Flash & 5/16 & 61/78 & 9/33 & 9/33 & 52/62 \\
\rowcolor{FLStripe}
DeepSeek V4 Pro & 2/15 & 63/79 & 9/43 & 7/43 & 46/53 \\
DeepSeek V4.1 Flash & 3/16 & 64/80 & 7/31 & 5/31 & 61/62 \\
\rowcolor{FLStripe}
GLM 5.3 & 6/15 & 64/79 & 5/30 & 3/30 & 51/59 \\
GLM 5.3 Flash & 5/14 & 56/75 & 8/39 & 6/39 & 46/51 \\
\rowcolor{FLStripe}
Hy3 & 0/14 & 36/67 & 13/65 & 9/65 & 27/33 \\
Kimi K3 & 4/15 & 66/83 & 5/34 & 4/34 & 56/59 \\
\rowcolor{FLStripe}
Qwen 3.8 Flash & 4/14 & 59/76 & 5/39 & 3/39 & 46/50 \\
GPT-5.6 Sol & 6/16 & 63/78 & 6/30 & 4/30 & 47/55 \\
\rowcolor{FLStripe}
GPT-6 Astra & 5/15 & 66/81 & 6/36 & 5/36 & 57/57 \\
Grok 4.6 & 1/14 & 57/75 & 10/42 & 9/42 & 33/46 \\
\rowcolor{FLStripe}
Hy4-Preview & 3/15 & 62/76 & 8/33 & 8/33 & 47/55 \\
\bottomrule\end{tabular}\end{table}

The macro-PCC analysis gives group-bootstrap intervals of 22.9\% [8.9, 38.5] for the strict cohort and 16.7\% [5.6, 29.9] without mixed items. These intervals summarize sensitivity to the sampled requirement groups. The resampling unit is the requirement group, using the finalized labels.

\end{document}